\documentclass[eqsecnum,twocolumn,tightenlines,floats,floatfix,prc]{revtex4}
\def\bea{\begin{eqnarray}}
\def\eea{\end{eqnarray}}

\def\sfrac#1#2{{\textstyle \frac{#1}{#2}}}

\def\bal{\begin{align}}
\def\eal{\end{align}}

\usepackage{graphics}
\usepackage{graphicx}
\usepackage{amsmath}
\usepackage{amssymb}
\usepackage{slashed}
\usepackage{longtable}
\usepackage{bm}

\def\J#1#2#3#4{ {#1} {\bf #2}, #3 (#4).}

\def\EPJC{{\it Eur.\ Phys.\ J\/} C}
\def\FBS{\it Few Body Syst.\/}

\def\NCL{\it Nuovo Cimento Lett.\/}

\def\NIM{\it Nucl.\ Instrum.\ Methods\/}
\def\NIMA{{\it Nucl.\ Instrum.\ Methods\/} A}
\def\NPA{{\it Nucl.\ Phys.\/} A}
\def\NPB{{\it Nucl.\ Phys.\/} B}
\def\PLB{{\it Phys.\ Lett.\/}  B}

\def\PR{\it Phys. Rev.\/}
\def\PRL{\it Phys.\ Rev.\ Lett.\/}
\def\PRD{{\it Phys.\ Rev.\/} D}
\def\PRC{{\it Phys.\ Rev.\/} C}

\def\ZPA{{\it Z.\ Phys.\/} A}

\def\PPSLA{\it Proc.\ Phys.\ Soc.\ London, Sec.\ A\/}
\def\ANPL{\it Ann.\ Phys.\ (Leipzig)\/} 
\def\JPSJ{\it J.\  Phys.\  Soc.\ Jpn.\/} 
\def\PPSL{\it Proc.\ Phys.\ Soc.\ London\/}
\def\NP{\it Nucl.\  Phys.\/}

\def\SPJETP{\it Sov.\ Phys.\ JETP\/}
\def\PS{\it Phys.\ Scripta\/}
\def\JP{\it J.\ de Physique\/}
\def\PM{\it Phil.\ Mag.\/}
\def\PRSA{\it Proc.\ Royal Soc.\ London, Sec.\ A\/}

\def\et{{\it et.al.}}

\begin{document} 


\phantom{0}
\hspace{5.5in}\parbox{1.5in}{ \leftline{JLAB-THY-08-777}
                \leftline{WM-08-101}
                \leftline{}\leftline{}\leftline{}\leftline{}
}
\title
{\bf Covariant spectator theory of $np$ scattering:\\  Phase shifts obtained from precision fits to data below 350 MeV }

\author{Franz Gross$^{1,2}$ and Alfred Stadler$^{3,4}$
\vspace{-0.1in}  }

\affiliation{
$^1$College of William and Mary, Williamsburg, Virginia 23185 \vspace{-0.15in}}
\affiliation{
$^2$Thomas Jefferson National Accelerator Facility, Newport News, VA 23606 \vspace{-0.15in}}
\affiliation{
$^3$Centro de F\'\i sica Nuclear da Universidade de Lisboa, 1649-003 Lisboa, Portugal,\vspace{-0.15in}}
\affiliation{
$^4$Departamento de F\'\i sica da Universidade de \'Evora, 7000-671 \'Evora, Portugal
}


\date{\today}

\begin{abstract} 

Using the covariant spectator theory (CST), we present two one boson exchange kernels that have been successfully adjusted to fit the  2007 world $np$  data (containing 3788 data)  below 350 MeV.   One model (which we designate WJC-1) has 27 parameters and  fits with a $\chi^2/N_\mathrm{data} = 1.06$.  The other model  (designated WJC-2)  has only 15 parameters and fits with a $\chi^2/N_\mathrm{data} = 1.12$.  Both of these models  also reproduce the experimental triton binding energy without introducing additional irreducible three-nucleon forces.    One result of this work is a new phase shift analysis, updated for all data until 2006, which is useful even if one does not work within the CST.  In carrying out these fits we have reviewed the entire data base, adding new data not previously used in other high precision fits  and restoring some data omitted in previous fits.  A full discussion and evaluation of the 2007 data base is presented.

\end{abstract}
 
\phantom{0}

\maketitle


\section{Introduction} 

This paper presents many details and new results from a recent application \cite{Gross:2007be} of the covariant spectator theory (CST) \cite{Gross:1969rv,
Gross:1982ny} to the description of low energy neutron-proton ($np$) scattering.   In this work the parameters of generalized one-boson-exchange (OBE) models are adjusted to obtain precision fits to the $np$ scattering data for lab energies $E_{\rm lab} \le$ 350 MeV.  The  OBE models fixed by the fits give a simple, manifestly covariant description of the nuclear force, a necessary starting point for the computation of many properties of interacting few-body systems.  These models will be particularly useful for the description of interactions where the two nucleon system has low relative momentum but recoils at GeV energies; in these cases a covariant approach based on a fit to low energy data is both necessary and effective.  Furthermore, following the procedure of Ref.~\cite{Gross:1987bu}, exchange currents consistent with these   OBE models can be easily determined and conserved currents defined.  With these extensions these models can be applied to the description of the electromagnetic interactions studied at Jefferson Laboratory  and elsewhere.

A brief overview of the theory is presented in Sec.~\ref{sec:overview}, where the parameters of the class of OBE models considered in this work are defined.  CST models of this type were first applied to the quantitative description of $np$ scattering in 1992 \cite{GVOH}, and except for a few important differences the theory is unchanged.  Details of the theory are reviewed in Appendices.  

We present two models motivated by quite different philosophies.  Both fit the data very well. The first, WJC-1 with  27 adjustable parameters, gives a high precision fit with a $\chi^2/N_{\rm data}=1.06$.  Here we allowed the masses of the heavy bosons and most of the coupling constants to vary in order to obtain the best fit possible.  For the second, WJC-2, we simplified the model as much as possible by fixing some of the meson masses and eliminating some of the less important degrees of freedom.  The goal was to see how good a fit could be achieved with {\it only\/} 15 essential parameters.   This fit  was less precise but still remarkably good, giving a  $\chi^2/N_{\rm data}=1.12$.   In Sec.~\ref{Sec:III} we compare the quality of these fits  to the 1993 Nijmegen phase shift analysis \cite{Stoks:1993tb}, the 1995 Argonne AV18 potential \cite{Wir95}, and the 2001 CD-Bonn potential \cite{Mac01}.  The data base we use is listed and discussed in Sec.~\ref{Sec:IV}.  It includes data from the original Nijmegen \cite{Stoks:1993tb} and Bonn analyses \cite{Mac01}, as well as additional data from the SAID on-line data base \cite{GWU}, the Nijmegen NN-OnLine data base \cite{NNonline}, and a few sets we have collected ourselves.  This data base  is completely up-to-date, including more data than used in any previous analysis.  The $\chi^2$ we obtain for Model WJC-1 is as good as other high precision fit, and both models require fewer parameters than  ever used before.  

To obtain such ``perfect'' fits it is necessary to reject certain sets of measurements that seem to be inconsistent with the bulk of the data.  We use a statistical selection criteria first introduced by the Nijmegen group \cite{Ber88}, and these are reviewed and discussed in detail in Sec.~\ref{Sec:IV}.  We show, using specific examples, how these selection criteria work.  Data reported to have systematic errors can be scaled during the fits, and we give an example of the impact of this scaling.

The phase shifts obtained from the fit are given and discussed in Sec.~\ref{Sec:V}.  We find significant difference between our phases and the famous Nijmegen phases \cite{Stoks:1993tb} obtained from the 1993 analysis.  

The CST has also been used to calculate the three-body wave function and the triton binding energy \cite{Sta97,Stadler:1997iu}.  (We have not yet included the Coulomb part of the $pp$ interaction, and hence cannot calculate $pp$ scattering or the binding energy of $^3$He).  In 1997, using a family of less precise models,  we found \cite{Sta97} that the correct triton binding energy emerged {\it automatically\/} from the model that gave the best fit to the two-body data, requiring no new mechanisms or assumptions.  In Sec.~\ref{Sec:VI} we show how this remarkable result continues to hold for these new high precision models, suggesting that it is a robust feature of the CST.  Sec.~\ref{Sec:VII} presents our conclusions. 

Details of the theory and the models have been developed in several long Appendices, which also review and  compile many results reported previously.  Appendix \ref{App:A0} gives a short introduction to all of Appendices.   Appendix \ref{App:A} discusses some of the implications of the CST prescription that one particle is on-shell.  We show there that (i) the equations satisfy the generalized Pauli principal, even though the equations {\it appear\/} to treat the two identical particles differently (because only one particle is on-shell), (ii) the equations give the {\it same\/} answer for the fully on-shell scattering amplitude, independent of which particle is on-shell (the convention used here is to place particle 1 on-shell), and (iii)  the new prescription used in this paper for removing spurious singularities from the kernel is simple and effective.  Appendix \ref{App:B} shows that the OBE models used in this paper are able to reproduce the spin and isospin structure of the most general {\it on-shell\/} $NN$ kernel, explaining why bosons of spin 2 and larger are not needed.  Appendix \ref{App:C} discusses the role of the nucleon form factor in removing (spurious) deeply bound states from the theory, and Appendix \ref{App:D} gives a detailed review of the helicity, angular momentum expansions, and symmetry relations used to reduce the equations to the simple form used for numerical solutions.

\section{Overview of the Theory} \label{sec:overview}

In the CST \cite{Gross:1969rv,
Gross:1982ny}, the two-body scattering amplitude $M$ is the solution of a covariant integral equation derived from field theory (sometimes referred to as the ``Gross equation'').  In common with many other equations, it has the form
\bea
M=V-VGM
\eea
where $V$ is the irreducible kernel (playing the role of a potential) and $G$ is the intermediate state propagator.  As with the Bethe-Salpeter (BS) equation \cite{Salpeter:1951sz}, if the kernel is exact and nucleon self energies are included in the propagators, iteration of  the CST equation generates the full Feynman series.  In cases where this series does not converge (nearly always!) the equation solves the problem nonperturbatively.  With the BS equation the four-momenta of all $A$ intermediate particles are subject only to the conservation of total four-momentum $P=\sum_{i=1}^A p_i$, so the integration is over $4(A-1)$ variables.  In the CST equation, all but one of the intermediate particles are restricted to their positive-energy mass shell, constraining $A-1$ energies (they become functions of the three-momenta) and leaving only $3(A-1)$ internal variables, the {\it same number\/} of variables as in nonrelativistic theory. Since the on-shell constraints are covariant, the resulting equations remain manifestly covariant even though all intermediate loop integrations reduce to three dimensions, which greatly simplifies their numerical solution and physical interpretation. This framework has been applied successfully to many problems, in particular also to the two- and three-nucleon system \cite{GVOH,Sta97,Stadler:1997iu}.

\begin{figure}
\centerline{
\mbox{
\includegraphics[width=3in]{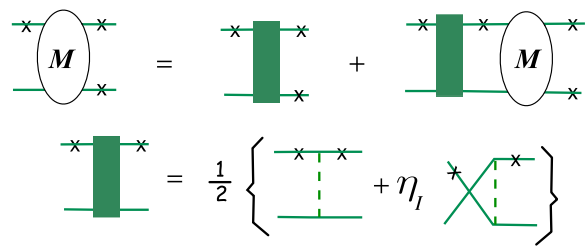}
}
}
\caption{\footnotesize\baselineskip=10pt (Color on line) Top line: diagramatic representation of the Covariant Spectator equation (\ref{eq:spec}) with particle 1 on-shell (the on-shell particle is labeled with a $\times$).  Second line: diagrammatic representation of the definition of the antisymmetrized kernel (\ref{eq:kernel}).}
\label{fig:spec}
\end{figure}


The specific form of the CST equation for the two-nucleon scattering amplitude $M$, with particle 1 on-shell in both the initial and final state, is derived in Ref.~\cite{GVOH} (referred to as Ref.~I below) and illustrated in Fig.~\ref{fig:spec}.  The equation is
\begin{align}
&M_{12}(p, p'; P)=
\overline {V}_{12}(p, p'; P)\cr
&-\int \frac{d^3 k}{(2\pi)^3} \frac{m}{E_k} 
\overline {V}_{12}(p,k;P)G_2(k,P)
{M}_{12}(k,p';P)\, ,\qquad \label{eq:spec}
\end{align}
where $P$ is the conserved total four-momentum, and $p, p'$, and $k$ are relative four-momenta related to the momenta of particles  1 and 2 by
$p_1=\sfrac12 P+p$, $p_2=\sfrac12 P-p$, $E_k=\sqrt{m^2+k^2}$ is the energy of the on-shell particle 1 in the cm system, and
\begin{align}
&M_{12}(p, p'; P)\equiv M_{\lambda\lambda',\beta\beta'}(p, p'; P)\cr
&\qquad=\bar u_\alpha({\bf p},\lambda){\cal M}_{\alpha\alpha';\beta\beta'}(p, p'; P)u_{\alpha'}({\bf p'},\lambda')\qquad
\end{align}
is the matrix element of the Feynman scattering amplitude ${\cal M}$ between positive energy Dirac spinors of particle 1.  The definitions of the nucleon spinors $u(p,\lambda)$ (with $\lambda$ the helicity of the nucleon) and the partial wave decomposition of the amplitude $M_{12}$ and $\overline {V}_{12}$      are given in Appendix \ref{App:D}.  The propagator for the off-shell particle 2 is
\bea
G_2(k,P)\equiv G_{_{\beta\beta'}}(k_2)=\frac{\left(m+\slashed{k}_2\right)_{_{\beta\beta'}}}{m^2-k_2^2-i\epsilon}\,H^2(k_2)
\eea
with $k_2=P-k_1$, $k_1^2=m^2$, and $H$ the form factor of the off-shell nucleon (related to its self energy), normalized to unity when $k_2^2=m^2$.  In this paper we use
\bea
H(p)=\left[\frac{(\Lambda_N^2-m^2)^2}{(\Lambda^2_N-m^2)^2+(m^2-p^2)^2}\right]^2\, .
\label{nuclff}
\eea
See Appendix \ref{App:C} for further discussion of the nucleon form factor $H$.
The indices 1 and 2 refer collectively to the two helicity or Dirac indices of particle 1, either  $\{\lambda\lambda'\}$ or $\{\alpha\alpha'\}$, and particle 2, $\{\beta\beta'\}$.

The covariant kernel $\overline V$ is explicitly antisymmetrized, as illustrated in the second line of Fig.~\ref{fig:spec}.  In its Dirac form it is
%
\begin{align}
&{\overline V}_{\alpha\alpha';\beta\beta'}(p,k;P)\cr
&=\sfrac12
\left[ V_{\alpha\alpha';\beta\beta'}(p,k;P)+\eta_I V_{\beta\alpha';\alpha\beta'}(-p,k;P)
\right] \, ,\qquad \label{eq:kernel}
\end{align}
%
%
where  the factor $\eta_I=\zeta(-)^{I+1}$ (with $I$=0 or 1 the isospin of the $NN$ state) accounts for the sign change due to the exchange of the isospin indices (which are suppressed in these formulae), and $\zeta=1$ for bosons and $-1$ for fermions. Hence, for fermions, the remaining amplitude has the symmetry $\eta_I=(-)^I$ under particle interchange $\{p_1,\alpha\}\leftrightarrow\{p_2,\beta\}$ as required by the generalized Pauli principle.  This symmetry insures that identical results emerge if a different particle is chosen to be on-shell in either the initial or final state.  Some details of the construction of this equation can be found in Appendix \ref{App:A}.

It is assumed that the kernel can be written as a sum of OBE contributions 
\bea
 V_{\alpha\alpha';\beta\beta'}(p,k;P)=\sum_b V^b_{12}(p,k;P) \label{2.7}
\eea
with individual boson contributions of the form
\bea
V^b_{12}(p,k;P)=\epsilon_b\delta\frac{\Lambda^b_1(p_1,k_1)\otimes \Lambda^b_2(p_2,k_2)}{m_b^2+|q^2|} f(\Lambda_b,q)
\label{OBE}
\eea
with $b=\{s, p, v, a\}$ denoting the boson type, $q=p_1-k_1=k_2-p_2=p-k$ the momentum transfer, $m_b$ the boson mass, $\epsilon_b$ a phase factor, and $\delta=1$ for isoscalar bosons and $\delta=\tau_1\cdot\tau_2=-1-2(-)^I$ for isovector bosons.  All boson form factors, $f$, have the simple form
\bea  
f(\Lambda_b,q)=\left[\frac{\Lambda_b^2}{\Lambda_b^2+|q^2|}\right]^4 \label{mesonff}
\eea
with $\Lambda_b$ the boson form factor mass.  The use of the absolute value $|q^2|$ 
amounts to a covariant redefinition of the propagators and form factors in the region $q^2>0$.  It is a significant new theoretical improvement that removes all singularities and can be justified by a detailed study of the structure of the  exchange diagrams, as discussed in detail in Appendix \ref{App:A3}.    The axial vector bosons are treated as contact interactions, with the structure as in (\ref{OBE}), but with the propagator replaced by a constant, $m_a^2+|q^2|\to m^2$, where the nucleon mass sets a convenient scale not related to a boson mass (the effective boson mass in a contact interaction is infinite).
The explicit forms of the numerator functions $\Lambda^b_1\otimes \Lambda^b_2$ can be inferred from Table \ref{tab:Ls}.
Note that $\lambda_p=0$ corresponds to pure pseudovector coupling, and that the definitions of the off-shell coupling parameters $\lambda$ or $\nu$ differ for each boson.

\begin{table}
\begin{minipage}{3.2in}
\caption{Mathematical forms of the $bNN$ vertex functions, with $\Theta(p)\equiv(m-\slashed{p})/2m$.
The vector propagator is $\Delta_{\mu\nu}=g_{\mu\nu}-q_\mu q_\nu/m_v^2$ with the boson momentum $q=p_1-k_1=k_2-p_2$.}
\label{tab:Ls}
\begin{tabular}{lccl}
$J^P (b)\quad$ & $\epsilon_b$& $\quad\Lambda_1\otimes\Lambda_2\quad$  & $\Lambda(p,k)$ or $\Lambda^\mu(p,k)$ \cr
\tableline
$0^+ (s)$ & $-$& $ \Lambda_1 \Lambda_2$ &
$g_s-\nu_s\left[\Theta(p)+\Theta(k)\right] $\cr
$0^- (p)$ & + &$ \Lambda_1 \Lambda_2$ &
$g_p\gamma^5$\cr
&&&$-g_p(1-\lambda_p)\left[\Theta(p)\gamma^5+\gamma^5\Theta(k)\right]$\cr
$1^- (v)$ & $+$ & $\Lambda_1^\mu \Lambda_2^\nu \Delta_{\mu\nu}$
& $g_v\left[\gamma^\mu +  \frac{\kappa_v}{2M}i\sigma^{\mu\nu}(p-k)_\nu\right]$\cr
&&&$+g_v\nu_v \left[\Theta(p)\gamma^\mu  +  \gamma^\mu \Theta(k) \right]$\cr
$1^+ (a)$ & $+$  & $\Lambda_1^\mu \Lambda_2^\nu g_{\mu\nu}$ & $g_a\gamma^5\gamma^\nu$
\end{tabular}
\end{minipage}
\end{table}

\begin{figure}[b]
\centerline{
\mbox{
\includegraphics[width=3in]{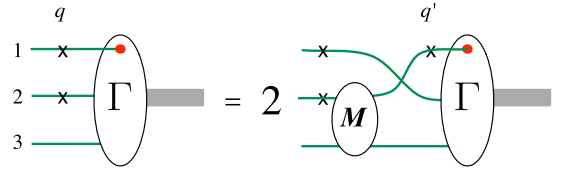}
}
}
\caption{\footnotesize\baselineskip=10pt (Color on line) Diagramatic representation of the Covariant Spectator equation for the three-body bound state vertex function $\Gamma$  with particles 1 and 2 on-shell (labeled with a $\times$).  Here particle 1 is the spectator to the last two-body interaction between particles 2 and 3, described by the scattering amplitude $M$ with particle 3 off-shell.  The spectator has three momentum $q$ after the two-body interaction and $q'$ before.}
\label{fig:3bspec}
\end{figure}


In the most general case the kernel is the sum of the exchange of pairs of pseudoscalar, scalar, vector, and axial vector bosons, with one isoscalar and one isovector meson in each pair.   If the external particles are all on-shell, we show in Appendix \ref{App:B}  that these 8 bosons give the {\it most general\/} spin-isospin structure possible (because the vector mesons have both Dirac and Pauli couplings, the required 10 invariants can be expanded in terms of only 8 boson exchanges), explaining why bosons with more complicated quantum numbers are not required.   Model WJC-1 allows the boson masses (except the pion) to vary, letting the data fix the best mass for each boson in each exchange channel.   Finally, charge symmetry is broken by treating charged and neutral pions independently, and by adding a one-photon exchange interaction, simplified by assuming the neutron coupling is purely magnetic, $i\sigma^{\mu\nu}q_\nu$,  and  that the remaining electromagnetic form factors $F_1$ and $F_2$ have the dipole form. To solve the CST $NN$ equation numerically, it  was expanded in a basis of partial wave helicity states as described in Ref.~I and Appendix \ref{App:D}.

The three-body CST equation, derived in Refs.~\cite{Gross:1982ny,Stadler:1997iu} and first solved numerically in Ref.~\cite{Sta97}, is illustrated in Fig.~\ref{fig:3bspec}.  Once the two-body amplitude is determined, the three-body vertex function and the three-body binding energy can be calculated {\it without any new parameters\/}.  

The best, short summary of the three-body theory can be found in Re.~\cite{Sta97}.  Here we wish to draw attention to only one feature of this theory.  Since the spectator (particle 1 in this case) is on-shell, the relativistic mass $W$ of the interacting two-body subsystem depends on $q$, the magnitude of the spectator three-momentum, through the relation
\bea
W^2=M_t^2+m^2-2M_t E_{q}\, ,
\eea
where $E_{q}$ is the spectator energy in the three-body rest system and $M_t$ is the triton mass.
Note that this mass is zero at the critical momentum
\bea
q_{\rm crit}=\sqrt{\frac{(M_t^2+m^2)^2}{4M_t^2}-m^2}Ê\simeq \frac{4}{3} m
\eea
where the later relation holds approximately because $M_t\simeq3m$.  Initially, as suggested by Fig.~\ref{fig:3bspec}, the spectator momentum is integrated over all possible values from 0 to $\infty$, and for $q'>q_{\rm crit}$ this would require knowledge of the two-body scattering amplitude in space-like regions where $W^2<0$.  This is surely beyond the region where the OBE description could be taken seriously. 

Fortunately the spectator theory presents its own solution to this problem.  As the spectator momentum $q$ approaches the critical value $q_{\rm crit}$ and the mass $W$ of the two-body system approaches zero, it can be shown \cite{Stadler:1997iu} that the three-body vertex function $\Gamma$ goes to zero as a high power of $W$,  providing a natural cutoff  that insures that the contributions from the region $W^2\simeq0$ (where $q$ is close to $q_{\rm crit}$) are very small.  In  applications the integral over $q$, initially extending from $0\to\infty$, is approximated by the covariant integral  over the finite interval $[0,q_{\rm crit}]$.  We will study these features in more detail in Appendix \ref{App:C}.

\begin{table}
\begin{minipage}{3.5in}
\caption{Values of the 27 parameters  for WJC-1 with 7 bosons and 2 axial vector contact interactions.  All masses and energies are in MeV; other couplings are dimensionless;  $G_b=g_b^2/(4\pi)$.  Parameters in {\bf bold} were varied during the fit; those labeled with an $^*$ were  constrained to equal the one above.  The triton binding energy is $E_t$ (with its experimental  value in parentheses). }
\label{tab:par1}
\begin{tabular}{lcccrcc}
$b$ & I & $\quad G_b\quad$ & $m_b$ & $\lambda_b$ or $\nu_b$  & $\quad\kappa_v\quad$ & $\Lambda_b$\cr
\tableline
$\pi^0$ & $\quad1\quad$ & {\bf 14.608}&134.9766 &{\bf 0.153}$\,\,$& --- &{\bf 4400}\cr
$\pi^\pm$ & $1$ & {\bf 13.703} &139.5702 & ${\bf -0.312}$$\,\,$& --- &4400$^*$\cr
$\eta$ & $0$ & {\bf 10.684} & {\bf 604} & ${\bf 0.622}$$\,\,$& --- &4400$^*$\cr
$\sigma_0$ & $0$ & {\bf 2.307} & {\bf 429} & ${\bf -6.500}$$\,\,$& --- &{\bf 1435}\cr
$\sigma_1$ & $1$ & {\bf 0.539} & {\bf 515} & ${\bf 0.987}$$\,\,$& --- &1435$^*$\cr
$\omega$ & $0$ & {\bf 3.456} & {\bf 657} & ${\bf 0.843}$$\,\,$& ${\bf 0.048}$ &{\bf 1376}\cr
$\rho$ & $1$ & {\bf 0.327} & {\bf 787} & ${\bf -1.263}$\,\,$$& ${\bf 6.536}$ &  1376$^*$ \cr
$h_1$ & $0$ & {\bf 0.0026} & --- & ---$\,\,\,\,\,\,$& ---$\,\,$ &1376$^*$\cr
$a_1$ & $1$ & ${\bf -0.436}$ & --- & ---$\,\,\,\,\,\,$& ---$\,\,$ &1376$^*$\cr
\tableline
 \multicolumn{7}{c}{$\Lambda_N={\bf 1656};\; E_t=-8.48\, (-8.48)$}\end{tabular}
\end{minipage}
\end{table}

\begin{table}
\begin{minipage}{3.5in}
\caption{Values of the 15 parameters for  WJC-2 with 7 bosons.  See the caption to Table \ref{tab:par1} for further explanation. }
\label{tab:par2}
\begin{tabular}{lcccrcc}
$b$ & I & $\quad G_b\quad$ & $m_b$ & $\lambda_b$ or $\nu_b$  & $\quad\kappa_v\quad$ & $\Lambda_b$\cr
\tableline
$\pi^0$ & $\quad1\quad$ & {\bf 14.038}&134.9766 & 0.0& --- &{\bf 3661}\cr
$\pi^\pm$ & $1$ &  14.038$^*$&139.5702 & $0.0$& --- &3661$^*$\cr
$\eta$ & $0$ & {\bf 4.386} & 547.51 & $0.0$& --- &3661$^*$\cr
$\sigma_0$ & $0$ & {\bf 4.486} & {\bf 478} & ${\bf -1.550}$& --- &3661$^*$\cr
$\sigma_1$ & $1$ & {\bf 0.477} & {\bf 454} & ${\bf 1.924}$& --- &3661$^*$\cr
$\omega$ & $0$ & {\bf 8.711} & 782.65 & $0.0$& $0.0$ &{\bf 1591}\cr
$\rho$ & $1$ & {\bf 0.626} & 775.50 & ${\bf -2.787}$& ${\bf 5.099}$ &  1591$^*$ \cr
\tableline
 \multicolumn{7}{c}{$\Lambda_N={\bf 1739};\; E_t=-8.50\, (-8.48)$}
\end{tabular}
\end{minipage}
\end{table}

\begin{table}
\begin{minipage}{3.5in}
\caption{Comparison of precision $np$ models and the 1993 Nijmegen phase shift analysis.  Our calculations are in bold face.  Number of data used in each fit is in parentheses.}
\label{tab:1}
\begin {tabular}{lcc|ccc} \multicolumn{3}{c}{models}&\multicolumn{3}{c}{$\chi^2/N_\mathrm{data}(N_\mathrm{data})$}\cr
\tableline
Reference & \#\footnote[1]{Number of parameters} & year\footnote[2]{Includes all data prior to this year. } & $\;\;$1993$\;\;$ & 2000& 2007 \cr
\tableline
PWA93\cite{Stoks:1993tb} &39\footnote[3]{For a fit to both $pp$ and $np$ data.} &1993 &  0.99(2514) & --- & ---\cr
&&&{\bf 1.09}\footnote[4]{Our fitting procedure uses the effective range expansion.  The Nijmegen $^3S_1$ parameters were taken from Ref.\ \cite{deSwart:1995ui}, but  as no $^1S_0$ parameters are available we used those of WJC-1.}(3010)&{\bf 1.11}(3336) &{\bf 1.12}(3788)\cr
Nijm I\cite{Sto94}&41\footnotemark[3]&1993&1.03\footnotemark[3](2514)&---&---\cr
AV18\cite{Wir95} & 40\footnotemark[3] & 1995 &1.06(2526) & --- & --- \cr
CD-Bonn\cite{Mac01} &43\footnotemark[3] & 2000 & --- &1.02(3058)&--- \cr
WJC-1& 27$\;$ & 2007 &{\bf 1.03}(3010)& {\bf 1.05}(3336) & {\bf 1.06}(3788) \cr
WJC-2 & 15$\;$ & 2007 & {\bf 1.09}(3010)& {\bf 1.11}(3336) & {\bf 1.12}(3788)
\end{tabular}
\end{minipage}
\end{table}


\section{Meson parameters and quality of the fits} \label{Sec:III}

Previous models of the kernel, such as models IA, IB, IIA, and IIB of Ref.~I \cite{GVOH} and the updated, $\nu$-dependent versions such as W16 used in \cite{Sta97}, had been obtained by fitting the potential parameters to the Nijmegen or VPI phase shifts. In a second step the $\chi^2$ to the observables was determined. The models presented in this paper were fit directly to the data, using a minimization program that can constrain two of the low-energy parameters  (the deuteron binding energy, $E_d=-2.2246$ MeV, and the $^1S_0$ scattering length, $a_0=-23.749$ fm, chosen to fit the very precise cross sections at near zero lab energy).  This was a significant improvement, both because the best fit to the 1993 phase shifts did not guarantee a best fit to the 2007 data base, and because the low-energy  constraints stabilized the fits.

The three-body binding energy is very sensitive to the off-shell coupling of the sigma meson, $\nu_\sigma$, but it turns out that the value of $\nu_\sigma$ determined by the best fit to the two-body data also gives an essentially perfect fit to the triton binding energy, as shown in Sec.~\ref{Sec:VI}.  This confirms the result first reported in Fig.~1 of Ref.~\cite{Sta97}.

The parameters obtained in the fits are shown in Tables \ref{tab:par1} and \ref{tab:par2}.  The $\chi^2/N_\mathrm{data}$  resulting from the fits are compared with results obtained from earlier fits in Table \ref{tab:1}.  The data base used in the fits is derived from the previous SAID \cite{GWU,SAID} and Nijmegen \cite{Stoks:1993tb}  analyses with some new data added. The current data set includes a total of 3788 data, 3336 of which are prior to 2000 and  3010 prior to 1993.  For comparison, the PWA93 was fit to 2514, AV18 to 2526, and CD-Bonn to 3058 $np$ data.   We restored some data sets  previously discarded because their $\chi^2$ were no longer outside of statistically acceptable limits, and this increased the $\chi^2$ slightly.  A full discussion of the data and our selection criteria are given in Sec.~\ref{Sec:IV}.  

\begin{figure*}
\centerline{
\mbox{
\includegraphics[width=5.in]{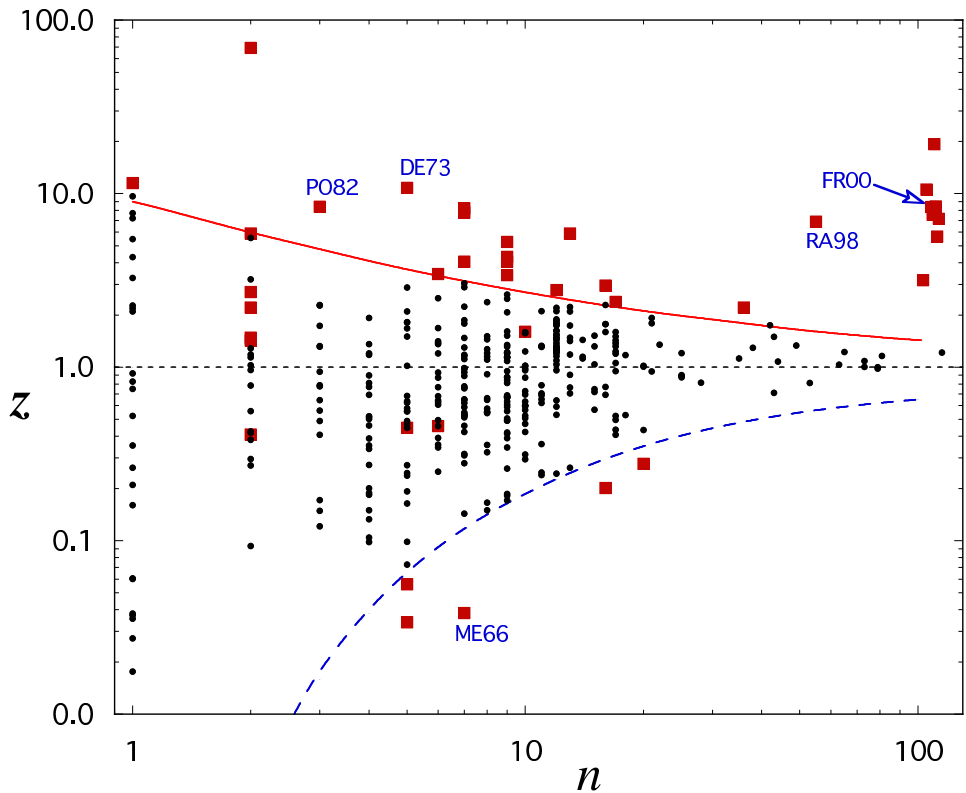}
}
}
\caption{\footnotesize\baselineskip=10pt (Color on line) Log-log scatter plot showing the value of $z=\chi^2/n$ for each data set with $n$ data.  The sets that are retained are represented by small circles; those rejected by larger boxes.  The $z_{\it max}$ (solid line) and $z_{\it min}$ (dashed line) limits given by Eq.~(\ref{zlimits}) are shown.  The five labeled data sets are discussed in Figs.~\ref{fig:cross},  \ref{fig:162}, and \ref{fig:320}.}
\label{fig:scatter}
\end{figure*} 

In both of our models the high-momentum cutoff is provided by the {\it nucleon\/} form factor and not the meson form factors.  Hence the very hard pion form factors merely reflect the fact that the nucleon form factors are sufficient to model the short range physics in the pion exchange channel.  The off-shell scalar couplings are perhaps the most uncommon features of these models.  They are clearly essential for the accurate prediction of three-body binding energies \cite{Sta97}.  It is gratifying to see that the pseudoscalar components of the pion couplings (proportional to $\lambda_p$) remain close to zero, even when unconstrained, and that effective masses of all the bosons remain in the expected range of 400-800 MeV.  

Aside from this, the parameters of WJC-2 are quite close to values expected from older OBE models of nuclear forces.  A possible exception is  the pion coupling constant, somewhat larger than the $g^2/(4\pi)=13.567$ found by the Nijmegen group.  The high-precision Model WJC-1 shows some novel features: (a) $g_{\pi^0}>g_{\pi^\pm}$, (b) large $g_\eta$, and (c) small $g_\omega$.   

During the fits we did not restrict the signs of $G_b=g_b^2/(4\pi)$, and the fact that they turn out to be positive is an important prediction of the OBE model.  The exception was the strength of the $a_1$ ``meson'' in Model WJC-1.  Since $G_a<0$, this requires reinterpreting this ``exchange'' as a contact interaction (allowed within the general framework of an OBE model) with its sign not fixed by theory.  This approach was further supported by the discovery that allowing the axial vectors to have finite masses did not significantly improve the fits.

Why do these OBE models work so well? We are reminded of the Dirac equation; it automatically includes the $p^4/(8m^3)$ energy correction that contributes to fine structure, the Darwin term (including the Thomas precession), the spin-orbit  interaction, and the anomalous gyromagnetic ratio.   Similarly, the CST automatically generates  relativistic structures hard to identify, and impossible to add to a nonrelativistic model without new parameters.

\begin{figure}
\centerline{
\mbox{
\includegraphics[width=3in]{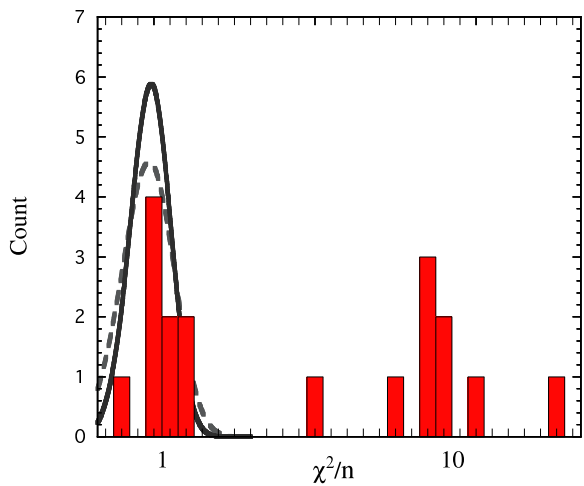}
}
}
\caption{\footnotesize\baselineskip=10pt (Color on line) Distribution in $\chi^2/n$ of data sets with $n> 50$.  The continuous curves are the theoretical distributions (normalized to this number of sets) with $n=100$ (solid curve) and $n=60$ (dashed curve).  (Both curves are given in order to show the dependence of the theoretical distributions (\ref{theordis}) on $n$.)}
\label{fig:distribution}
\end{figure} 

\section{Selection of Data} \label{Sec:IV}

The data used in the fits were originally obtained from R.~A.~Arndt's SAID program \cite{SAID}, kept up-to-date by the George Washington University \cite{GWU} $NN$ on-line data base.  These were then compared with the data tables used in the 1993 Nijmegen phase shift analysis \cite{Stoks:1993tb}, with the additional data used by CD-Bonn \cite{Mac01}, and with the Nijmegen group's on-line data base \cite{NNonline}.   We also added a few data sets that had either been overlooked, or were too recent to be included in any of these other data sets.  We discussed details of the data selection and rejection (and other issues) with several members of the Nijmegen group \cite{Nijdiscussions}.  We believe that our new 2007 data set is the most complete available at the present time.  

The full data file included some data that were never published in refereed journals, and, following the accepted practice,  these were excluded from consideration right from the beginning.  The set of published data includes 3788  data used in our fits, listed in Table \ref{tab:datakeep}, and an additional 1180 published data that we did not use, listed in Table \ref{tab:dataskip}.   There are two principal reasons for excluding published data.   Some data were extracted from deuteron or other few-body targets and might be subject to unknown theoretical errors associated with this extraction.  These data are labeled with a ``c'' in the comment column of Table \ref{tab:dataskip}.  In agreement with previous practice these data were excluded;  fortunately the data set is now so complete that it is no longer necessary to use such data.  Other data have improbably large (or small) statistical errors (i.e. $\chi^2$), and following the practice first introduced by the Nijmegen group this data  is also excluded.

Considerable time and effort was spent examining this last criterion in detail, and an independent decision  about whether or not to exclude each data set was made.  In doing so, the same criterion originally introduced by the Nijmegen group \cite{Ber88} was used.   The heart of the data selection process is to evaluate whether or not each data set is consistent with the rest of the data.  If a particular data set has an error that is statistically ``too large''  ot ``too small,'' then this set is highly unlikely to be correct, and it is justified to exclude the set from the analysis.

If the data satisfy a gaussian distribution, it is pointed out in Ref.~\cite{Ber88} that the statistical distribution of $z\equiv \chi^2/n$ for $n$ data will satisfy the following normalized distribution
\bea
{\cal P}_{n}(z)=\frac{n(nz/2)^{n/2-1}}{2\Gamma(n/2)}\exp\left(-nz/2\right)\, , \label{theordis}
\eea
with  expectation value $z=1$ and and variance $\delta z= 2/\sqrt{n}$.   

We adopt the Nijmegen criteria that the error is ``too large'' or ``too small''  if the probability  that such a measurement could be obtained is less than 0.27\%.  This corresponds to the ``$3\sigma$'' criterion, obtained by considering the probability that a measurement lies beyond the 3$\sigma$ limit of a gaussian distribution, either too large or too small.  For a measurement with expected value of zero, this probability is obtained by integrating the normalized distribution over the regions that are ``too large'' or ``too small''  by $3\sigma$
\bea
{\cal P}_{3\sigma}=2\int_{3\sigma}^\infty dx \;N\exp\left(-\frac{x^2}{2\sigma^2}\right)=0.0027
\eea
The 3$\sigma$ criterion thus leads to both minimum and maximum allowed values of $z$ that depend on the number of data in each set.  These are given by
\begin{align}
&0.0027=\int_{z_{\rm max}(n)}^\infty dz\,{\cal P}_n(z)\nonumber\\
&0.0027=\int^{z_{\rm min}(n)}_0 dz\,{\cal P}_n(z)\, . \label{zlimits}
\end{align}

Fig.~\ref{fig:scatter} shows a scatter plot of $z=\chi^2/n$ versus the number of measurements $n$ for each of the 393 published sets listed in Tables \ref{tab:datakeep} and \ref{tab:dataskip}.  Those sets included in the analysis (from Table \ref{tab:datakeep})  are represented by a dot, and those excluded (from Table \ref{tab:dataskip})  by a small square.  The maximum and minimum $z$ allowed by the criteria of  Eq.~(\ref{zlimits}) are also shown in the figure.  If  all of these data sets were statistically consistent with each other, we
\phantom{might expect at most {\it  one\/} set to lie either above or below}   
\setlength\LTleft{0pt}
\setlength\LTright{0pt}
\begin{longtable*}{crcccccccccc}
\caption{List of all data used in the fitting. Here $n_1$ is the number of energy or angular measurements in the set and $n_2$ is the number of data points used in the fit (including a data point for systematic error whenever specified, and reduced in some cases by rejection of measurements at the particular angles or energies given in the footnote).   For each set with a systematic error specified by ``sys'' (column 7), the $\chi^2_{\rm sys}$ and scale factor (as defined in the text) are also given.  When given, the comment code consists of three letters, as discussed in the text.}\\
 $\quad E_{\rm lab}\quad $ & $\quad$ Ref. $\quad$ & Type &  $\theta$ & $\;\;  n_1\;\; $ &  $\;\;  n_2\;\; $ &  sys & $\chi^2_{\rm sys}$ & scale &  $\qquad\chi^2_t\qquad$ &   $\quad\chi^2_t/n\quad$ & comments\\[0.1pc]
 \hline\hline\\[0.3pc]
 \endfirsthead
 \caption{(continued)}\\ $\quad E_{\rm lab}\quad $ & $\quad$ Ref. $\quad$ & Type &  $\theta$ & $n_1$ & $n_2$ &  sys & $\chi^2_{\rm sys}$ & scale &  $\quad\chi^2_t\quad$ &   $\quad\chi^2_t/n\quad$ & comments\\[0.1pc]
 \hline\hline\\[0.3pc]
 \endhead
 \endfoot
   0.0       & DI75\cite{DI75} & SGT  &  --- &   1 &    1 & \multicolumn{3}{c}{no systematic error}  &    2.18 &  2.18 & NLx  \\[0.1pc]
   0.0       & HO71\cite{HO71} & SGT  &  --- &   1 &    1 & \multicolumn{3}{c}{no systematic error}  &    0.06 &  0.06 & NLx  \\[0.1pc]
   0.0       & KO90\cite{KO90} & SGT  &  --- &   1 &    1 & \multicolumn{3}{c}{no systematic error}  &    9.65 &  9.65 & xLx  \\[0.1pc]
   0.0       & FU76\cite{FU76} & SGT  &  --- &   1 &    1 & \multicolumn{3}{c}{no systematic error}  &    5.48 &  5.48 & NLx  \\[0.1pc]
   0.1-  0.6 & AL55\cite{AL55} & SGT  &  --- &   5 &    5 & \multicolumn{3}{c}{no systematic error}  &    2.24 &  0.45 & NLx  \\[0.1pc]
   0.5-  3.2 & EN63\cite{EN63} & SGT  &  --- &   2 &    2 & \multicolumn{3}{c}{no systematic error}  &    2.06 &  1.03 & NLS  \\[0.1pc]
   0.5- 24.6 & CL72\cite{CL72} & SGT  &  --- & 114 &  115 &   0.1\%  &   9.88 & 1.003 &  139.48 &  1.21 & RRS  \\[0.1pc]
   0.8- 20.0 & CL69\cite{CL69} & SGT  &  --- &  17 &   15 & \multicolumn{3}{c}{no systematic error}  &   11.20 &  0.75 & NLS \footnotemark[ 1]\footnotetext[ 1]{    1.2    9.9
} \\
\end{longtable*}
 ------------------
%
\setlength\LTleft{0pt}
\setlength\LTright{0pt}
\begin{longtable*}{crcccccccccc}
\newpage
\caption{List of all data not used in the fitting.}\\
 $\quad E_{\rm lab}\quad $ & $\quad$ Ref. $\quad$ & Type &  $\theta$ & $\;\;  n_1\;\; $ &  $\;\;  n_2\;\; $ &  sys & $\chi^2_{\rm sys}$ & scale &  $\qquad\chi^2_t\qquad$ &   $\quad\chi^2_t/n\quad$ & comments\\[0.1pc]
 \hline\hline\\[0.3pc]
 \endfirsthead
 \caption{(continued)}\\ $\quad E_{\rm lab}\quad $ & $\quad$ Ref. $\quad$ & Type &  $\theta$ & $n_1$ & $n_2$ &  sys & $\chi^2_{\rm sys}$ & scale &  $\quad\chi^2_t\quad$ &   $\quad\chi^2_t/n\quad$ & comments\\[0.1pc]
 \hline\hline\\[0.3pc]
 \endhead
 \endfoot
   0.5-  2.0 & PO82\cite{PO82} & SGT  &  --- &   3 &    3 & \multicolumn{3}{c}{no systematic error}  &   25.26 &  8.42 & xLxb \\[0.1pc]
  14.1 & SU67\cite{SU67} & DSG  &    11.9- 92.8 &   16 &   16 & \multicolumn{2}{c}{float} & 0.997 &    3.22 &  0.20 & NLSs \\[0.1pc]
  16.9 & TO88\cite{TO88} & P    &   136.5-166.5 &    4 &    5 &   1.0\%  &   0.00 & 1.000 &    0.17 &  0.03 & NLSs \\[0.1pc]
  29.9 & FI90\cite{FI90} & DSG  &   104.6-164.9 &    5 &    6 &  10.0\%  &   0.01 & 1.008 &   20.67 &  3.45 & NLSb \\[0.1pc]
  31.5 & BE97\cite{BE97} & DSG  &    40.0-140.0 &    6 &    7 &   5.0\%  &   0.01 & 0.995 &   54.28 &  7.75 & RLRb \\[0.1pc]
  50.0 & FI80\cite{FI80} & P    &   108.0-174.0 &    4 &    5 &   2.0\%  &   0.00 & 1.000 &    0.28 &  0.06 & NLSs \\[0.1pc]
  58.5 & BE97\cite{BE97} & DSG  &    40.0-140.0 &    6 &    7 &   5.0\%  &   0.00 & 1.002 &   57.50 &  8.21 & RLRb \\[0.1pc]
  63.1 & KI80\cite{KI80} & DSG  &    39.4-165.8 &   19 &   17 &   3.0\%  &   0.01 & 0.998 &   40.44 &  2.38 & NLSb \\
\end{longtable*}
%
%
\noindent might expect at most {\it  one\/} set to lie either above or below  the $3\sigma$ limits (i.e. 393$\times$ 0.0027 $\simeq1$), and these would most likely lie close to the boundaries, where there are already several sets.    The plot shows graphically  that most of the data sets excluded for statistical reasons lie way outside of the 3$\sigma$ limits, and it appears to be clearly justified to discard them.

Applying these criteria to our complete data set, with $n=3788$, gives 3$\sigma$ rejection limits of $0.937 \le z \le 1.065$.  Our best fit just lies within these limits, allowing us to conclude that the overall fit itself satisfies the 3$\sigma$ criterion.

\begin{figure*}
\centerline{
\mbox{
\includegraphics[width=6in]{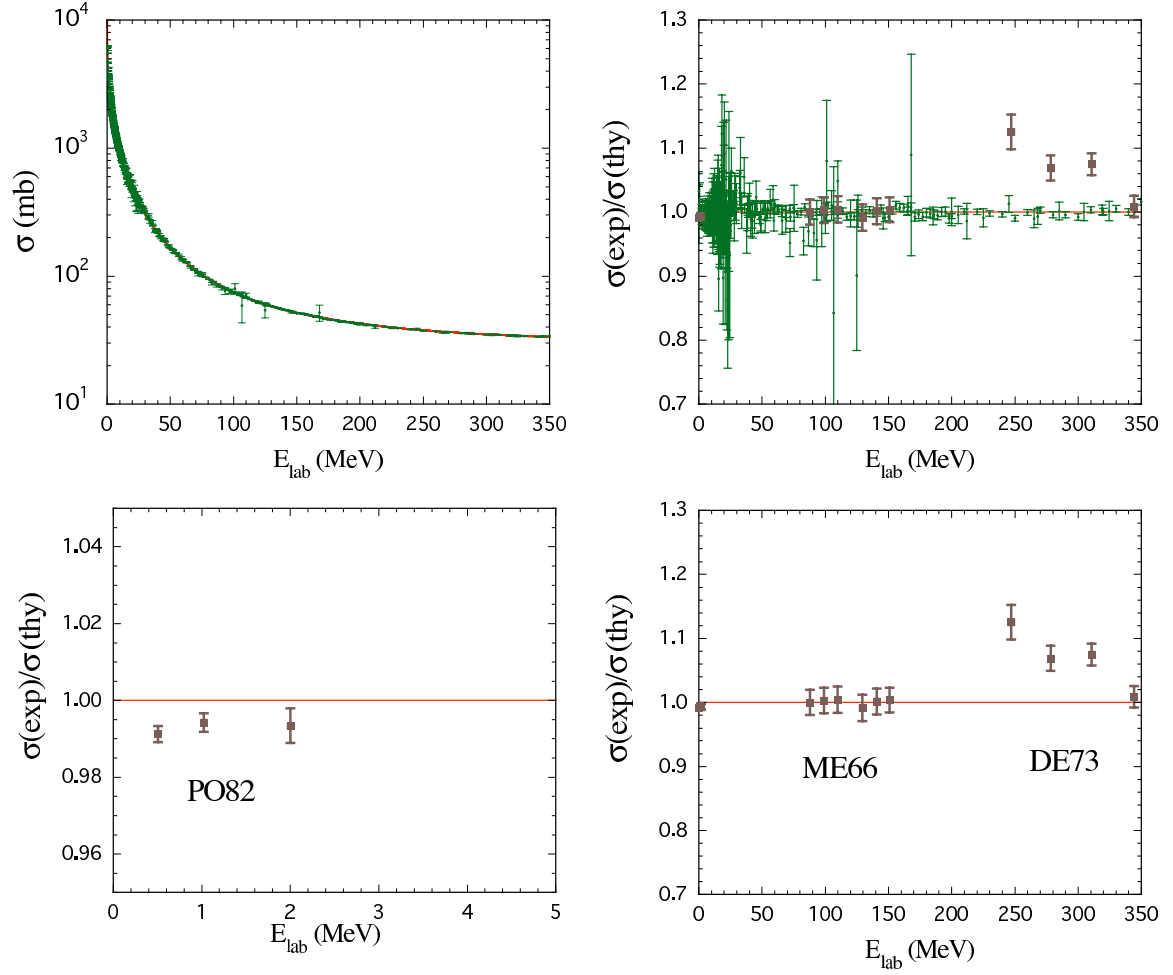}
}
}
\caption{\footnotesize\baselineskip=10pt (Color on line) Total cross section as a function of lab energy.  The lower two panels  compare the three rejected cross section data sets, PO82 \cite{PO82}  (shown in the lower left panel on an expanded energy scale, and also in the lower right panel as three indistinguishable points near zero), ME66 \cite{ME66}, and DE73 \cite{DE73} with theory.  The very small $\chi^2$ for ME66 is due to an overestimate of the errors (not consistent with the expected statistical fluctuations)  while the other data sets disagree strongly with the fit (theory), and all three sets lie well outside the boundaries (see their locations on Fig.~\ref{fig:distribution}).  The solid line is the fit (referred to as the theory). }
\label{fig:cross}
\end{figure*} 

\begin{figure*}
\centerline{
\mbox{
\includegraphics[width=6in]{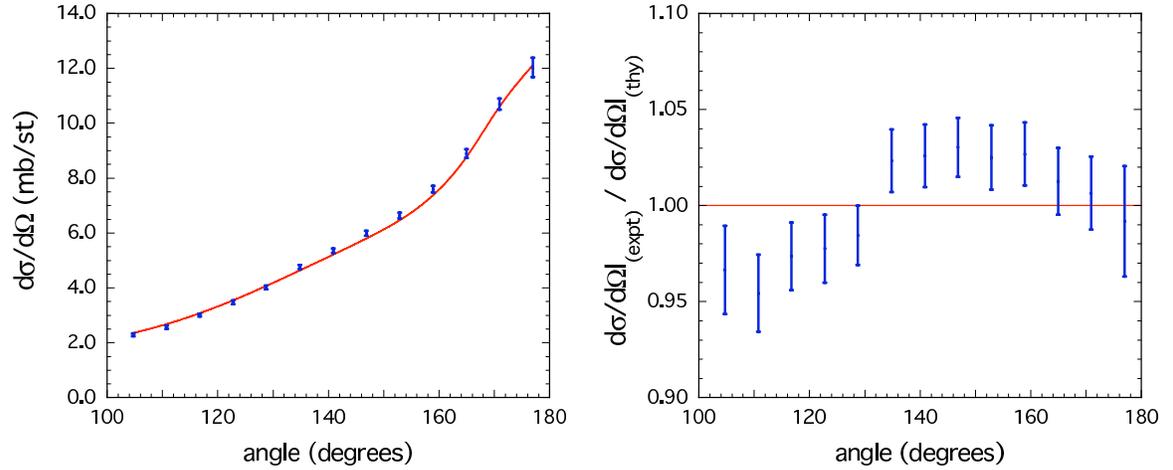}
}
}
\caption{\footnotesize\baselineskip=10pt (Color on line)  Example of the quality of the fit to some recent  194 MeV differential cross section data  \cite{SA06}.}
\label{fig:SA06}
\end{figure*} 

\begin{figure*}
\vspace*{-0.5in}
\centerline{
\mbox{
\includegraphics[width=4.9in]{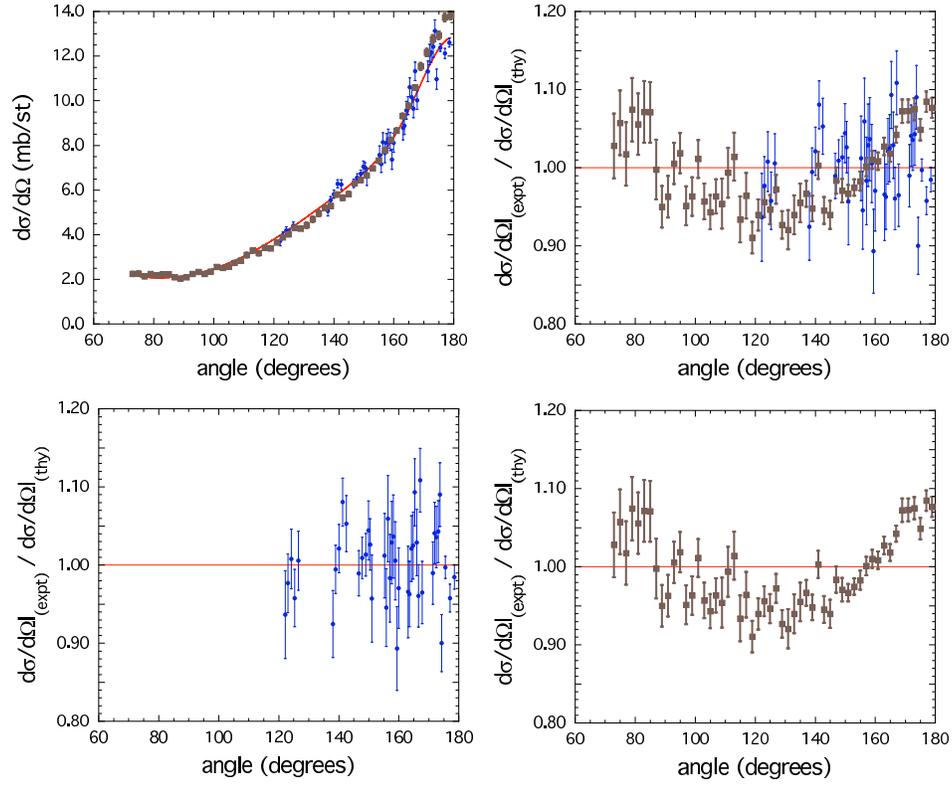}
}
}
\caption{\footnotesize\baselineskip=10pt (Color on line) Differential cross section measurements at 162 MeV \cite{BO78,RA98}.  The upper two panels show both data sets, the lower-left panel shows only the data set BO78 \cite{BO78} which is kept, the bottom-right panel only the data set RA98 \cite{RA98} which is excluded.  RA98 seems to have some unexplained systematic error, particularly at the backward angles, giving it the wrong shape and a large $\chi^2/n$ of 6.9.}
\label{fig:162}
\end{figure*}


\begin{figure*}
\vspace*{-0.5in}
\centerline{
\mbox{
\includegraphics[width=4.9in]{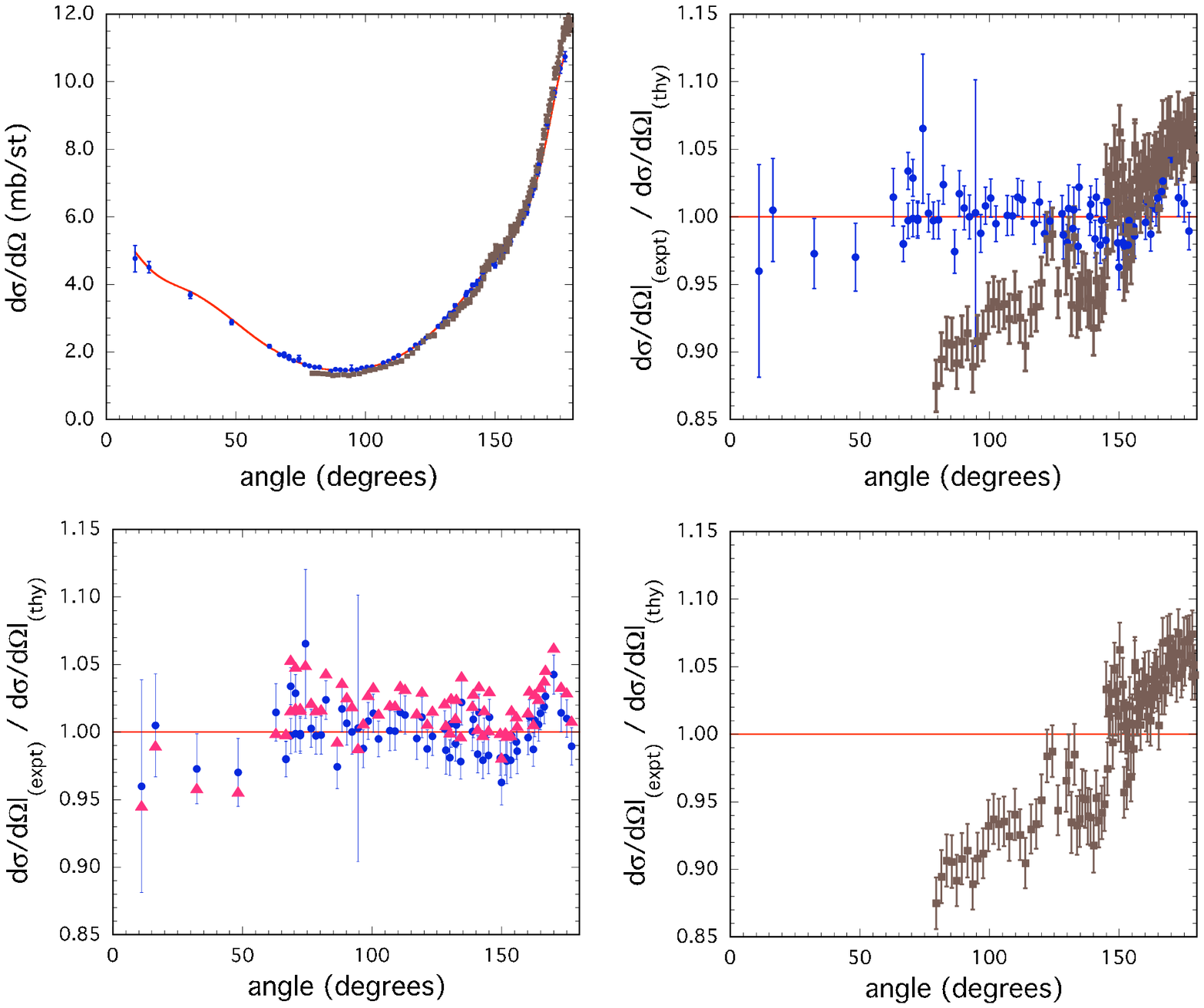}
}
}
\caption{\footnotesize\baselineskip=10pt (Color on line) Differential cross section measurements at 319 MeV \cite{KE82} and  320.1 MeV \cite{FR00}.  The data set FR00 \cite{FR00} shown in the bottom-right panel is excluded; do to some large unexplained systematic error it  has the wrong shape and a large $\chi^2/n$ of 8.4.  The lower-left panel shows the effect of scaling on the KE82 data \cite{KE82}, as discussed in the text.  In this panel the triangles are {\it unscaled\/} data not shown in the upper panels; the small circles are scaled data shown also in the upper panels. }
\label{fig:320}
\end{figure*} 

The comment column of Table \ref{tab:dataskip} gives a four character symbol that details information about the data that have been skipped.  The {\it first\/} character is either N, R, or x, where N denotes a data set that was used in the Nijmegen/Machleidt fits \cite{Stoks:1993tb,Mac01}, R a set that was rejected, and x a set that was not listed.  The second character is either L, R, or x, where L denotes a data set that is listed in NN-OnLine, the Nijmegen online data base \cite{NNonline}, R a set listed there but labled for rejection, and x a set that is not listed.   The third character is either S, R, or x, where S denotes a data set that is listed in the 2005 SAID data base \cite{GWU}, R a set listed there but labled for rejection, and x a set that is not listed.  Finally, the forth character is either b, s, c, or ?, where   b means that its $\chi^2$ is too large (as discussed above), s means that the $\chi^2$ is too small, c means that the measurement is from a composite target (and hence subject to unknown theoretical errors), and ? means that the data set is questionable for other reasons.  Hence, for example the 50 MeV polarization measurements of FI80 have the label NLSs, meaning that they are listed in all the data bases, but we have rejected them because their $\chi^2$ is too small (below the minimum line in Fig.~\ref{fig:scatter}).  Similarily, the 340 MeV differential cross section measurements of FR00 carries the notation RLRb, meaning that it was rejected by Machleidt and SAID, listed without comment in the NN-OnLine data base, and rejected  here because its $\chi^2$ is too large (above the maximum line in Fig.~\ref{fig:scatter}).

The same code is used for comments, when available, for data sets used in the fit (shown in Table~ \ref{tab:dataskip}).  For example, the cross section measurements from 0.5 to 24.6 MeV of CL72 contain the comment  RRS, meaning that they were rejected in the Nijmegen/Machleidt fits and  NN-OnLine, and listed without comment in SAID.  We have kept these data because their $\chi^2/{n_2}=1.21$ lies within the statistically acceptable range for a data set with 115 points, although this decision clearly increases the $\chi^2$ of the overall fit. 

Next, we take a brief look at the statistical distribution of the 18 large data sets with more that 50 points each.  Together these total 1607 measurements, and also include the bulk of the rejected data (926 out of 1180).  The distribution of these 18 sets is shown in Fig.~\ref{fig:distribution}, which compares the histogram of distributions with the theoretical distributions for data sets with $n=100$ and $n=60$ points.  The figure shows clearly how the 9 sets with large $\chi^2$ lie way beyond the region that is probable, while the 9 sets that have been used in the fit satisfy a reasonable distribution (but still skewed slightly toward $\chi^2$ that are too large).  If we had only these large data sets to work with, it might not be clear that we have rejected the right data, but  it is important to realize that the overall fit is largely fixed by the large number of data in {\it smaller\/} sets, which number 3107 out of all the 3788 data which are kept.

The rejection of data is couched in statistical terms, but examination of actual data sets shows that the ``judgement'' of statistics agrees with one's intuitive notions.  To get a feeling for how rejected data compare with the {\it whole\/} data set, and to see how good the fits really are, we look at a few examples illustrated in Figs.~\ref{fig:cross}, \ref{fig:SA06}, \ref{fig:162}, and \ref{fig:320}.  Figure~\ref{fig:cross} shows the fit to total cross section data.  The rejected data has not been included in the upper left panel, showing how close the data is to the fit.  The lower two panels show the three rejected cross section data sets, and one can easily see why the $\chi^2$ of the sets PO82 and DE73 is too high.  On the other hand, the set ME66 illustrates a situation in which the $\chi^2$ is too low: the actual scatter of the data around the theoretical line is much smaller than is to be expected from the size of experimental error bars.  In this case the data seem perfectly consistent with the fit, but it seems that there is something wrong with the error estimates so that the set cannot be used to properly constrain the fit.  As an example of the overall quality of the fit, Fig.~\ref{fig:SA06} shows how consistent the new 194 MeV differential cross section measurements \cite{SA06} are with the rest of the data.   Finally, Figs.~\ref{fig:162} and \ref{fig:320} show measurements of the differential cross sections in the backward directions at 162 and 320 MeV.  In each of these cases, one data set is consistent with the fit, and one is inconsistent.  The reasons are similar in both cases; the inconsistent data sets seem to have some unexplained angular dependent systematic error in the backward direction which disagrees with the rest of the data base (as represented by the fit).  The data sets at 162 and 320 MeV are certainly inconsistent with each other, and we emphasize that we can only decide which of these sets to include and which to exclude because of the presence of {\it all of the other data\/}.

\begingroup
\squeezetable
\begin{table*}
\caption{Isoscalar and $^1S_0$ $np$ phase shifts  for Model WJC-1.}
\begin{tabular}{rrrrrrrrrrrrrrrrr}
 $E_{\rm lab}$  &   $^1S_0$ & $^1P_1$ & $^3S_1$ & $^3D_1$ & $\qquad\epsilon_1$ & $\qquad^3D_2$  & $^1F_3$ & $^3D_3$ & $^3G_3$ & $\epsilon_3$ & $\quad^3G_4$  & $^1H_5$ & $^3G_5$ & $^3I_5$ & $\epsilon_5$ & $\quad^3I_6$ \cr \hline
$      1$ & $  62.071$ & $  -0.196$ & $ 147.644$ & $   0.007$ & $   0.111$ & $   0.006$ & $   0.000$ & $  -0.005$ & $   0.005$ & $  -0.001$ & $   0.000$ & $   0.000$ & $  -0.003$ & $   0.003$ & $   0.000$ & $   0.000$ \cr
$      5$ & $  63.608$ & $  -1.561$ & $ 117.998$ & $  -0.163$ & $   0.685$ & $   0.231$ & $  -0.012$ & $  -0.009$ & $   0.011$ & $   0.012$ & $   0.001$ & $   0.000$ & $  -0.008$ & $   0.007$ & $  -0.001$ & $   0.000$ \cr
$     10$ & $  59.885$ & $  -3.202$ & $ 102.408$ & $  -0.664$ & $   1.172$ & $   0.880$ & $  -0.068$ & $  -0.008$ & $   0.013$ & $   0.082$ & $   0.014$ & $  -0.002$ & $  -0.011$ & $   0.011$ & $   0.001$ & $   0.000$ \cr
$     25$ & $  50.631$ & $  -6.701$ & $  80.398$ & $  -2.836$ & $   1.780$ & $   3.855$ & $  -0.432$ & $   0.033$ & $  -0.029$ & $   0.566$ & $   0.176$ & $  -0.032$ & $  -0.025$ & $   0.014$ & $   0.037$ & $   0.012$ \cr
$     50$ & $  39.911$ & $ -10.289$ & $  62.498$ & $  -6.567$ & $   2.072$ & $   9.285$ & $  -1.155$ & $   0.308$ & $  -0.231$ & $   1.652$ & $   0.744$ & $  -0.169$ & $  -0.072$ & $   0.000$ & $   0.209$ & $   0.093$ \cr
$     75$ & $  31.944$ & $ -12.907$ & $  51.264$ & $  -9.809$ & $   2.257$ & $  13.988$ & $  -1.770$ & $   0.784$ & $  -0.545$ & $   2.681$ & $   1.470$ & $  -0.353$ & $  -0.132$ & $  -0.037$ & $   0.459$ & $   0.243$ \cr
$    100$ & $  25.468$ & $ -15.139$ & $  42.838$ & $ -12.578$ & $   2.469$ & $  17.699$ & $  -2.270$ & $   1.345$ & $  -0.923$ & $   3.563$ & $   2.238$ & $  -0.544$ & $  -0.187$ & $  -0.094$ & $   0.736$ & $   0.436$ \cr
$    125$ & $  19.957$ & $ -17.149$ & $  36.004$ & $ -14.968$ & $   2.722$ & $  20.473$ & $  -2.689$ & $   1.910$ & $  -1.336$ & $   4.298$ & $   3.003$ & $  -0.725$ & $  -0.230$ & $  -0.165$ & $   1.017$ & $   0.653$ \cr
$    150$ & $  15.137$ & $ -19.001$ & $  30.222$ & $ -17.058$ & $   3.005$ & $  22.448$ & $  -3.062$ & $   2.426$ & $  -1.764$ & $   4.903$ & $   3.744$ & $  -0.891$ & $  -0.257$ & $  -0.246$ & $   1.290$ & $   0.882$ \cr
$    175$ & $  10.825$ & $ -20.723$ & $  25.155$ & $ -18.906$ & $   3.315$ & $  23.775$ & $  -3.413$ & $   2.865$ & $  -2.192$ & $   5.396$ & $   4.451$ & $  -1.039$ & $  -0.269$ & $  -0.335$ & $   1.550$ & $   1.118$ \cr
$    200$ & $   6.899$ & $ -22.332$ & $  20.573$ & $ -20.561$ & $   3.657$ & $  24.585$ & $  -3.757$ & $   3.214$ & $  -2.611$ & $   5.799$ & $   5.119$ & $  -1.173$ & $  -0.268$ & $  -0.429$ & $   1.796$ & $   1.355$ \cr
$    225$ & $   3.291$ & $ -23.843$ & $  16.370$ & $ -22.059$ & $   4.020$ & $  24.986$ & $  -4.105$ & $   3.470$ & $  -3.015$ & $   6.127$ & $   5.747$ & $  -1.292$ & $  -0.255$ & $  -0.527$ & $   2.027$ & $   1.590$ \cr
$    250$ & $  -0.050$ & $ -25.271$ & $  12.471$ & $ -23.431$ & $   4.394$ & $  25.064$ & $  -4.464$ & $   3.635$ & $  -3.400$ & $   6.395$ & $   6.331$ & $  -1.400$ & $  -0.233$ & $  -0.628$ & $   2.243$ & $   1.822$ \cr
$    275$ & $  -3.160$ & $ -26.629$ & $   8.834$ & $ -24.698$ & $   4.767$ & $  24.885$ & $  -4.838$ & $   3.712$ & $  -3.762$ & $   6.614$ & $   6.872$ & $  -1.498$ & $  -0.203$ & $  -0.729$ & $   2.444$ & $   2.050$ \cr
$    300$ & $  -6.065$ & $ -27.930$ & $   5.441$ & $ -25.868$ & $   5.120$ & $  24.504$ & $  -5.228$ & $   3.708$ & $  -4.099$ & $   6.794$ & $   7.368$ & $  -1.590$ & $  -0.168$ & $  -0.831$ & $   2.631$ & $   2.272$ \cr
$    325$ & $  -8.807$ & $ -29.181$ & $   2.183$ & $ -26.967$ & $   5.477$ & $  23.962$ & $  -5.635$ & $   3.630$ & $  -4.408$ & $   6.943$ & $   7.822$ & $  -1.676$ & $  -0.130$ & $  -0.932$ & $   2.804$ & $   2.488$ \cr
$    350$ & $ -11.403$ & $ -30.390$ & $  -0.962$ & $ -28.005$ & $   5.833$ & $  23.293$ & $  -6.058$ & $   3.486$ & $  -4.690$ & $   7.067$ & $   8.233$ & $  -1.758$ & $  -0.091$ & $  -1.032$ & $   2.966$ & $   2.698$ \cr
 
 \end{tabular}
\label{tab:scalarph} 
\end{table*}
\endgroup

\begingroup
\squeezetable
\begin{table*}
\caption{Isovector (except the$^1S_0$) $np$ phase shifts  for Model WJC-1.}
\begin{tabular}{rrrrrrrrrrrrrrrrr}
 $E_{\rm lab}$  &  $^3P_0$ & $^3P_1$ & $\quad^1D_2$ & $\quad^3P_2$ & $^3F_2$ & $\epsilon_2$ & $^3F_3$  & $\quad^1G_4$ & $^3F_4$ & $\quad^3H_4$ & $\epsilon_4$ & $^3H_5$  & $\quad^1I_6$ & $^3H_6$ & $\quad^3J_6$ & $\epsilon_6$ \cr \hline
$      1$ & $   0.206$ & $  -0.100$ & $   0.001$ & $   0.017$ & $   0.007$ & $  -0.003$ & $   0.000$ & $   0.000$ & $  -0.004$ & $   0.004$ & $   0.000$ & $   0.000$ & $   0.000$ & $  -0.003$ & $   0.003$ & $   0.000$ \cr
$      5$ & $   1.630$ & $  -0.869$ & $   0.039$ & $   0.259$ & $   0.018$ & $  -0.049$ & $  -0.004$ & $   0.000$ & $  -0.009$ & $   0.009$ & $  -0.001$ & $   0.000$ & $   0.000$ & $  -0.006$ & $   0.006$ & $   0.000$ \cr
$     10$ & $   3.614$ & $  -1.918$ & $   0.147$ & $   0.744$ & $   0.033$ & $  -0.176$ & $  -0.024$ & $   0.002$ & $  -0.012$ & $   0.013$ & $  -0.004$ & $   0.000$ & $   0.000$ & $  -0.009$ & $   0.009$ & $  -0.001$ \cr
$     25$ & $   8.036$ & $  -4.570$ & $   0.640$ & $   2.672$ & $   0.122$ & $  -0.714$ & $  -0.183$ & $   0.030$ & $  -0.003$ & $   0.023$ & $  -0.038$ & $  -0.010$ & $   0.002$ & $  -0.014$ & $   0.014$ & $  -0.003$ \cr
$     50$ & $  10.605$ & $  -7.868$ & $   1.627$ & $   6.010$ & $   0.342$ & $  -1.540$ & $  -0.568$ & $   0.125$ & $   0.077$ & $   0.048$ & $  -0.159$ & $  -0.064$ & $   0.016$ & $  -0.015$ & $   0.022$ & $  -0.021$ \cr
$     75$ & $  10.117$ & $ -10.587$ & $   2.671$ & $   8.754$ & $   0.577$ & $  -2.104$ & $  -0.948$ & $   0.243$ & $   0.231$ & $   0.084$ & $  -0.310$ & $  -0.149$ & $   0.042$ & $  -0.008$ & $   0.030$ & $  -0.053$ \cr
$    100$ & $   8.348$ & $ -13.066$ & $   3.706$ & $  10.820$ & $   0.788$ & $  -2.443$ & $  -1.283$ & $   0.370$ & $   0.442$ & $   0.129$ & $  -0.463$ & $  -0.247$ & $   0.074$ & $   0.007$ & $   0.041$ & $  -0.095$ \cr
$    125$ & $   6.053$ & $ -15.416$ & $   4.685$ & $  12.318$ & $   0.957$ & $  -2.617$ & $  -1.575$ & $   0.502$ & $   0.694$ & $   0.179$ & $  -0.610$ & $  -0.349$ & $   0.110$ & $   0.029$ & $   0.054$ & $  -0.142$ \cr
$    150$ & $   3.564$ & $ -17.675$ & $   5.577$ & $  13.378$ & $   1.071$ & $  -2.675$ & $  -1.835$ & $   0.639$ & $   0.969$ & $   0.232$ & $  -0.747$ & $  -0.448$ & $   0.147$ & $   0.060$ & $   0.068$ & $  -0.191$ \cr
$    175$ & $   1.028$ & $ -19.863$ & $   6.364$ & $  14.109$ & $   1.124$ & $  -2.653$ & $  -2.071$ & $   0.779$ & $   1.257$ & $   0.288$ & $  -0.872$ & $  -0.543$ & $   0.185$ & $   0.098$ & $   0.084$ & $  -0.240$ \cr
$    200$ & $  -1.470$ & $ -21.985$ & $   7.037$ & $  14.594$ & $   1.112$ & $  -2.576$ & $  -2.295$ & $   0.922$ & $   1.549$ & $   0.344$ & $  -0.984$ & $  -0.632$ & $   0.223$ & $   0.143$ & $   0.101$ & $  -0.289$ \cr
$    225$ & $  -3.891$ & $ -24.046$ & $   7.594$ & $  14.890$ & $   1.036$ & $  -2.462$ & $  -2.511$ & $   1.068$ & $   1.835$ & $   0.398$ & $  -1.085$ & $  -0.715$ & $   0.262$ & $   0.193$ & $   0.118$ & $  -0.337$ \cr
$    250$ & $  -6.216$ & $ -26.051$ & $   8.037$ & $  15.036$ & $   0.896$ & $  -2.325$ & $  -2.726$ & $   1.214$ & $   2.112$ & $   0.450$ & $  -1.175$ & $  -0.792$ & $   0.301$ & $   0.249$ & $   0.137$ & $  -0.384$ \cr
$    275$ & $  -8.438$ & $ -28.004$ & $   8.374$ & $  15.062$ & $   0.697$ & $  -2.173$ & $  -2.944$ & $   1.361$ & $   2.375$ & $   0.497$ & $  -1.253$ & $  -0.864$ & $   0.341$ & $   0.310$ & $   0.156$ & $  -0.428$ \cr
$    300$ & $ -10.542$ & $ -29.911$ & $   8.612$ & $  14.988$ & $   0.441$ & $  -2.015$ & $  -3.165$ & $   1.506$ & $   2.622$ & $   0.540$ & $  -1.321$ & $  -0.930$ & $   0.381$ & $   0.374$ & $   0.176$ & $  -0.471$ \cr
$    325$ & $ -12.542$ & $ -31.776$ & $   8.760$ & $  14.839$ & $   0.135$ & $  -1.854$ & $  -3.393$ & $   1.649$ & $   2.852$ & $   0.576$ & $  -1.380$ & $  -0.993$ & $   0.421$ & $   0.442$ & $   0.196$ & $  -0.512$ \cr
$    350$ & $ -14.435$ & $ -33.601$ & $   8.824$ & $  14.627$ & $  -0.218$ & $  -1.694$ & $  -3.627$ & $   1.790$ & $   3.064$ & $   0.605$ & $  -1.429$ & $  -1.052$ & $   0.462$ & $   0.511$ & $   0.216$ & $  -0.552$ \cr
 
\end{tabular}
\label{tab:vectorph} 
\end{table*}
\endgroup

Finally, we discuss systematic errors and how they are treated.  As specified by the experimentalists, data may have a specified systematic error, no systematic error (absolute measurements), or an arbitrarily large systematic error (floated data).  In all cases the $\chi_t^2$ for a data set can be written
\bea
\chi_t^2=\sum_{i=1}^{n}\frac{(Z\,o_i-t_i)^2}{(Z\,\delta o_i)^2} + \frac{(Z-1)^2}{(\delta_{\rm sys})^2}
\label{syserror}
\eea
where $o_i$ and $t_i$ are the measured and calculated value of the observable at point $i$, $\delta o_i$ and $\delta_{\rm sys}$ are the statistical errors at point $i$ and the systematic error, and $Z$ is a factor by which the data and errors are scaled to improve agreement with theory.  The last term in (\ref{syserror}) is denoted $\chi^2_{\rm sys}$.   The value of $Z$ is chosen to minimize $\chi_t$.  Data with no systematic error cannot be scaled ($Z=1$, so  $\chi^2_{\rm sys}$ is zero),  data that floats could be treated using  ({\ref{syserror}) with $\delta_{\rm sys}=\infty$ so that  $\chi^2_{\rm sys}=0$ no matter what the value of $Z$, and data with a specified systematic error generaly fit the theory best if $Z\ne 1$ giving a value of  $\chi^2_{\rm sys}>0$ (as shown in the tables).  In this case the term  $\chi^2_{\rm sys}$ is a new contribution to the overall error and is counted as a new data point.

The lower-left panel of Fig.~\ref{fig:320}  illustrates how data with systematic error are adjusted to improve the fit.  At 319 MeV, KE82 \cite{KE82} measured the  differential cross section over an angular range from 11.1 to 94.5 degrees with an estimated systematic error of 2\%, and over an angular range from 66.7 to 177 degrees with an estimated systematic error of about 4\%.  These errors permit us to scale these two data sets independently in order to get the best fit to the data, and  as reported in Table \ref{tab:datakeep}, the result scales the data in the first range by $Z=$1.015, and in the second by $Z=$0.981.  These shifts  give a small additional $\chi_{\rm sys}^2$ of 0.55 and 0.24, respectively.  In the figure the solid triangles show the data {\it before\/} scaling, and the solid circles (centered on the error bars) show the data {\it after\/} scaling.  Close examination of the figure shows how these small shifts improve the overall fit.

\section{Phase shifts and low-energy parameters} \label{Sec:V}

Numerical values of the phase shifts for Model WJC-1 are given in Tables \ref{tab:scalarph} and \ref{tab:vectorph}, and the phase shifts for both models are compared to the Nijmegen 1993 phases in Figs.~\ref{fig:Jlt2}, \ref{fig:Jgt2}, and \ref{fig:Jcoupled}.

Note that the $J=4$ phases are very similar, but there are significant differences (a few degrees in many cases) between the three sets of phases for $J\le3$ (except that all three sets give a nearly identical $\epsilon_3$).   For all but the $^3P_0$ and $^1P_1$  phases, there is a tendency for our two models to agree (or at least have the same shape) and differ from the Nijmegen phases.  This is especially true of the $^1S_0$, $^3S_1$ -- $^3D_1$, $^1D_2$, $^3D_2$, $^3P_2$ -- $^3F_2$,  $^1F_3$, and $^3D_3$ -- $^3G_3$ phases, where our two models are much closer (identical in some cases) than the separation from the Nijmegen phases.  In other cases this trend is less clear.   The $^3P_1$ and $\epsilon_1$ phases are less close together and only depart from Nijmegen above 100 MeV, while WJC-1 and WJC-2 straddle the the Nijmegen $^3F_3$ and $\epsilon_2$ phases, showing a clearly different trend only above 200 MeV.   This universal pattern is broken by the $^3P_0$ and $^1P_1$  phases. The Nijmegen $^3P_0$ phase is very close to WJC-1 up to a cross over point about 200 MeV, where it then tracks Model WJC-2.  The differences in the $^1P_1$ phases are small, but in the neighborhood of 200 MeV, the WJC-1 and Nijmegen models are very close together and clearly distinct from WJC-2.

\begin{figure*}
\centerline{
\mbox{
\includegraphics[width=6in]{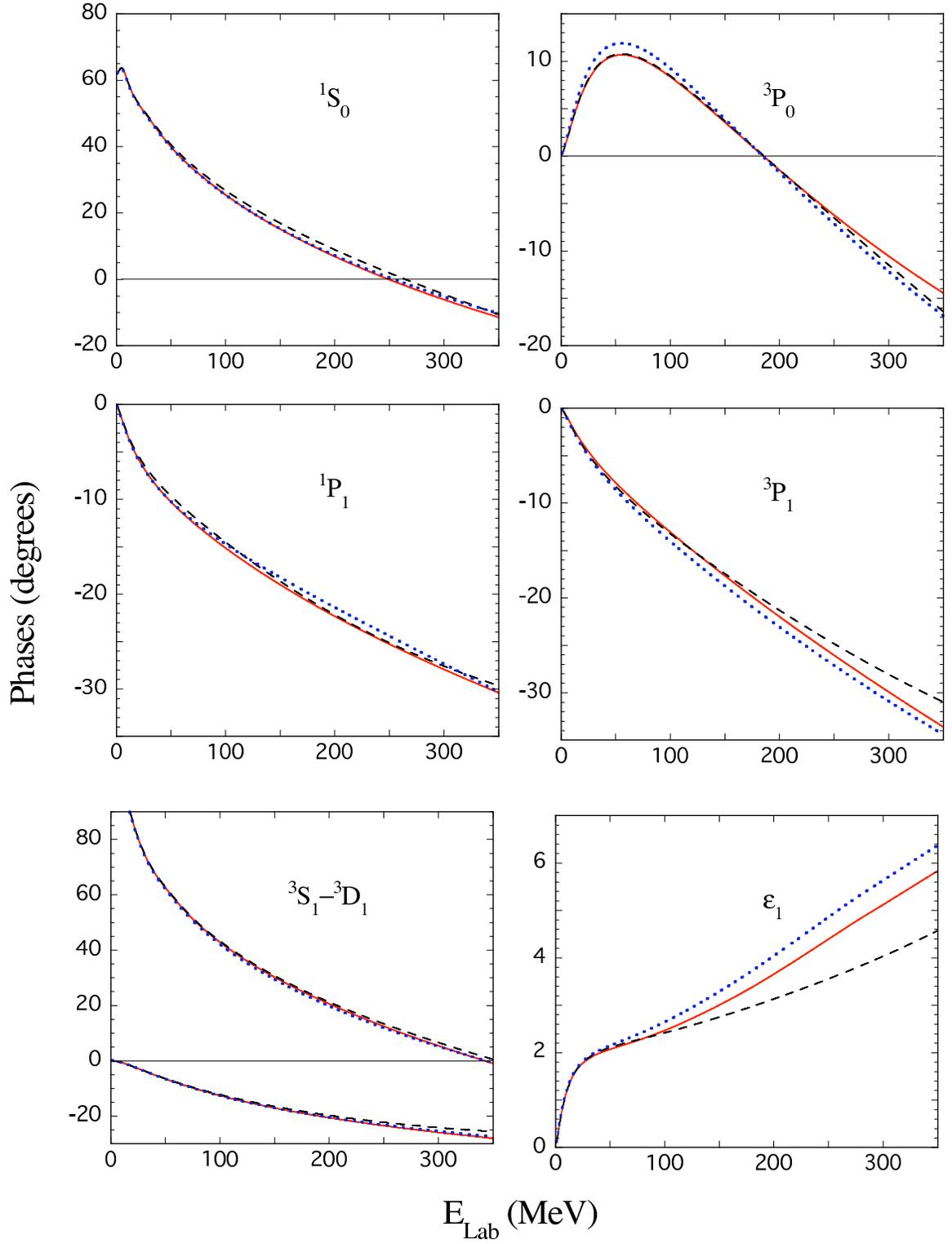}
}
}
\caption{\footnotesize\baselineskip=10pt (Color on line) Phase shifts and mixing parameters for all states with $J\le1$.  Models WJC-1 (solid line) and WJC-2 (dotted line) are compared to the Nijmegen phases (dashed line).}
\label{fig:Jlt2}
\end{figure*} 

\begin{figure*}
\centerline{
\mbox{
\includegraphics[width=6in]{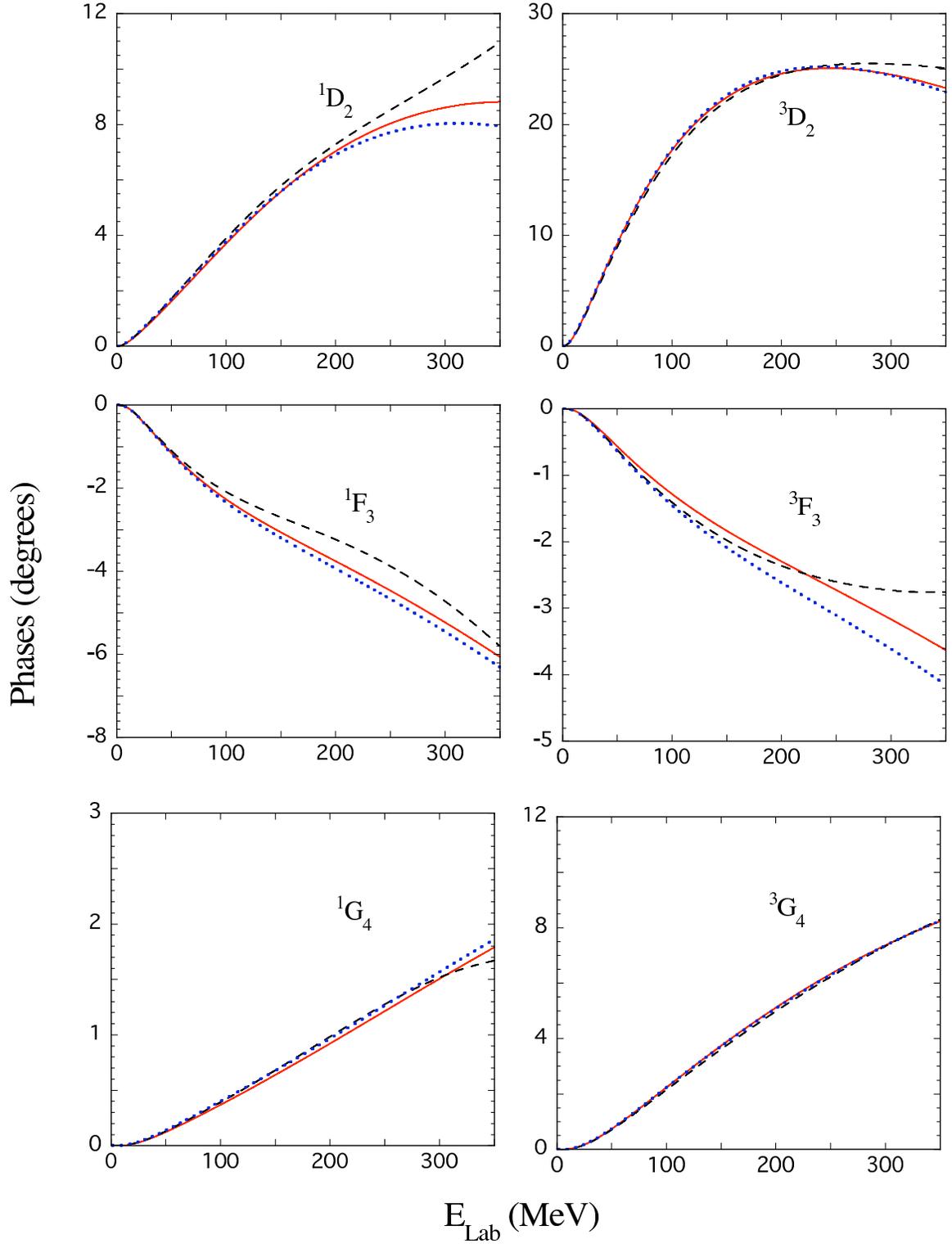}
}
}
\caption{\footnotesize\baselineskip=10pt (Color on line) Singlet an uncoupled triplet phase shifts  for all states with $2\le J\le4$.  Curves are drawn as in Fig.~\ref{fig:Jlt2}.}
\label{fig:Jgt2}
\end{figure*} 

\begin{figure*}
\centerline{
\mbox{
\includegraphics[width=6in]{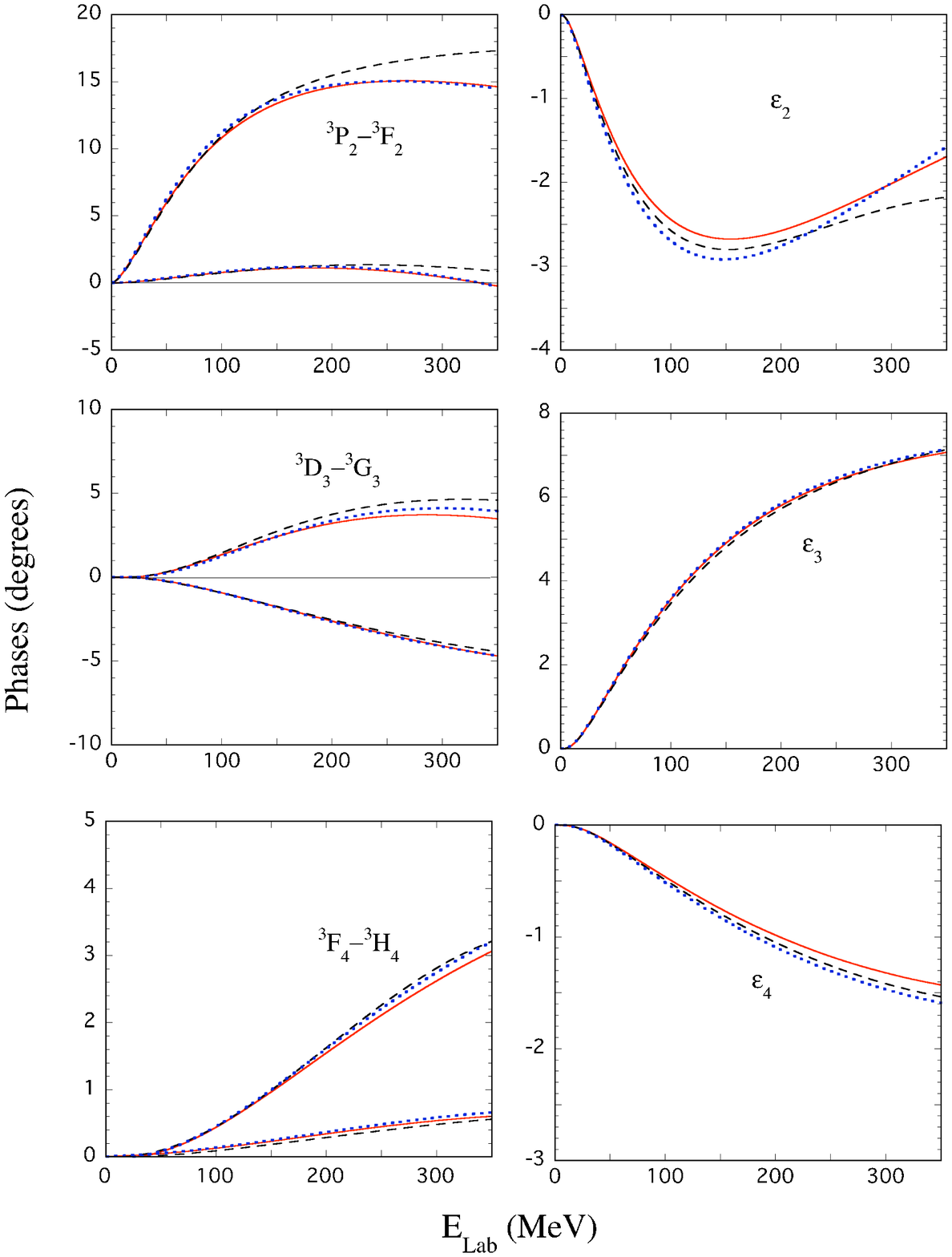}
}
}
\caption{\footnotesize\baselineskip=10pt (Color on line) Triplet coupled phase shifts and mixing parameters for states with $2\le J\le4$. Curves are drawn as in Fig.~\ref{fig:Jlt2}.}
\label{fig:Jcoupled}
\end{figure*} 

We were surprised that the phase shifts for Model WJC-1 (which we regard as a new, accurate $NN$ phase shift analysis) were not in closer agreement with the Nijmegen phases.   In the beginning we thought it would be sufficient to fit our model to the Nijmegen phases, and then calculate the $\chi^2$ in a second step, without further fitting.  Our early models did not give a very good fit to the Nijmegen phases, and we assumed that our higher $\chi^2/N_{\rm data}\sim2$ was do to this deficiency.  Later, our fits to the Nijmegen phases improved, and we also developed the capability to fit the data directly.  We then discovered that, starting from a good fit to the Nijmegen phases and fitting the data in a second step, not only lead to significant improvement in the $\chi^2$, but also to a region away from the best fit to the Nijmegen phases.  Eventually, as we acquired more experience and skill with the fits, we realized that it was counterproductive to fit the Nijmegen phases too accurately; improving the accuracy of the fit to the phases only lead us away from the best fit to the data.  When we started using the WJC-1 phases as a first step in future fits (such as Model WJC-2), this problem vanished and a good fit to the phases assured a good fit to the data.  We conclude that the WJC-1 phases are more accurate, and that any discrepancy between the fits to the phases and the fits to the data will be greatly reduced if the WJC-1 phases are used.

\section{The three-body binding energy} \label{Sec:VI}

The covariant OBE models of the $NN$ interaction presented in this paper posses a remarkable property: they can explain the three-body binding energy of the triton, naturally and without additional assumptions.   This result, first reported in Ref.~\cite{Sta97}, might have appeared to be an accident.   Now that we also see it for the more accurate models reported here, we believe it to  be a robust feature of the covariant spectator theory, for which there might be a simple explanation  (but at this time we have not found it).

\begin{figure}
\centerline{
\mbox{
\includegraphics[width=3.5in]{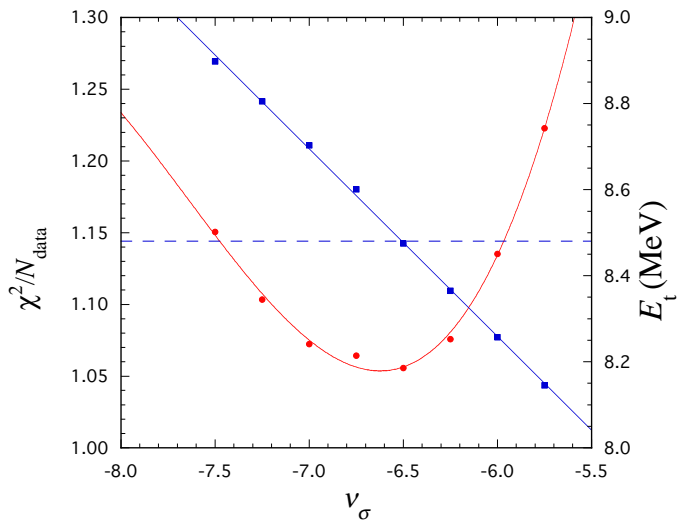}
}
}
\caption{\footnotesize\baselineskip=10pt (Color on line) The family of WJC-1 models with $\nu_{\sigma_0}$ constrained to various fixed values.  The left-hand axis shows the best $\chi^2/N_{\rm data}$ that can be found for each value of $\nu_{\sigma_0}$ (the data shows some scatter with respect to the solid line, which is a cubic fit to the 8 cases shown), and the right hand axis shows the triton binding energy ($E_t$ in MeV) for each member of the family.  Note that the correct binding energy (shown by the dashed horizontal line) is obtained for the value of $\nu_{\sigma_0}$ that also gives the best fit to the data.  } 
\label{fig:WJC1family}
\end{figure} 

\begin{figure}
\centerline{
\mbox{
\includegraphics[width=3.5in]{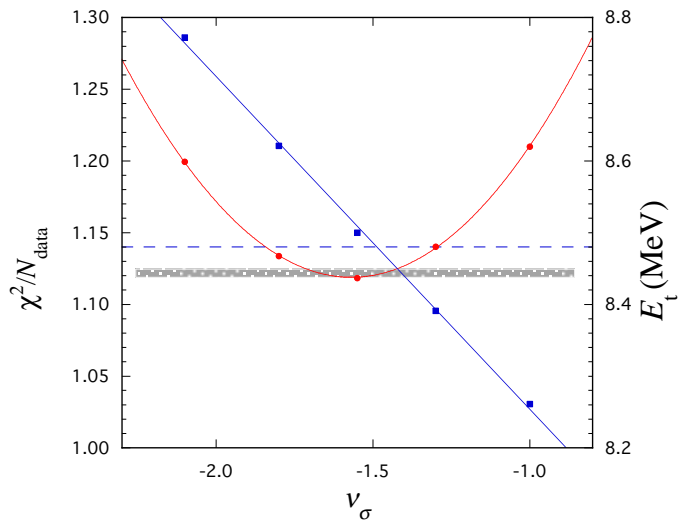}
}
}
\caption{\footnotesize\baselineskip=10pt (Color on line) The family of WJC-2 models with $\nu_{\sigma_0}$ constrained to various fixed values.  Here the fit to $\chi^2/N_{\rm data}$ is a quadratic, and the binding energy for the best case, 8.50, is shifted slightly from the experimental value.  However, shifting $\nu_{\sigma_0}$ by about 0.05 should give exact agreement with the binding energy, and this shift changes $\chi^2/N_{\rm data}$ by less that 0.003 (one standard deviation for about 3500 data, shown as a shaded band on the figure).  (See the caption to Fig.~\ref{fig:WJC1family} for further explanation.)  } 
\label{fig:WJC2family}
\end{figure} 

The result is shown in Figs.~\ref{fig:WJC1family} and \ref{fig:WJC2family}.   For both models we have found that both the triton binding energy $E_t$ and the quality of the fit (as measured by $\chi^2/N_{\rm data}$) are particularly sensitive to the off-shell coupling $\nu_{\sigma_0}$ of the $\sigma_0$ meson.  For Model WJC-1, the best fit gave $\nu_{\sigma_0}=-6.5$ (cf. Table~\ref{tab:par1}) and this is confirmed by fixing $\nu_{\sigma_0}$ at various values and refitting (by allowing {\it all\/} of the parameters {\it except} $\nu_{\sigma_0}$ to vary).  The trition binding energy is approximately linear in $\nu_{\sigma_0}$, and the figure shows that the value of $\nu_{\sigma_0}$ that gives the experimental value of $-8.48$ {\it also\/} gives the best fit to the two-body data.  An identical conclusion holds for Model WJC-2, as shown in Fig.~\ref{fig:WJC2family}.   

Nonrelativistic calculations of the triton binding cannot reproduce the experimental results without adding a three-body force.  How can our results be consistent with this well known observation?  The answer, discussed first in Ref.~\cite{Sta97} and later in various conference talks, depends on how the three-body force is defined.  

As an example, consider two successive emissions (or absorptions) of a scalar meson from an off-shell nucleon.  The interactions along the nucleon line will include cross terms of the form
\bea
&&\Lambda(p_3,p_2)S(p_2)\Lambda(p_2,p_1)\nonumber\\
 &&\qquad=-\frac{g_\sigma\nu_\sigma}{2m}\bigg\{(m-\slashed {p}_2) \frac{H^2(p_2)}{m-\slashed{p}_2} +\frac{H^2(p_2)}{m-\slashed{p}_2}(m-\slashed {p}_2)\bigg\}\nonumber\\
 &&\qquad=- \frac{g_\sigma\nu_\sigma}{m} H^2(p_2)\, . \label{contact}
\eea
This is equivalent to a contact interaction [with the form factor $H^2(p_2)$] as illustrated in Fig.~\ref{fig:contact}.  Successive applications of this affect will generate an infinite number of  multi-loop contributions to two and three-body forces.  A few of the simplest cases are shown in Figs.~\ref{fig:2body} and \ref{fig:3body}.

\begin{figure}
\centerline{
\mbox{
\includegraphics[width=3.5in]{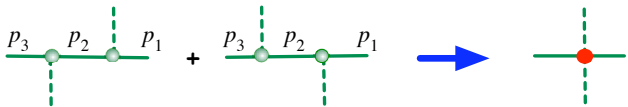}
}
}
\caption{\footnotesize\baselineskip=10pt (Color on line) Diagrams showing the collapse of off-shell interactions into an effective contact interaction, as derived in Eq.~(\ref{contact}).  } 
\label{fig:contact}
\end{figure} 

\begin{figure}
\centerline{
\mbox{
\includegraphics[width=3.0in]{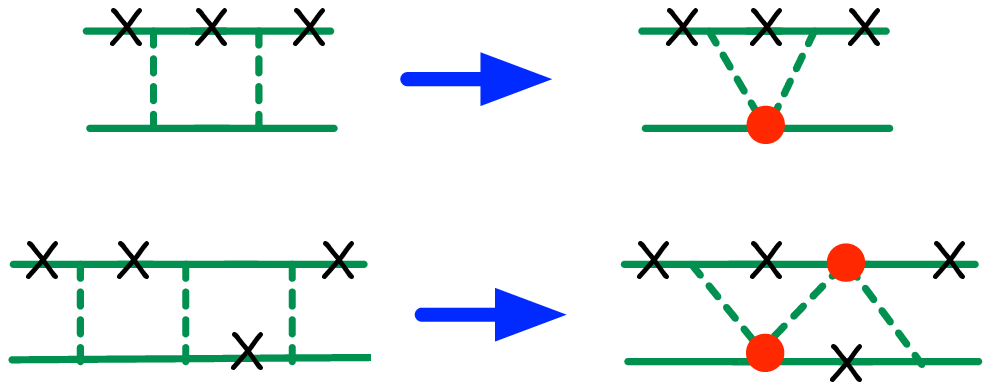}
}
}
\caption{\footnotesize\baselineskip=10pt (Color on line) Examples, from the two-body sector, of one and two-loop diagrams  generated by iteration of {\it off-shell OBE} couplings.  The off-shell couplings {\it automatically\/} generate these diagrams, but the same result could be obtained from a theory {\it without\/} off-shell couplings if these diagrams (and an infinite number of others) were added explicitly to the two-body force.  The lines marked with an $\times$ are on-shell particles and in each case a mechanism similar to that shown in Fig.~\ref{fig:contact} collapses the off-shell propagation to a point.  } 
\label{fig:2body}
\end{figure} 

\begin{figure}
\centerline{
\mbox{
\includegraphics[width=3.0in]{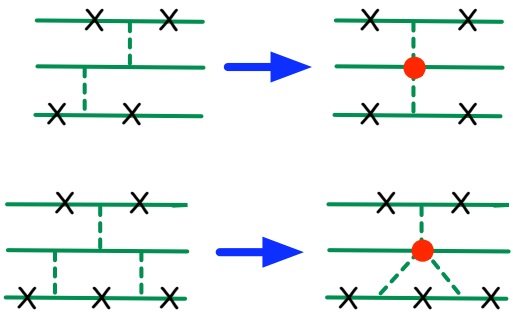}
}
}
\caption{\footnotesize\baselineskip=10pt (Color on line) Examples, from the three-body sector, of no-loop and one-loop diagrams generated by iteration of {\it off-shell OBE} couplings.  The off-shell couplings {\it automatically\/} generate these diagrams, but the same result could be obtained from a theory {\it without\/} off-shell couplings if these diagrams (and an infinite number of others) were added explicitly to the three-body force.  The lines marked with an $\times$ are on-shell particles and in each case a mechanism similar to that shown in Fig.~\ref{fig:contact} collapses the off-shell propagation to a point.   } 
\label{fig:3body}
\end{figure} 

Clearly, off-shell OBE couplings, when iterated to all orders, generate an infinite series of effective two and three-body force diagrams (and $n$-body forces for the $n$-body problem) involving loops and effective contact interactions.  If {\it all\/} of the diagrams so generated could be calculated explicitly, and added as separate two and three-body forces, then it would be possible to remove the off-shell couplings from the OBE kernels without altering any results.  There is a correspondence theorem, or a duality relation, which can be stated as follows: {\it a\/} pure {\it OBE theory with off-shell OBE couplings is equivalent to another theory with an infinite number of two and three-body forces but\/} no {\it off-shell OBE couplings.\/}    So the existence of three-body forces depends on the structure of the two-body interactions, and cannot be uniquely defined.

In conclusion: Figs.~\ref{fig:WJC1family} and \ref{fig:WJC2family} show that the effective two and three-body forces that depend on $\nu_{\sigma_0}$  are related in such a way that a {\it single\/} value of $\nu_{\sigma_0}$ gives {\it both\/} two-body loop contributions that give the best fit to the two-body data {\it and\/} three-body forces that fit the triton binding energy.   This result is a robust consequence of the spectator theory, but its origin is not understood.

\section{Conclusions and outlook} \label{Sec:VII}

In this paper we use the covariant spectator theory (CST) with a simple one-boson exchange (OBE) kernel to fit  $np$ scattering data for laboratory energies below 350 MeV.  We present two precision fits to the data.  One model, designated WJC-1, has 27 parameters and fits the 2007 data base  with a $\chi^2/N_{\rm  data}$ = 1.06 for $N_{\rm data}=3788$ data.  A second model, with many parameters fixed at physical values, has only 15 free parameters and fits with a $\chi^2/N_{\rm  data}$ = 1.12, as good as the fit of the 1993 Nijmegen phase shifts to the 2007 data base.  Both of these models have a simple one-boson exchange structure without any special partial-wave-dependent parameters, and have far fewer parameters than have been needed for previous high precision fits.  The fit from our best model WJC-1 automatically produces a new, accurate $NN$ phase shift analysis, useful even outside of the context of the CST.

In carrying out this study, we have updated the $np$ database by adding published data up through 2006 not previously included in any other fits, and doing an independent evaluation of which data are to be excluded and which are to be retained.  As a result, our database includes more data than used by the Nijmegen group in their famous 1993 partial wave analysis, or by the Idaho group in their construction of the CD-Bonn potential.   

Using the three-body CST equations, the binding energy of the triton can be calculated  from the $NN$ scattering amplitude.  A remarkable feature of this calculation is that the {\it correct\/} binding energy emerges automatically from the best fit to the two-body data, {\it without need for any additional three-body forces\/}.  This result is due to the presence of off-shell couplings for the scalar meson exchanges that are part of the kernel.  The same result could be obtained using a kernel without such off-shell couplings, provided an infinite number of  two-body loop diagrams of a particular structure were added to the two-body kernel, and an infinite number of three-body force diagrams of a corresponding structure were added to the  three-body equations.  
The off-shell scalar couplings are therefore a remarkably efficient way to unify and improve both the description of the two-and three-body sytems without departing from a kernel with a simple OBE structure.  

The next task is to see if it is possible to construct  a nonrelativistic, phase equivalent potential, that, when inserted into a Schr\"odinger equation, will give the same phase shifts as those of model WJC-1.  This will require several new ideas, but we believe that this should be possible.  Along the way we will learn more precisely what are the nature of the ''relativistic corrections''  that account for the success of the CST.

The OBE structure of the $NN$ kernel allows for a comparative simple construction of {\it consistent\/} (but not unique) electromagnetic interaction currents, and this work can therefore be extended to the description of electromagnetic scattering from the deuteron.  The new off-shell scalar exchange couplings will generate a new kind of isoscalar exchange current that can be tested in elastic electron-deuteron scattering (deuteron form factors).  We expect new exchange current contributions to the deuteron quadrupole moment, which may shed some light on failure of current potential models to explain this important low-energy parameter.

Finally,  note that all the tools needed  for an accurate relativistic  calculation of the three-body scattering problem are now in hand.    A first step might be to study elastic $nd$ scattering and see if this relativistic approach can help with the $A_y$ puzzle.  

In conclusion, we believe that the availability of these precision fits should make it possible to extend the great success of precision nonrelativistic few-body physics into a relativistic domain.

\acknowledgements

This work is the conclusion of an effort extending over more that a decade, supported initially by the DOE  through grant  No.~DE-FG02-97ER41032, and recently supported by Jefferson Science Associates, LLC under U.S. DOE Contract No.~DE-AC05-06OR23177. A.\ S.\ was supported by FCT under grant No.~POCTI/ISFL/2/275 and thanks the Jefferson Lab Theory Group for the hospitality extended to him during several visits while this work was performed.  We also acknowledge prior work by R. Machleidt and J.W. Van Orden, who wrote some earlier versions of parts of the $NN$ code.  The data analysis used parts of the SAID code supplied to us by R.\ A.\ Arndt.  Helpful conversations with the the Nijmegen group (J. J. de Swart, M. C. M. Rentmeester, and R.G.E. Timmermans) and with R.\ Schiavilla are gratefully acknowledged.

\appendix

\section{Overview}\label{App:A0}

Many features of the two-body theory are described in detail in the following appendices.

The form of the partial wave expansion of Eq.~(\ref{eq:spec}) is derived in Appendix \ref{App:D}.  Some features of the on-shell prescription of the CST are discussed in Appendix \ref{App:A}, and in particular, we discuss the removal of spurious singularities from the kernel in Sec.~\ref{App:A3}.  This latter issue has been a problem with the CST for many years, and in this work we feel we have found a satisfactory solution.  The simplest one channel CST (used here) also has spurious singularities when the mass of the two-body system $W\to0$.  These lead to the existence of deeply bound states which are an artifact of the one channel approximation.  It is shown in Appendix \ref{App:C} that these bound states have no effect on any of the results of this paper, and can in fact be removed by artificially setting the kernel to zero for small masses $W$.

The OBE model appears to provide no rule for how many bosons to include in the kernel, and in the absence of such a rule seems to have little predictive power.  Strictly speaking this is objection cannot be answered, since the OBE model is, at its foundation, merely a phenomenology.  Still, we show in Appendix \ref{App:B} that the exchange of only scalar, pseudoscalar, vector (with both Dirac and Pauli couplings), and axial vector mesons, one of each isospin, is sufficient to describe the most general spin and isospin structure of an on-shell $NN$ kernel (but not the most general functional dependence, of course).  This provides a partial answer to this objection.

\section{On-shell prescription} \label{App:A}

In this appendix we discuss the nature of the requirement that one of the two nucleons is restricted to its positive energy mass-shell (by convention, nucleon 1).   If the particles are identical, identical results are obtained when nucleon 2 is on-shell.  In this appendix we represent the nucleon energy $\sqrt{m^2+{\bf k}^2}$ by $E(k)$ (instead of $E_k$ used in the rest of the paper).

\subsection{Non-identical particles}

If the particles are distinguishable, the spectator formalism places the heaver of the two (assumed to be particle 1 in this paper) on-shell.  The OBE  kernel is then
\bea
V({\bf p}, {\bf p}'; P)=\frac{N_1(p_1,p'_1)\,N_2(p_2,p'_2)}{m_{{\rm ex}}^2-(p- p')^2} \label{2.1}
\eea
where, in the rest frame with $P=\{W,{\bf 0}\}$,
\bea
&&p_1=\{E_1(p),{\bf p}\}\nonumber\\
&&p_2=\{W-E_1(p),-{\bf p}\}\; . \label{2.2}
\eea
with $E_1(p)=\sqrt{m_1^2+p^2}$.  The relative momentum in the rest frame is therefore 
\bea
p = \sfrac12(p_1-p_2)=\{\textstyle{E_1(p)-\frac{1}{2}}W, {\bf p}\}\, .\label{B4}
\eea
The kernel is written as a function of the relative three-momenta only, since the prescription that particle 1 is on-shell fixes the relative energy, and it cannot be changed.  In spite of this, the kernel is covariant because the mass shell constraint that defines the relative energy is covariant  

The denominator of (\ref{2.1}) is manifestly covariant, and depends on the energy difference $(p_0-p'_0)^2$, which is referred to as the retardation factor.  This factor plays an important role which we now discussed in detail. 

 To demonstrate the importance and role of retardation,  the denominator will be evaluated  in a frame where the total momentum is $P=\{P_0,{\bf 0}_\perp, P_\parallel\}$.   Explicitly,
\bea
D&=&\omega^2_{{\bf p}-{\bf p'}}-\left[E_1(p+{\textstyle \frac{1}{2}}P_\parallel)-E_1(p'+{\textstyle \frac{1}{2}}P_\parallel)\right]^2 ,\qquad	
\eea
where $\omega_{\bf q}=\sqrt{m_{{\rm ex}}^2+{\bf q}^2} $ is the mass-shell energy of the exchanged meson and $E_1(p+\sfrac12 P_\parallel)$ is a shorthand notation for $\sqrt{m_1^2+p_\perp^2+(p_\parallel+\sfrac12 P_\parallel)^2}$.
Assuming that the components of ${\bf p}$ and ${\bf p}'$ are much smaller than $E_{\parallel}\equiv E_1(\sfrac12 P_\parallel)$, and keeping terms to order $p^2$ only, the denominator may be expanded
\bea
D&\simeq& m_{{\rm ex}}^2 +({\bf p}_\perp-{\bf p}'_\perp)^2+(p_\parallel -p'_\parallel)^2\nonumber\\
&&-\Bigg\{E_{\parallel}\left[1+\frac{P_\parallel\,p_\parallel}{
2E^2_{\parallel}}+{\cal O}\left(\frac{p^2}{E^2_{\parallel}}\right)\right] \nonumber\\
&&\qquad-
E_{\parallel}\left[1+\frac{P_\parallel\,p'_\parallel}{
2E^2_{\parallel}}+{\cal O}\left(\frac{p'^2}{E^2_{\parallel}}\right)\right]  
\Bigg\}^2\nonumber\\
&\simeq&m_{{\rm ex}}^2 +({\bf p}_\perp-{\bf p}'_\perp)^2
+\frac{1}{\gamma^2}(p_\parallel -p'_\parallel)^2\nonumber\\
&=&\omega_{\bf p-p'}^2-\frac{\gamma^2-1}{\gamma^2}(p_\parallel-p'_\parallel)^2\
\label{B5}
\eea
where 
\bea
E_1({\textstyle \frac{1}{2}}P_\parallel)=E_{\parallel}=\gamma\,m_1\, .
\eea
Note that the retardation factor suppresses the dependence of kernel on the parallel component of $p$ and $p'$.  In the limiting case when $\gamma\to\infty$ the kernel (and hence the wave function) will not depend on the parallel component, leading to a coordinate space   wave function with a $\delta(r_\parallel)$ dependence.  This describes a wave function contracted into a thin disk in the direction of motion, as intuitively expected.  

\begin{figure*}
\centerline{
\mbox{
\includegraphics[width=5in]{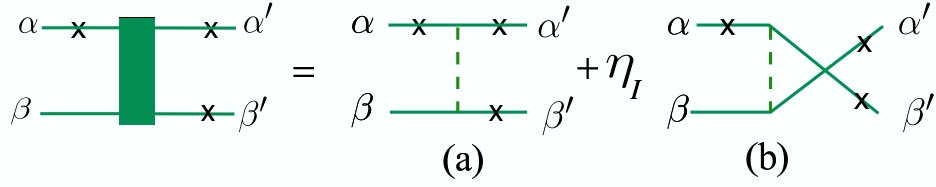}
}
}
\caption{\footnotesize\baselineskip=10pt (Color on line) Diagrammatic representation of the symmetrized OBE kernel given in Eq.~(\ref{VB}) (with the factor of 1/2 suppressed). Particle 1 in the final state is on-shell (denoted by the $\times$) and the particles in the initial state, with relative four-momentum $p^*$, are {\it both\/}  on-shell. With our convention, Feynman diagrams are written so that the end of each external line is always labeled with the same four-momentum and Dirac index ($\alpha,\alpha',\beta,\beta'$); this requires crossing the initial nucleon lines in the exchange term (b).  Diagrammatically, the symmetry relation (\ref{sym0}) corresponds to crossing the initial nucleon lines and interchanging the $\alpha'$ and $\beta'$ indices, giving the same two diagrams multiplied by the factor $\eta_I$. }
\label{fig1}
\end{figure*} 

If $m_2=m_1$, the same result emerges from the time ordered formalism (which, however, is not manifestly covariant).  In the time ordered formalism the kernel obtained from the two time orderings of the meson exchange (ignoring the numerators) is 
\bea
V_{_{\rm TO}}({\bf p},{\bf p}';P_\parallel)&=&\left[2\omega\left(\omega+
E_{1-} + E_{1+}'-{\cal E}\right)\right]^{-1}\nonumber\\
&&+\left[2\omega\left(\omega+
E_{1-}' + E_{1+}-{\cal E}\right)\right]^{-1}\qquad
\eea
where $E^{(\prime)}_{1\pm}=E_1(p^{(\prime)}\pm\sfrac12 P_\parallel)$, ${\cal E}=(W^2+P_\parallel^2)^{1/2}$ (with $W$ the mass of the two-body system), and $\omega=\omega_{{\bf p}-{\bf p}'}$.  Assuming that $p$ and $p'$ are much smaller than $E_\parallel$, that $2E_\parallel-{\cal E}\sim {\cal O}(p^2)$, and expanding the energy factors to order $p$ (neglecting terms of order $p^2$) gives
\bea
V_{_{\rm TO}}({\bf p},{\bf p}';P_\parallel)&\simeq&\left[2\omega\left(\omega-
\frac{P_\parallel}{2E_\parallel} (p_\parallel-p'_\parallel)\right)\right]^{-1}\nonumber\\
&&+\left[2\omega\left(\omega+
\frac{P_\parallel}{2E_\parallel} (p_\parallel- p'_\parallel)\right)\right]^{-1}\nonumber\\
&\simeq&\left[\omega^2-\frac{\gamma^2-1}{\gamma^2}(p_\parallel-p'_\parallel)^2\right]^{-1}\, ,
\eea
in agreement with the Spectator result (\ref{B5}).
This result has been obtained recently for QED in 1+1 dimension by J\"arvinen \cite{Jarvinen:2004pi}.

It can be shown that the Spectator results do {\it not\/} agree with the time ordered formalism  if $m_2\ne m_1$, or if the binding energy is not small. 

We think that the retardation factors may be partly responsible for the success of the OBE approximation to the CST theory.  We have seen that the retardation factors present in time-ordered theory are similar (at least for nonrelativistic energies), so a time-ordered calculation might also enjoy similar success if its retardation factors were retained.

\subsection{Identical particles}

The extension of the Spectator theory to identical particles is less than straightforward, and is perhaps one reason why it has not been used more widely.   Here we review and describe the justification and motivation for the prescription used previously, which has been found to have many advantages for applications.

\subsubsection{Kernel for both particles on-shell in the initial state}

Begin with the case when both particles in the initial state are on-shell.  In this case the prescription  is unique, and free from problems.  We antisymmetrize the kernel in the initial state, as shown in Fig.~\ref{fig1}.  The explicit form of the kernel is
\bea
&&V^1_{\alpha\alpha',\beta\beta'}({\bf p}, {\bf p^*};P)=\frac{1}{2}\Bigg\{ \frac{N_{\alpha\alpha'}(p_1, p^*_1)\,
N_{\beta\beta'}(p_2, p^*_2)}
{m_{\rm ex}^2-(p-p^*)^2}\nonumber\\
&&\qquad\qquad+\eta_I\;\frac{N_{\alpha\beta'}(p_1, p^*_2)\, 
N_{\beta\alpha'}(p_2, p^*_1)}
{m_{\rm ex}^2-(p+p^*)^2} \Bigg\}\, , \label{VB}
\eea
where $\eta_I$ is a phase depending on the isospin of the $NN$ channel under consideration, the superscript 1 denotes the fact that particle 1 is on-shell in the final state, and $p^*$ is the relative four-momentum when {\it both\/} particles in the initial state are on-shell.  In the two-body rest system, where $P=\{W,{\bf 0}\}$,
\bea
p^*&=&\{0, {\bf p^*}\}\, ;\qquad |{\bf p^*}|^2=\sfrac14 W^2-M^2\nonumber\\
p^*_1&=&\{\sfrac12 W, {\bf p^*}\}\nonumber\\
p^*_2&=&\{\sfrac12 W, -{\bf p^*}\} \label{star}
\eea
The antisymmetrized kernel satisfies the relation
\bea
V^1_{\alpha\alpha',\beta\beta'}({\bf p}, {\bf p^*};P)=\eta_I V^1_{\alpha\beta',\beta\alpha'}({\bf p}, -{\bf p^*};P)\, . \label{sym0}
\eea
%
\begin{figure}
\centerline{
\mbox{
\includegraphics[width=3.0in]{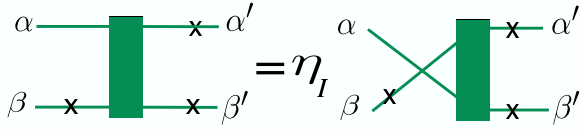}
}
}
\caption{\footnotesize\baselineskip=10pt (Color on line) Diagrammatic representation of the interchange rule for the symmetrized kernel, Eq.~(\ref{sym2}).  Here the solid box represents the full kernel, which is the sum of the two diagrams as shown in Fig.~\ref{fig1}.  The left-hand diagram is $V^2$ and the right hand diagram is $V^1$ (with indices and momenta exchanged). }
\label{fig2}
\end{figure} 

Note that, since the particles are identical, we could just as well have started with the case when  particle 2 is on-shell in the final state (instead of particle 1).  To distinguish this case from the standard case (particle 1 on shell) we make the substitution $p\to\hat p$ so that in the rest system
\bea
\hat p_1&=&\{W-E(p),{\bf p}\}=\sfrac12 P+\hat p\nonumber\\
\hat p_2&=&\{E(p),-{\bf p}\}=\sfrac12 P-\hat p\nonumber\\
\hat p&=&\sfrac12(\hat p_1-\hat p_2)=\{\sfrac12 W-E(p), {\bf p}\}\, . \label{2.12}
\eea
We observe that $\hat p$ differs from $p$ only in the sign of $p_0$, so they are identical when $p_0=0$ ({\it both\/} particles are on shell).  The kernel for particle 2 on-shell can be obtained from (\ref{VB}) if $\hat p$ is substituted for $p$ everywhere.   Explicitly, 
\bea
&&V^2_{\alpha\alpha',\beta\beta'}({\bf p}, {\bf p^*};P)=\frac{1}{2}\Bigg\{ \frac{N_{\alpha\alpha'}(\hat p_1, p^*_1)\,
N_{\beta\beta'}(\hat p_2, p^*_2)}
{m_{\rm ex}^2-(\hat p-p^*)^2}\nonumber\\
&&\qquad\qquad +\eta_I\;\frac{N_{\alpha\beta'}(\hat p_1, p^*_2)\, 
N_{\beta\alpha'}(\hat p_2, p^*_1)}
{m_{\rm ex}^2-(\hat p+p^*)^2} \Bigg\}\, , \label{VB1}
\eea
where the superscript 2 refers to particle 2 on-shell.  Under the interchange of ${\bf p}\to -{\bf p}$, the momenta are mapped as follows
\bea
\hat p_1&\to& p_2\nonumber\\
\hat p_2&\to& p_1\nonumber\\
\hat p&\to& -p\, .
\eea
This implies that 
\bea
V^2_{\alpha\alpha',\beta\beta'}({\bf p}, {\bf p^*};P)=\eta_I V^1_{\beta\alpha',\alpha\beta'}(-{\bf p}, {\bf p^*};P)\, . \label{sym2}
\eea
%
\begin{figure*}
\centerline{
\mbox{
\includegraphics[width=5in]{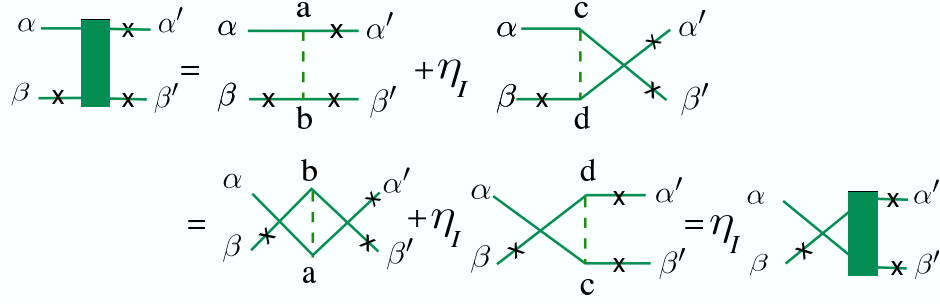}
}
}
\caption{\footnotesize\baselineskip=10pt (Color on line) Drawing showing how the interchange rule of Fig.~\ref{fig2} [and Eq.~(\ref{sym2})] can be derived diagrammatically.  All Feynman diagrams are equal to their  twisted versions.  In this example we start with $V^2$, equal to the two boson exchange diagrams shown in the first row.  Each diagram is then twisted  [obtained, in this example, by exchanging the vertices $a\leftrightarrow b$ and $c\leftrightarrow d$ while leaving the labeling of all external particles unchanged], as shown in the second row, and then the two twisted diagrams are  collected into $\eta_I V^1$ (with a change in the sign of the final state three-momentum and the final Dirac indices exchanged, as suggested by the labeling of the diagram).}
\label{fig3}
\end{figure*} 

This is the symmetry we want to preserve in the following discussion:  {\it in the rest frame, the amplitude for particle 1 on-shell can be obtained from that for particle 2 on-shell by interchanging Dirac indices, multiplying by the phase $\eta_I$, and changing the sign of the relative three-momentum.\/}   This interchange rule, illustrated in Fig.~\ref{fig2}, is the proper way to apply the Pauli principle to a state with one particle off-shell and one particle on-shell.

It is instructive to see how the interchange rule can be derived directly from the Feynman diagrams of Fig.~\ref{fig1}.  This is illustrated in Fig.~\ref{fig3}.  Here we use the fact that, if the labeling of the external particles is unchanged, any Feynman diagram can be ``twisted'' [i.e. all of the internal vertices on the top of the diagram exchanged with those on the bottom] without changing its value.

\subsubsection{Calculation of the half on-shell scattering amplitude}

Armed with the kernel (\ref{VB}), the {\it half on-shell scattering amplitude\/} (defined to be the scattering amplitude with both particles on-shell in the {\it initial\/} state) can be calculated.  As a first (incorrect) attempt, consider the uncoupled integral equations for the amplitudes $M^1$ and $M^2$ shown in Fig.~\ref{fig4}.   These equations define the scattering amplitudes as the infinite series generated by the initial on-shell interaction $V^{1}$ or $V^{2}$  followed successive exchanges of the unsymmetrized kernels $V^{11}$ (for $M^1$) or $V^{22}$ (for $M^2$), where
\bea
V^{11}_{\alpha\alpha',\beta\beta'}({\bf p}, {\bf k};P)&=&\frac{N_{\alpha\alpha'}(p_1, k_1)\,
N_{\beta\beta'}(p_2, k_2)}
{m_{\rm ex}^2-(p-k)^2} \nonumber\\
V^{22}_{\alpha\alpha',\beta\beta'}({\bf p}, {\bf k};P)&=&\frac{N_{\alpha\alpha'}(\hat p_1, \hat k_1)\,
N_{\beta\beta'}(\hat p_2, \hat k_2)}
{m_{\rm ex}^2-(\hat p-\hat k)^2} \, , \qquad\label{direct}
\eea
 where $p_1$, etc., were previously defined in Eq.~(\ref{2.2}) and $\hat p_1$, etc., were defined in Eq.~(\ref{2.12}), and the $k$'s and $\hat k$'s are similarily defined in terms of ${\bf k}$ instead of ${\bf p}$.  Note that $V^{22}$ has particle 2 on-shell in both the initial and final state, while particle 1 is on-shell in $V^{11}$.  

\begin{figure*}
\centerline{
\mbox{
\includegraphics[width=4.5in]{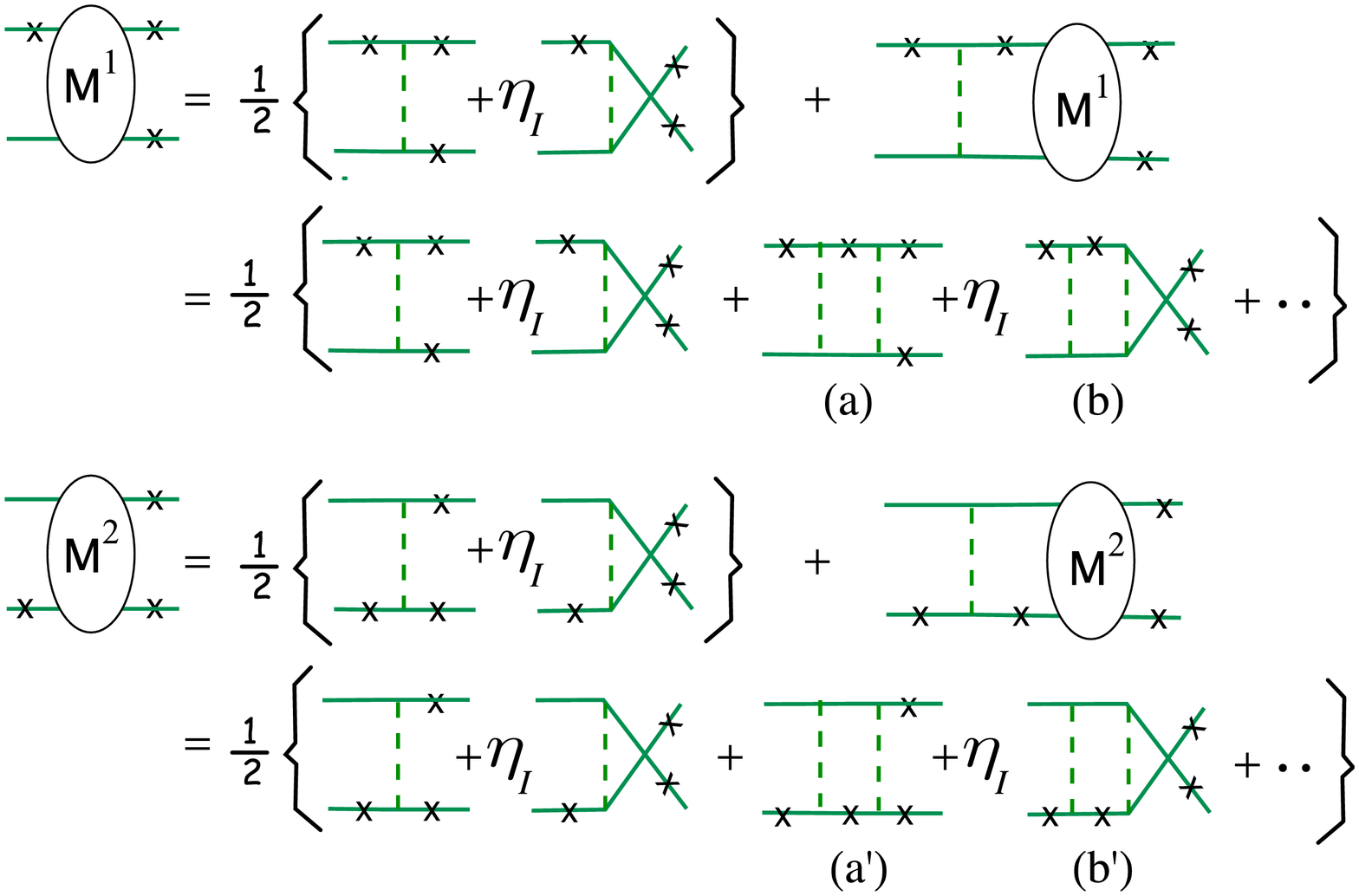}
}
}
\caption{\footnotesize\baselineskip=10pt (Color on line) Diagrammatic representation of (incorrect) uncoupled integral equations for the half on-shell scattering amplitudes $M^1$ and $M^2$.  The proposed equations are given on the first and third lines. The second and forth lines give the scattering amplitudes iterated to 4th order.  These equations satisfy the reflection property (\ref{sym2}), but are unsatisfactory because {\it when both final state particles are on-shell\/} diagrams (a) and (b) are not equal to diagrams (a') and (b'). }
\label{fig4}
\end{figure*} 

\begin{figure*}
\centerline{
\mbox{
\includegraphics[width=5.9in]{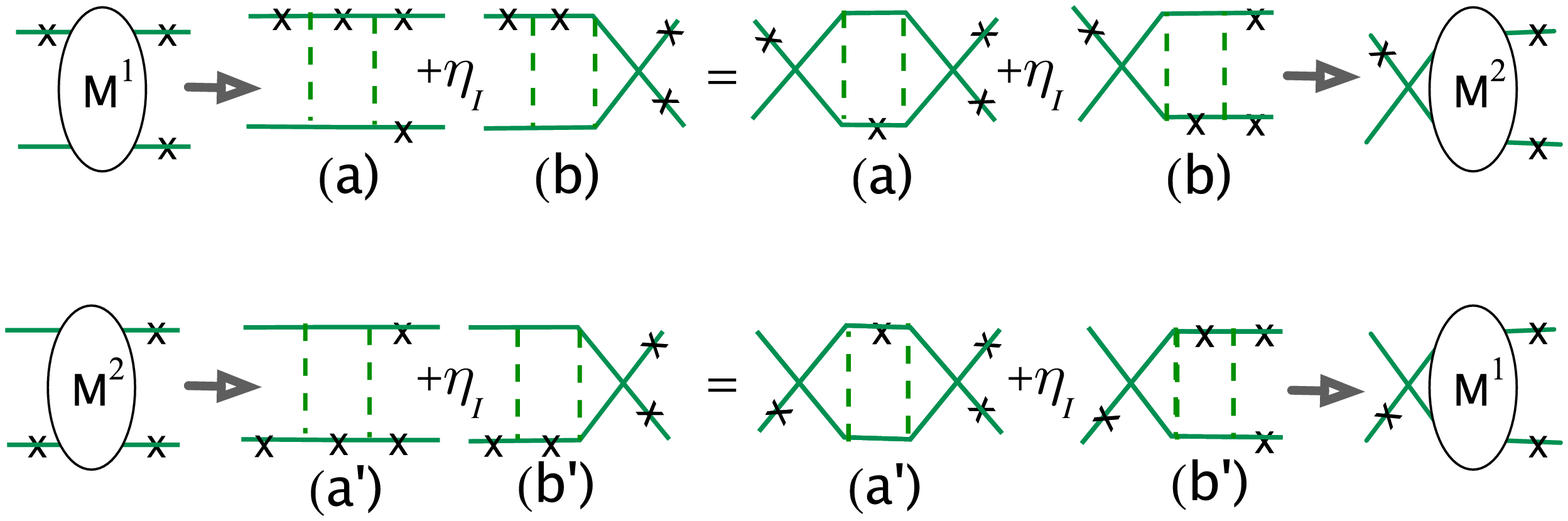}
}
}
\caption{\footnotesize\baselineskip=10pt (Color on line) Top row: diagrams (a) and (b) from the 4th order expansion of $M^1$; bottom row: diagrams (a') and (b') from the 4th order expansion of $M^2$ (in both cases the overall factor of $\frac12$ has been suppressed).  The right shows the twisted versions of each diagram on the left.  Comparison of the two top-left diagrams with the two bottom-right ones shows that the symmetry (\ref{sym2}) is indeed satisfied (and similarly for the top-right and bottom-left).  However, when both final state particles are on shell, these amplitudes are not equal; for example the leftmost figures (a) and (a') differ by which particle is on-shell inside the box. }
\label{fig5}
\end{figure*} 

\begin{figure*}
\centerline{
\mbox{
\includegraphics[width=5.0in]{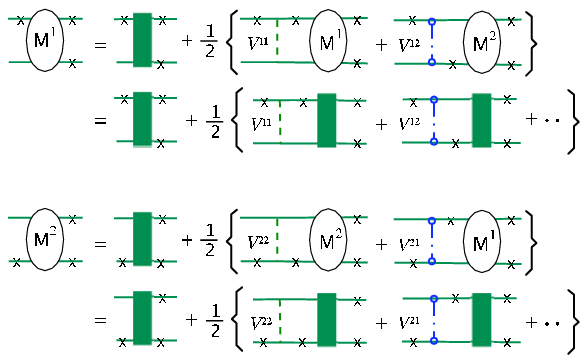}
}
}
\caption{\footnotesize\baselineskip=10pt (Color on line) Diagrammatic representation of coupled integral equations for the half on-shell scattering amplitudes $M^1$ and $M^2$.  The equations are given on the first and third lines. The second and forth lines give the scattering amplitudes iterated to 4th order.  These equations satisfy the reflection property (\ref{sym2}), and also give identical results when both final state particles are on-shell, provided that, for this on-shell point, $V^{22}=V^{12}$ and $V^{11}=V^{21}$.}
\label{fig6}
\end{figure*} 

The advantage of the equations in Fig.~\ref{fig4} is that the kernels (\ref{direct}) are free of singularities for all values of the three momenta (this will be discussed in more detail below).   Furthermore, by twisting the diagrams as demonstrated in Fig.~\ref{fig3}, it is easy to see that these two amplitudes are related by the generalized Pauli principle (\ref{sym2}).  This is illustrated in Fig.~\ref{fig5}.

The problem with this prescription is subtle.  When both particles in the final state are on-shell,  we obtain two different answers, depending on whether or not we start with $M^1$ or $M^2$.  These differences are illustrated in Fig.~\ref{fig5}.  Explicitly, consider the leftmost diagrams (a) and (a').  If both final particles are on shell, these boxes are
\begin{widetext}
\bea
M^{\rm (a)}_{\alpha\alpha',\beta\beta'}({\bf p'^*}, {\bf p^*};P_0)&=&\int\frac{d^3k}{(2\pi)^3} 
\frac{ V^{1}_{\alpha\gamma,\beta\delta}({\bf p'^*}, {\bf k};P)\Lambda_{\gamma\gamma'}(k_1) \Lambda_{\delta\delta'}(k_2)
V^{1}_{\gamma'\alpha',\delta'\beta'}({\bf k}, {\bf p^*};P)}
{2E(k)W[2E(k)-W]}
\nonumber\\
M^{\rm (a')}_{\alpha\alpha',\beta\beta'}({\bf p'^*}, {\bf p^*};P_0)&=&\int\frac{d^3k}{(2\pi)^3} 
\frac{ V^{2}_{\alpha\gamma,\beta\delta}({\bf p'^*}, {\bf k};P)\Lambda_{\gamma\gamma'}(k_1) \Lambda_{\delta\delta'}(k_2)
V^{2}_{\gamma'\alpha',\delta'\beta'}({\bf k}, {\bf p^*};P)}
{2E(k)W[2E(k)-W]}\, ,\qquad
\eea
where $P_0=\{W,{\bf 0}\}$ and here we use the notation $\Lambda(k)=m+\not\!k$.
For a simple scalar meson exchange, for example, these become
\bea
M^{\rm (a)}_{\alpha\alpha',\beta\beta'}({\bf p'^*}, {\bf p^*};P_0)&=& \int_k
\frac{ (m+E(k)\gamma^0-{\bf k}^i\gamma^i)_{\alpha\alpha'} (m +[W-E(k)]\gamma^0+{\bf k}^i\gamma^i)_{\beta\beta'}}
{2E(k)-W}
\nonumber\\
M^{\rm (a')}_{\alpha\alpha',\beta\beta'}({\bf p'^*}, {\bf p^*};P_0)&=&\int_k
\frac{ (m+[W-E(k)]\gamma^0-{\bf k}^i\gamma^i)_{\alpha\alpha'} (m +E(k)\gamma^0+{\bf k}^i\gamma^i)_{\beta\beta'}}
{2E(k)-W}\, ,\qquad
\eea
\end{widetext}
where
\bea
\int_k\equiv\int\frac{d^3k}{(2\pi)^3} \frac{1}{2E(k)W \Delta({\bf p'^*},{\bf k}) \Delta({\bf k},{\bf p^*})}\, ,
\eea
and the denominators of the meson propagators are $\Delta({\bf p},{\bf k})=\omega^2_{{\bf p}-{\bf k}}-[E(p)-E(k)]^2$.  When the two final particles are on shell, the difference between these two should be zero, but instead it is
\bea
&&\!\!\!\!\Delta M\equiv M^{\rm (a')}-M^{\rm (a)}=\nonumber\\
&&\int_k\Big\{ (m-{\bf k}^i\gamma^i)_{\alpha\alpha'} \gamma^0_{\beta\beta'}-\gamma^0_{\alpha\alpha'}
(m+{\bf k}^i\gamma^i)_{\beta\beta'} \Big\}\, .\qquad\quad
\eea
To be convinced that this is not zero, pick the special case of forward scattering ${\bf p'^*}={\bf p^*}\equiv{\bf p}$ and evaluate between positive energy spinors with spins $\lambda_1=\lambda'_1$ and  $\lambda_2=\lambda'_2$.  The only component of ${\bf k}$ that does not integrate to zero in the forward direction is that in the ${\bf p}$ direction, and we may substitute ${\bf k}^i\to({\bf k}\cdot{\bf p}) {\bf p}^i/p^2$, giving
\bea
\bar u_\alpha({\bf p},\lambda_1) \bar u_\beta(-{\bf p},\lambda_2)\;\Delta M \;
u_{\alpha'}({\bf p},\lambda_1) u_{\beta'}(-{\bf p},\lambda_2) \nonumber\\
= \int_k\frac{2{\bf k}\cdot{\bf p} E(p)}{m^2}\ne 0\, .\qquad
\eea
(As written, these integrals do not converge, but this is readily corrected by adding form factors, which are identical for both terms and so do not change the results.  In order to simplify the equations, these form factors have been omitted.)

The way to correct this problem is to couple the two equations together, as shown in Fig.~\ref{fig6}.  Here the off-shell exchange kernels $V^{12}$ and $V^{21}$ are drawn using a dot-dashed line with a small circle at each end, in anticipation of the fact that their definition could differ from the standard meson propagators used in $V^{11}$ and $V^{22}$.   When both final state particles are on-shell, the amplitudes $M^1$ and $M^2$ will now be equal, provided the following conditions are satisfied
\bea
V^{12}_{\alpha\alpha',\beta\beta'}({\bf p^*}, {\bf k};P)&=& V^{22}_{\alpha\alpha',\beta\beta'}({\bf p^*}, {\bf k};P) \nonumber\\
V^{21}_{\alpha\alpha',\beta\beta'}({\bf p^*}, {\bf k};P)&=& V^{11}_{\alpha\alpha',\beta\beta'}({\bf p^*}, {\bf k};P)\, , \label{sym3}
\eea
where, as before, the magnitude of ${\bf p^*}$ is given by the mass shell condition (\ref{star}).

 The definition of these kernels implied by field theory (Model B of Ref.~I) is 
\bea
V^{12}_{\alpha\alpha',\beta\beta'}({p}, {\bf k};P)&=&\frac{N_{\alpha\alpha'}(p_1, \hat k_1)\,
N_{\beta\beta'}(p_2, \hat k_2)}
{m_{\rm ex}^2-(p-\hat k)^2}  \nonumber\\
V^{21}_{\alpha\alpha',\beta\beta'}({\bf p}, {\bf k};P)&=&\frac{N_{\alpha\alpha'}(\hat p_1, k_1)\,
N_{\beta\beta'}(\hat p_2, k_2)}
{m_{\rm ex}^2-(\hat p-k)^2}\, . \qquad\label{exchange}
\eea
When the two final particles are on shell, the variables with and without hats are equal, so the constraints (\ref{sym3}) are automatically satisfied [compare Eq.~(\ref{direct})].  However, the definition (\ref{exchange}) is not ideal because off-shell these kernels have singularities.

\subsubsection{Singularities in the exchange kernels and their removal} \label{App:A3}

To see how these singularities arise, look at the exchange denominator for $V^{21}$ in the rest frame.  If we define
\bea
q_{\rm ex}^2(z)&=&(\hat p-k)^2
\nonumber\\
&=&[W-E(p)-E(k)]^2-{\rm p}^2-{\rm k}^2+2{\rm p}{\rm k}z\qquad
\eea
where ${\rm p}=|{\bf p}|$, ${\rm k}=|{\bf k}|$, and  $z=({\bf p}\cdot{\bf k})/{\rm pk}$ is the cosine of the angle between ${\bf p}$ and ${\bf k}$, the denominator can be written
\bea
m_{\rm ex}^2 -q_{\rm ex}^2(z) &=& m_{\rm ex}^2 +({\bf p}-{\bf k})^2-[W-E(p)-E(k)]^2\nonumber\\
&=&[\omega_{{\bf p}-{\bf k}}+W-E(p)-E(k)] \nonumber\\
&&\times[\omega_{{\bf p}-{\bf k}}-W+E(p)+E(k)]\, .\qquad
\eea
When ${\bf p}-{\bf k}=0$, for example, this denominator is zero at
\bea
W=2E(p)\pm m_{\rm ex}\, .
\eea
One of these zeros appears only when $W\ge 2m+m_{\rm ex}$, corresponding to the singularity associated with meson production.  This is a physical singularity and is avoided by working at energies below the production threshold.  The second zero at $W = 2E(p)-m_{\rm ex}$ 
occurs at all physical energies for values of the three-momentum
\bea
p^2\ge \frac{(W+m_{\rm ex})^2}{4}-m^2 \simeq (368\; {\rm MeV})^2\, ,
\eea
where 368 MeV is the threshold for pion exchange when $W=2m$.   
This is an unphysical singularity.  In Ref.~I it was shown that this singularity cancels exactly when the kernel is calculated to all orders.  Hence, the cancellation of the imaginary part is easily implemented order by order by simply dropping it and treating the singularity as a principal value.  In Ref.~I, one of the models studied (referred to as Model B) evaluated these principal values numerically.  This direct  calculation of the principal values was somewhat complicated, numerically inaccurate, and hard to extend to calculations of electromagnetic currents and the three-body system.  The effort would be justified if the treatment of these singularities were physically significant, but since they are cancelled by higher order terms in the kernel, it is desirable to find a way to remove them, order by order, justÊas the imaginary parts are removed.  

\begin{figure}
\centerline{
\mbox{
\includegraphics[width=3.0in]{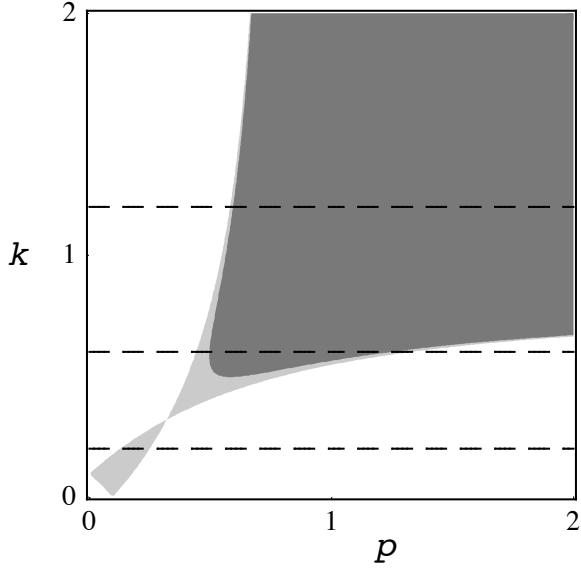}
}
}
\caption{\footnotesize\baselineskip=10pt  Here $p$ and $k$ are in units of $m$, with $W=2.1\, m$.  The dark shaded area is the region where $q_{\rm ex}^2(1)>m^2_{\rm ex}$, with $m_{\rm ex}$ equal to the pion mass.  The light shaded area is the region where $m^2_{\rm ex}>q_{\rm ex}^2(1)>0$, and the white area has $0>q_{\rm ex}^2(1)$ [except for the tiny triangular region near the origin where $q_{\rm ex}^2(1) \geq q_{\rm ex}^2(-1)>0$].  The singularities occur along the boundary between the dark and light shaded regions.  The three horizontal lines mark $k=0.2$, 0.6, and 1.2. }
\label{fig7}
\end{figure} 

To explore the nature of these singularities, and to see how they could be removed, it is sufficient to consider the $S$-wave projection of the singular meson propagator in $V^{21}$,
\bea
V_{\rm ex}(p,k)\equiv \int_{-1}^1 dz\; \frac{1}
{m_{\rm ex}^2-q_{\rm ex}^2(z)}  \, . \label{Vexc}
\eea
Using the principal value prescription for the $z$ integration, the function $V_{\rm ex}$ is
\bea
V_{\rm ex}(p,k)=\frac{1}{2{\rm pk}}\ln
\frac{|m_{\rm ex}^2-q_{\rm ex}^2(1)|}
{|m_{\rm ex}^2-q_{\rm ex}^2(-1)|}\, . \label{Vex}
\eea
For comparison, the $S$-wave projection of the direct kernel (denoted $V_{\rm direct}$) is proportional to the same function with $q^2_{\rm ex}(z)\to q^2_{\rm dir}(z)=[E(p)-E(k)]^2-\omega_{\bf p-k}^2$.  This propagator is not singular, as mentioned above.

The locus of the singularities of $V_{\rm ex}$ in the $p,k$ plane is shown in Fig.~\ref{fig7}.   In this example the momenta are expressed in units of the nucleon mass, $W=2.1\,m$, and the exchanged mass is the pion mass.  The three lines at fixed $k_i$ mark regions where $V_{\rm ex}$ has 0, 2, or 1 singularities in $p$. The functions $F(p)=V_{\rm ex}(p,k_i)$ at these three fixed values of $k_i$ are shown in Fig.~\ref{fig8}, with the smooth functions $V_{\rm direct}(p,k_i)$ shown for comparison.  The singularities are sharp, narrow spikes that clearly represent unphysical behavior.  For small $p$ (below the region of singularities) $V_{\rm direct}(p,k)\simeq V_{\rm ex}(p,k)$.  In the singular region,  $V_{\rm direct}(p,k_i)$ gives roughly a $p$-averaged value of $V_{\rm ex}(p,k_i)$ up to $p\simeq 1$, and at larger $p$ it looks like  $V_{\rm direct}(p,k_{i})$ gives roughly a $k$-averaged value of $V_{\rm ex}(p,k_{i})$.   In all, it looks like $V_{\rm direct}(p,k)\simeq$ an average value of $V_{\rm ex}(p,k)$ over the entire region of $k,p$ space. 

\begin{figure}
\centerline{
\mbox{
\includegraphics[width=3.5in]{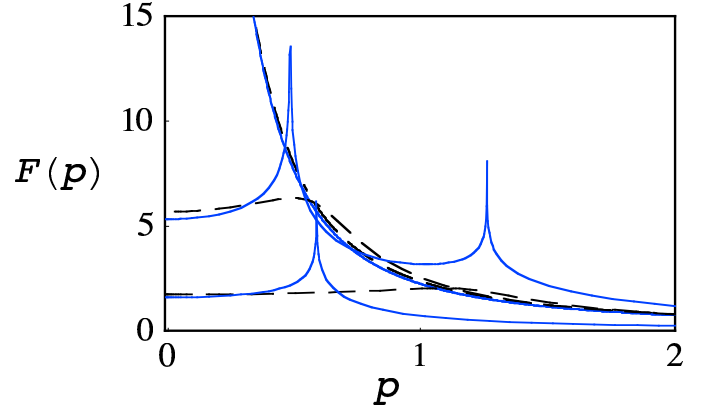}
}
}
\caption{\footnotesize\baselineskip=10pt (Color on line) Plots of the 6 dimensionless functions $F(p)=V_{\rm ex}(p,k_i)$ (solid lines) and $F(p)=V_{\rm direct}(p,k_i)$ (dashed lines) for the dimensionless values of $k_{i}=0.2, 0.6$ and 1.2 as shown in Fig.~\ref{fig7}.  The curves can be distinguished by looking at their behavior at small $p$, which is roughly proportional to $1/k_{i}$.}
\label{fig8}
\end{figure} 

In the work of Ref.~I, the two cases shown in Fig.~\ref{fig8} were both studied.  Model A replaced the {\it denominators\/} of the exchanged terms with the denominators of the direct term, leaving the numerators unchanged.  Specifically Model A employed the following prescriptions for the exchange kernels $V^{12}$ and $V^{21}$
\bea
V^{12}_{\alpha\alpha',\beta\beta'}({\bf p}, {\bf k};P)\to
V^{12\;A}_{\alpha\alpha',\beta\beta'}({\bf p}, {\bf k};P)\qquad\nonumber\\
= \frac{N_{\alpha\alpha'}(p_1, \hat k_1)\, 
N_{\beta\beta'}(p_2, \hat k_2)}
{m_{\rm ex}^2-(p+\hat k)^2} \, \nonumber\\
V^{21}_{\alpha\alpha',\beta\beta'}({\bf p}, {\bf k};P)\to
V^{21\;A}_{\alpha\alpha',\beta\beta'}({\bf p}, {\bf k};P)\qquad\nonumber\\
= \frac{N_{\alpha\alpha'}(\hat p_1, k_1)\, 
N_{\beta\beta'}(\hat p_2, k_2)}
{m_{\rm ex}^2-(\hat p+ k)^2} \, , \label{VA}
\eea
where, in the rest frame the relative four-momentum $\hat k$ is defined as in Eq.~(\ref{2.12}).  Note that 
\bea
p=p_1-\sfrac12P=\{E(p)- \sfrac12W,{\bf p}\}=\hat p-2\frac{\hat p\cdot P\; P}{P^2}\,  \qquad\,\,\nonumber\\
\hat p=\sfrac12P-\hat p_2=\{ \sfrac12W-E(p),{\bf p}\}=p-2\frac{p\cdot P\; P}{P^2}\, . \qquad\label{ks}
\eea
These equalities show how $\hat p$ ($p$) can be expressed in terms of $p$ ($\hat p$).   In the rest frame the denominators of (\ref{VA}) are now 
\bea
m_{\rm ex}^2-(p+\hat k)^2&=&\omega_{{\bf p}+{\bf k}}^2-[E(p)-E(k)]^2\nonumber\\
m_{\rm ex}^2-(\hat p+ k)^2&=&\omega_{{\bf p}+{\bf k}}^2-[E(p)-E(k)]^2\, ,\qquad
\label{denoms}
\eea
equal (except for the sign of ${\bf p}$) to the denominators of the direct term.  

A major advantage of this prescription is that it satisfies the important reflection property (\ref{sym2}) and the constraints (\ref{sym3}).  In fact, when either the initial or final state is on-shell, the exchange terms in (\ref{VA}) are indistinguishable from the field theory forms (\ref{exchange}).

However, to complete a program of electromagnetic few body calculations requires that exchange (or interaction) currents be found that are consistent with the interaction kernel.  Unfortunately, finding interaction currents consistent with Model A is far from straightforward.  The momentum carried by the exchange term in Model A is not related to the momentum transferred at the vertices (i.e. $\hat p +k \ne k_1-p_2$ for example) and hence the exchange term cannot be easily related to any kind of OBE mechanism.  The Model A exchange term is a phenomenological 4-point function, and while it is possible to find interaction currents consistent with a 4-point function, the lack of a meson exchange structure means that field theory is not very helpful is guiding its construction. Ultimately this current must be found phenomenologically, with resulting ambiguities, and much of the value of the connection between field theory and the Spectator theory is lost.  An additional related problem is that it is not clear how to use Model A to define amplitudes when {\it both\/} particles in the final (or initial state) are off-shell, and such amplitudes are needed for complete electromagnetic calculations.  While it is certainly possible to use Model A for a calculation of $NN$ and $NNN$ wave functions, scattering amplitudes, and simple electromagnetic observables, these shortcomings lead to the consideration of other options. 

{\bf Model C:}  In applications to electromagnetic interactions, where it is necessary to know both the interaction current and the extension of the kernel to cases where {\it both\/} final (or initial) nucleons are off-shell, there are significant advantages in retaining the basic OBE structure of the kernel.  The essential feature of the OBE structure is that there is a meson propagator that depends {\it only\/} on the square of the four-momentum $q^2$, and that $q$ is equal to the momentum transferred between the nucleons.  If the kernel has this form, it is known how to calculate consistent interaction currents.   Furthermore, if the functional form of the propagator is altered so that it has no singularities for any real value of $q^2$, it is straightforward to use it even when both nucleons are off-shell.

\begin{figure}
\centerline{
\mbox{
\includegraphics[width=3.5in]{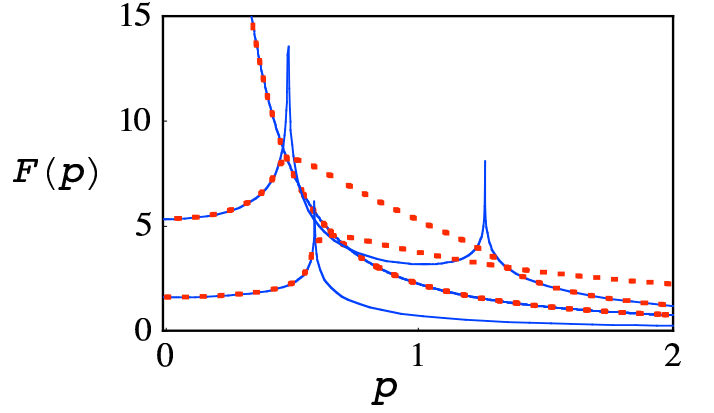}
}
}
\caption{\footnotesize\baselineskip=10pt (Color on line) Plots of the 6 dimensionless functions $F(p)=V_{\rm ex}(p,k_i)$ (solid lines) and $F(p)=V^C_{\rm ex}(p,k_i)$ (dotted lines) for the dimensionless values of $k_{i}=0.2, 0.6$ and 1.2 as shown in Fig.~\ref{fig7}.  The curves can be distinguished by looking at their behavior at small $p$, which is roughly proportional to $1/k_{i}$.}
\label{fig8a}
\end{figure} 

After some consideration of these issues, we settled on a very simple and straightforward prescription that satisfies all of the requirements outlined above.  Simply stated, the prescription is to replace the four momentum transfer in all OBE expressions by the negative of its absolute value
\bea
q^2\to -|q^2|\, .
\eea
Since $q^2$ is always negative in the direct terms, this will not alter the direct terms at all, preserving the basic results of the Spectator theory when the particles are not identical so that only one particle is on-shell.  For the exchange terms, the illustrative integral (\ref{Vexc}) becomes
\bea
V^C_{\rm ex}(p,k)\equiv \int_{-1}^1 dz\; \frac{1}
{m_{\rm ex}^2+|(\hat p-k)^2|}  \, . \label{Vexd}
\eea
If $-q^2=B(p,k)-2pkz$, with $B(p,k)=p^2+k^2-q_0^2$, $q^2$ will change sign whenever
\bea
-2pk<B(p,k)<2pk\, .
\eea
Hence the integral $V^C$ becomes
\bea
V^C_{\rm ex}(p,k)= \frac{1}{2pk}\ln R\, ,\label{Vabs}
\eea
where the form of $R$ depends on on the sign of $q^2$, with
\bea
R=\begin{cases}{\displaystyle\frac{m_{\rm ex}^2+B+2pk}{m_{\rm ex}^2+B-2pk}} & 2pk<B\cr &\cr
{\displaystyle\frac{\left(m_{\rm ex}^2+2pk\right)^2-B^2} {m_{\rm ex}^4}}&-2pk<B<2pk\quad\cr & \cr
{\displaystyle\frac{m_{\rm ex}^2-B+2pk}{m_{\rm ex}^2-B-2pk}} & B<-2pk\, .
\end{cases}
\eea
Note that both $V^C$ and $dV^C/dp$ (with $k$ held constant) are continuous.  These functions are compared with $V_{\rm ex}$ in Fig.~\ref{fig8a}.   They interpolate between the singularities, just as Model A did, but, as discussed above, the construction of exchange currents for Model C is more straightforward than for Model A.   All of the results presented in this paper use this Model C prescription.

\section{General form of the on-shell $NN$ kernel} \label{App:B}

In this Appendix we show that the 8 meson exchanges  used in this calculation are sufficient to describe the {\it most general spin and isospin structure\/} of the $NN$ kernel in the case when all of the external nucleons are on their mass-shell. 

As is well known, the most general 4 $\times$ 4 Dirac matrix can be expanded in terms of the  16 bi-linear covariants, $\bm{1}$, $\gamma^\mu$, $\sigma^{\mu\nu}$, $\gamma^5\gamma^\mu$, and $\gamma^5$.  Requiring that the $NN$ kernel be covariant, gives the most general expansion
\bea
{\cal V}_{12}&=& F_s \bm{1}_1\bm{1}_2 + F_v (\gamma^\mu)_1 (\gamma_{\mu})_2 + F_t (\sigma^{\mu\nu})_1(\sigma_{\mu\nu})_2 \nonumber\\
&&+ F_a (\gamma^5\gamma^\mu)_1 (\gamma^5\gamma_\mu)_2 + F_p \gamma^5_1\gamma^5_2\, ,  \label{B1}
\eea
where (suppressing the nucleon spin indices) ${\cal O}_i=\bar u(p_i) {\cal O} u(k_i)$ are the nucleon matrix elements of the operators.  If the particles are off shell, there are a great many more possible terms \cite{Tjon:1985cv}.

The on-shell OBE  kernel has the form 
\begin{align}
&{\cal V}^{\rm OBE}_{12}= f_s \bm{1}_1\bm{1}_2 + f_a (\gamma^5\gamma^\mu)_1 (\gamma^5\gamma_\mu)_2 + f_p \gamma^5_1\gamma^5_2 \nonumber\\
&\qquad+f_v \left[\gamma^\mu+\frac{\kappa_v}{2m}i\sigma^{\mu\alpha}q_\alpha\right]_1 \left[\gamma_\mu-\frac{\kappa_v}{2m}i\sigma_{\mu}{}^{\beta}q_\beta\right]_2\, ,
\end{align}
where $q=p_1-k_1=k_2-p_2$ and we have used the fact that the $q^\mu q^\nu/m_v^2$ term in the propagator  of the vector meson reduces to zero when the nucleons are on shell.  Using the well known Gordon decomposition for an on shell particle
\bea
i\sigma^{\mu\nu}(p_i-k_i)_\nu=2m\gamma^\mu -Q_i^\mu\, ,
\eea
where $Q_i\equiv p_i+k_i$, we can transform the OBE term into the form
\begin{align}
&{\cal V}^{\rm OBE}_{12}= f'_s \bm{1}_1\bm{1}_2 + f_a (\gamma^5\gamma^\mu)_1 (\gamma^5\gamma_\mu)_2 + f_p \gamma^5_1\gamma^5_2 \nonumber\\
&\quad+f'_v (\gamma^\mu)_1 (\gamma_\mu)_2 -f_v\frac{\kappa_v(1+\kappa_v)}{2m}\left[(\slashed{Q}_2)_1\bm{1}_2+\bm{1}_1(\slashed{Q}_1)_2\right]\, , \label{B4}
\end{align}
where
\bea
f'_s&=&f_s+f_v\frac{\kappa_v^2}{4m^2}\; Q_1\cdot Q_2\nonumber\\
f'_v&=&f_v(1+\kappa_v)^2\, .
\eea
Note that this expansion has 5 terms. To show that it has the most general spin dependence possible, we need only show that the most general expansion (\ref{B1}) can be cast into this form.  

To this end we use the identity (recalling that, in our notation, $\epsilon_{0123}=1$ and $\gamma^5=i\gamma^0\gamma^1\gamma^2\gamma^3$)
\bea
\epsilon^{\mu\nu\alpha\beta}\sigma_{\alpha\beta}=2i\gamma^5\sigma^{\mu\nu}\, .
\eea
Contracting both sides with $p_i-k_i$ and taking the on-shell matrix element (so the Dirac equation can be used) gives
\bea
\epsilon^{\mu\nu\alpha\beta}(\sigma_{\alpha\beta})_i(p_i-k_i)_\nu=-2\gamma^5_iQ_i^\mu\, .
\eea
Now, using this identity for both particles 1 and 2, multiplying the two terms together by contracting the free index $\mu$, reducing the expresions using the Dirac equation, and rearranging terms gives a very useful identity
\begin{align}
(\slashed{Q}_2)_1\bm{1}_2&+\bm{1}_1(\slashed{Q}_1)_2\nonumber\\
=&\frac{Q_1\cdot Q_2}{2m}\Big[\bm{1}_1\bm{1}_2+\gamma^5_1\gamma^5_2 \Big] 
\nonumber\\&
+2m(\gamma^\mu)_1 (\gamma_{\mu})_2
-\frac{q^2}{4m}(\sigma^{\alpha\beta})_1(\sigma_{\alpha\beta})_2\, ,
\end{align}
allowing the invariant  $\left[(\slashed{Q}_2)_1\bm{1}_2+\bm{1}_1(\slashed{Q}_1)_2\right]$ to be expressed  in terms of the invariants of Eq.~(\ref{B1}).  Substituting this into Eq.~(\ref{B4}) gives the following correspondence
\bea
F_s&=&f_s-f_v\kappa_v \,\frac{Q_1\cdot Q_2}{4m^2}\nonumber\\
F_v&=&f_v(1+\kappa_v)\nonumber\\
F_t&=&f_v\kappa_v(1+\kappa_v)\frac{q^2}{8m^2}\nonumber\\
F_a&=&f_a\nonumber\\
F_p&=&f_p-f_v\kappa_v(1+\kappa_v)\,\frac{Q_1\cdot Q_2}{4m^2}\, .
\eea
Hence every term in the most general spin and isospin expansion (\ref{B1}) can be expressed in terms of OBE parameters.  

However, the OBE assumption imposes severe constraints on the functional form of the $F_i$; an arbitrary form is not possible.   In particular, the model assumes $\kappa_v$ is a constant, so 
the ratio $F_t/F_v$ is proportional to $q^2$, and $F_t=0$ at $q^2=0$.   The OBE model also assumes the $f_i$ depend only on $q^2$ (and not on the energy squared $s$), so only $F_s$ and $F_p$ have an energy dependence through their dependence on $Q_1\cdot Q_2$.

\section{Nucleon form factor and removal of singularities at small $W$} \label{App:C}

\begin{figure}
\centerline{
\mbox{
\includegraphics[width=2.5in]{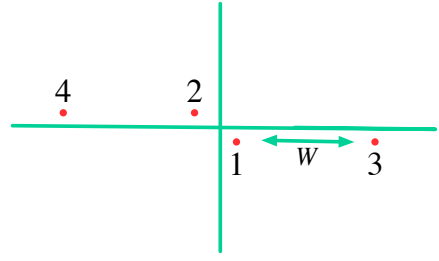}
}
}
\caption{\footnotesize\baselineskip=10pt (Color on line) The four poles in the complex $p_0$ plane arising from the term $[(m^2-p_1^2)(m^2-p_2^2)]^{-1}$.  The CST (with particle 1 on-shell) keeps only the pole at $p_0=E(p)-\sfrac12 W$ (\#1).  When $W\to0$ the pole at $p_0=E(p)+\sfrac12 W$ (\#3) cannot be neglected.  The full description in this case requires the two-channel spectator equation.  }
\label{fig12}
\end{figure} 

To provide needed convergence, the nucleon propagator is multiplied by a form factor, $H^2(p)$. [As suggested by the notation, this form factor is the square of the function $H(p)$; the reasons for this will be discussed below.]  The dressed propagator is therefore
\bea
S_D(p)=H^2(p)S_0(p)=\frac{H^2(p)}{m-\not\!p}\, .
\eea
In the cm frame, the propagator of the off-shell particle (taken to be particle 2) is
\bea
S_2(p)=\frac{H^2(p_2)(m+\not\!p_2)}{W[2E(p)-W]}
\, .\label{prop}
\eea
The singularity at $W=2E(p)$ gives rise to the elastic scattering cut, but the singularity at $W=0$ is unphysical.   This singularity is due to the presence the negative energy pole in the propagator of particle 2 (see Fig.~\ref{fig12}, where this pole is labeled 3), which is very distant from the physical scattering region, and becomes important only when $W\to0$ where it pinches the positive energy pole from particle 1 (labeled pole 1 in the figure).  When $W=0$ these two poles coalesce into a double pole, giving a finite result.  

The singularity at $W=0$ is therefore removed by including the contributions from both poles 1 and 3, which doubles the number of channels needed in the calculation.  The spectator theory that includes channels from both pole 1 and 3 is referred to as the two-channel spectator theory, and has been used for the description of the pion as a bound state of a massive constituent quark and antiquark, where an accurate description of states with masses near zeroÊ is required  \cite{Gross:1991te,Savkli:1999me}. 

In these calculations two-body scattering near $W=0$ plays a role in the three-body spectator equation when the spectator three-momentum $q\to q_{\rm crit}$, as discussed briefly in Sec.~\ref{sec:overview} above.  In this work we have used the nucleon form factor given in Eq.~(\ref{nuclff}) above, and with this simple choice the two-body scattering near $W=0$ gives a number of spurious deeply bound states.  These deeply bound states  are ``spurious'' because they would not exist if the propagator (\ref{prop}) did not have a singularity at $W=0$, and this singularity could be removed by using the two channel spectator theory. 

 All states with binding energies greater than $-1800$ MeV (corresponding to values of $W\agt78$ MeV) are tabulated in Tables \ref{tab:bound} and \ref{tab:bound2}, and there may be more in the region $0<W<78$ MeV.   Except for the deuteron, model WJC-1 has  no states with binding energies greater than $-1500$ and model WJC-2 has none with binding energies greater than $-1200$ MeV.  But  the convergence of the three-body integrals is such that, numerically, states with masses $W\alt m$ make no contribution at all.  Hence a realistic description of two-body scattering for binding energies greater than about $- 940$ MeV is sufficient, and since neither model has any spurious states in this region, the results are independent of their existence.

\begin{table}
\caption{Deeply bound states for Model WJC-1.   There are no states with binding energies greater than $-1500$ MeV  except for the deuteron at a binding energy of $-2.22$ MeV. }
\label{tab:bound}
\begin{ruledtabular}
\begin{tabular}{c|c|c|c}
channel & states &  channel  & states \\ [0.05in]
\tableline
$^1S_0$ & no bound states &$^3P_0$ & no bound states\cr
$^1P_1$ & 2 states $<-1600$ &$^1G_4$ & no bound states \cr
$^3P_1$ & 3 states $<-1500$ &$^3G_4$& 2 states $<-1600$\cr
$^3S_1 - {}^3D_1$ & 1 state $<-1700$   & $^3F_4 - {}^3H_4$ & 1 state $<-1700$\cr
$^1D_2$ & 1 state $<-1600$ & $^1H_5$ & 1 state $<-1700$\cr
$^3D_2$ & 1 state $<-1600$  & $^3H_5$ & no bound states \cr
$^3P_2 - {}^3F_2$ & no bound states & $^3G_5 - {}^3I_5$ & 1 state $<-1700$ \cr
$^1F_3$ & 1 state $<-1600$  & $^1I_6$ & 1 state $<-1700$\cr
$^3F_3$ & 2 states $<-1600$ & $^3I_6$ & 1 state $<-1700$ \cr
$^3D_3 - {}^3G_3$ & 1 state $<-1700$ & $^3H_6- {}^3K_6$  & 1 state $<-1700$  \cr
\end{tabular}
\end{ruledtabular}
\end{table}

\begin{table}
\caption{Deeply bound states for Model WJC-2.   There are no states with binding energies greater than $-1200$ MeV  except for the deuteron at a binding energy of $-2.22$ MeV. }
\label{tab:bound2}
\begin{ruledtabular}
\begin{tabular}{c|c|c|c}
channel & states &  channel  & states \\ [0.05in]
\tableline
$^1S_0$ & no bound states &$^3P_0$ & 2 states $<-1400$\cr
$^1P_1$ & 3 states $<-1200$ &$^1G_4$ & 2 states $<-1600$\cr
$^3P_1$ & 3 states $<-1400$ &$^3G_4$& 2 states $<-1600$\cr
$^3S_1 - {}^3D_1$ & 1 state $<-1600$   & $^3F_4 - {}^3H_4$ & no bound states\cr
$^1D_2$ & 1 state $<-1500$ & $^1H_5$ & 2 states $<-1600$\cr
$^3D_2$ & 1 state $<-1400$  & $^3H_5$ & no bound states \cr
$^3P_2 - {}^3F_2$ & 2 states $<-1600$ & $^3G_5 - {}^3I_5$ & 1 state $<-1700$ \cr
$^1F_3$ & 1 state $<-1500$  & $^1I_6$ & no bound states \cr
$^3F_3$ & 2 states $<-1600$ & $^3I_6$ & 2 states $<-1600$ \cr
$^3D_3 - {}^3G_3$ & 1 state $<-1700$ & $^3H_6- {}^3K_6$ & no bound states  \cr
\end{tabular}
\end{ruledtabular}
\end{table}


To understand this result, consider the two-body subsystem of particles 2 and 3 in the 
three-body equation, displayed in Fig.\ \ref{fig:3bspec} in Sec.\ \ref{sec:overview}.
In the three-body rest frame, where the four-momenta
of the three particles satisfy $k_1+k_2+k_3=(M_t,\bm{0})$, the
total four-momentum of the pair is $P_{23}=k_2+k_3$, and the four-momentum of the 
on-shell spectator particle 1 is  $k_1=(E_{k_1},\bm{k}_1) \equiv (E_q,-\bm{q})$, such that $\bm{P}_{23}=\bm{q}$. 
We choose $\bm{q}$ along the positive $\hat{z}$ direction, 
and the relative three-momentum $\bm{p}$ of the pair particles in the  $\hat{x}\hat{z}$ plane, 
oriented at an angle $\theta$ with respect to $\bm{q}$. Given the invariant mass $W$ of the pair, its 
four-momentum is $P_{23}=( {\cal E}_q,\bm{q} ) $, and the relative momentum $p=\frac{1}{2}(k_2-k_3)$
becomes 
\begin{equation}
p  =  \left(E_{\frac{1}{2}\bm{q}+\bm{p}}-\frac{1}{2}{\cal E}_q , p \sin \theta,0, p \cos\theta \right)  \, ,
\end{equation}
with ${\cal E}_q=\sqrt{W^2+\bm{q}^2}$ and 
$E_{\frac{1}{2}\bm{q}+\bm{p}}= \sqrt{m^2+( \frac{1}{2}\bm{q}+\bm{p})^2}$.

The boost $B$ is chosen to bring the two-body subsystem to its rest frame, such that
\begin{equation}
\tilde{P}_{23}=B P_{23}=\left( W, 0,0,0 \right) \, .
\end{equation}
The same boost is now applied to the relative momentum $p$ and yields the relative pair momentum 
in its rest frame, where the two-body scattering amplitude is actually calculated 
when the three-body equation is solved,
\bea
\tilde{p} & = & Bp  = 
\left(
\begin{array}{cccc} 
\frac{ {\cal E}_q}{W} & 0 & 0 & -\frac{q}{W} \\
0 & 1 & 0 & 0 \\
0 & 0 & 1 & 0 \\
-\frac{q}{W} & 0 & 0 &\frac{ {\cal E}_q}{W} \end{array}
\right) p 
\nonumber \\
& = &
\left( \begin{array}{c}
 \frac{{\cal E}_q}{W} \left( E_{\frac{1}{2}\bm{q}+\bm{p}} -\frac{1}{2}{\cal E}_q \right) -\frac{q}{W} p \cos \theta \\
p \sin\theta \\
0 \\
-\frac{q}{W}  \left( E_{\frac{1}{2}\bm{q}+\bm{p}} -\frac{1}{2}{\cal E}_q \right) + \frac{{\cal E}_q}{W} p \cos\theta
\end{array}
\right)
\eea
The magnitude of the relative three-momentum squared becomes
\begin{equation}
\tilde{\bm p}^2=
p^2 \sin^2\theta 
+
\left[ 
 \sqrt{1+\frac{q^2}{W^2} }
\left(p \cos\theta + \frac{q}{2}\right) 
-\frac{q}{W} 
E_{ \frac{1}{2} \bm{q }+\bm{p} } 
\right]^2 .
\end{equation}
%
Near the critical momentum, when $W\to 0$, this reduces to 
\begin{equation}
\tilde{\bm p}^2 \to p^2 \sin^2\theta
+ \frac{q}{W}
 \left[ p \cos\theta + \frac{q}{2}
-E_{ \frac{1}{2} \bm{q }+\bm{p} }   
\right]^2 \, ,
\end{equation}
and because 
\begin{equation}
E_{ \frac{1}{2} \bm{q }+\bm{p} } > p \cos\theta + \frac{q}{2} \, ,
\end{equation}
the term inside the brackets cannot vanish, and therefore $\tilde{\bm p}^2$ 
diverges.
Since the two-body scattering amplitude goes to zero as a high power of the 
magnitude of the relative 
three-momentum, this strongly suppresses the high $q$ (or low $W$) 
contributions. 
It may be surprising that this happens even when  ${\bm p}$ is perpendicular 
to the direction of the boost. If both particles were on-shell, the energy 
component of
the relative momentum would be zero and indeed no change of magnitude of the 
three-vector would occur as a result of the boost. 
However, since one particle is off-shell, a non-zero energy component mixes 
with the three-vector components and changes their magnitude for all angles 
$\theta$.

If the nucleon form factor is to be interpreted as a self-energy, the form factor $H(p)$ can be a covariant function of $p^2$ only.  However, since the light cone $p^2=0$ is covariant under all Poincar\'e transformations, this function may have a different functional form in each of the three regions invariant under Lorentz transformations: (i) the forward light cone defined by $p_0>|{\bf p}|>0$, (ii) the space-like region (sometimes referred to as the ``now'' region) defined by $p^2<0$, and (iii) the backward light cone defined by $p_0<-|{\bf p}|<0$.  

Bearing this in mind, a nucleon form factor could be chosen to greatly reduce the interaction for all energies below $W<m$, insuring that the singularities  at $W=0$ are suppressed and there are no spurious bound states.  
The simplest way to eliminate the deeply bound states is to use the form factor to cut off the interaction for all $W < W_{\rm crit}$, where we could choose $W_{\rm crit}\simeq1200$ MeV.  As our discussion shows, this would not alter {\it any of the results in this paper\/}.  The only phenomenological objection to such a choice is that the electromagnetic exchange currents are usually difficult to calculate for a nucleon form factor that depends on $W^2$ as well as $p^2$.  However, if the $W^2$ dependence is a sharp cutoff, then the exchange currents will also be zero for all $W < W_{\rm crit}$, and the problem is avoided. We have not used this possibility in this paper; it may be investigated in subsequent work.

In summary: the previous discussion shows that modifying  the nucleon form factor by adding a cutoff 
\bea
H_0=H(p)\;\left[W^2-(1200\; {\rm MeV})^2\right]
\eea
is a covariant change that will  remove all of the spurious bound states, but will otherwise not change {\it any\/} of the other results in this paper.  Hence the spurious bound states present no problem at all.

\section{Computation of the OBE kernel and reduction of the two-body equations}  \label{App:D}

\subsection{Overview}

In this appendix, we work out the detailed form of the partial wave expansion of Eq.~(\ref{eq:spec}). 

The first step is to define the helicity spinors in the $\hat x\hat z$ scattering plane, and this is done in Sec.~\ref{sec:spinors}.  Here we define the states for both $\rho$-spin  (where $\rho=+$ is the positive energy spinor, $u$, and $\rho=-$ is the negative energy spinor, $v$).  The spinors for particle 1 and 2 are related by a rotation about the $\hat y$ axis (and a phase).  

Next, in Sec.~\ref{sec:offG} the off-shell propagator is decomposed into a sum of two terms, one with the off-shell particle in a positive $\rho$-spin state and the other in a negative $\rho$-spin state.  There are therefore two channels, one channel describing the propagation of two positive $\rho$-spin states, $\{\rho_1\rho_2\}=\{++\}$, and one the propagation of  a positive and a negative $\rho$-spin state, $\{\rho_1\rho_2\}=\{+-\}$.  
Using the properties of the rotation group, the two-body states for all $\{\rho_1\rho_2\}$ and arbitrary $(\theta,\phi)$ are defined, and the partial wave expansions of the states is given in Sec.~\ref{sec:ang}.  Section~\ref{sec:sym} gives the symmetries of the two-body states under parity and particle exchange.  The exchange operator relates states with different $\rho$-spins to each other.  Equipped with the partial wave expansions from Sec.~\ref{sec:ang}, the expansions of the two-body kernel are developed in Sec.~\ref{sec:me}.   The partial wave matrix elements can be written in terms of a simple integral in the $\hat x\hat z$ plane.  Using this formalism, the two-body partial wave equations are given in Sec.~\ref{sec:wave}.

The partial wave equations (\ref{E25}) are not efficient for solution, since they mix states of different parities.  In Sec.~\ref{sec:sep} these equations are separated into the three independent scattering states with good parity and exchange symmetry: spin singlet, uncoupled triplet (referred to as ``triplet'' below), and coupled triplet (referred to as ``coupled'' below).  The final result, Eq.~(\ref{E49}), involves four coupled channels for all states with total angular momentum $J\ge 1$ and two channels of the special cases with $J=0$.

\subsection{Nucleon helicity spinors}
\label{sec:spinors}

The nucleon helicity spinors are defined as in previous references.  The four-component helicity spinors can be written as a direct product of a two-component spinor in Dirac rho-space and a two-component spinor in spin 1/2 space.  For particle 1 (in the sense of Jacob and Wick \cite{Jacob:1959at}) they are
\begin{align}
u^+_1({\bf p},\lambda)&\equiv u({\bf p},\lambda)=u_1({\bf p},\lambda)=N_+(p\lambda)\otimes \chi_{_\lambda}(\theta) 
\nonumber\\
u^-_1({\bf p},\lambda)&\equiv v(-{\bf p},\lambda)=v_1({\bf p},\lambda)
=N_-(p\lambda)\otimes\chi_{_\lambda}(\theta) \qquad
\label{A1}
\end{align}
where the rho-space spinors are 
\bal
N_+(p\lambda)&=\left(\begin{array}{c} \cosh\sfrac12 \zeta \\ [0.1in]
2\lambda\sinh \sfrac12\zeta
\end{array}\right) \nonumber\\
N_-(p\lambda)&=\left(\begin{array}{c} -2\lambda\sinh \sfrac12\zeta \\ [0.1in]
\cosh\sfrac12 \zeta
\end{array}\right),
\end{align}
with $p=|{\bf p}|$ and $\tanh \zeta = p/E_p$.  For momenta limited to the $\hat x\hat z$ plane, the spin 1/2 spinors are
\begin{align}
\chi_{_{1/2}}(\theta)&=R_y(\theta)\left(\begin{array}{c} 1 \\ 0\end{array}\right)=
\left(\begin{array}{c} \cos\sfrac12\theta \\ [0.1in]
\sin \sfrac12\theta
\end{array}\right) \nonumber\\
\chi_{_{-1/2}}(\theta)&=R_y(\theta)\left(\begin{array}{c} 0 \\ 1\end{array}\right)=\left(\begin{array}{c} -\sin\sfrac12\theta \\ [0.1in]
\cos \sfrac12\theta
\end{array}\right), \label{A3}
\end{align}
where $R_y(\theta)$ is the active rotation through angle $\theta$ about the $\hat y$ axis.
These definitions are identical to those given in Eq.~(A9) of Ref.~\cite{GVOH}, Eq.~(4.23) of Ref.~\cite{Stadler:1997iu}, and Eq.~(A4) of Ref.~\cite{Adam:2002cn}.  The spinors (\ref{A1}) will be collectively denoted
\bea
u^\rho_1({\bf p},\lambda)&=&N_\rho(p\lambda)\otimes \chi_{_\lambda}(\theta) \equiv
u_1^\rho([p,\theta],\lambda)\nonumber\\
&=&{\cal R}_y(\theta)u^\rho_1([p,0],\lambda) \label{A4}
\eea
where the $\rho$-spin is either + or $-$.

Similarily,  the helicity spinors for particle 2 are obtained by a rotation from those for particle 1.  
Following the conventions of Jacob and Wick the rotation is
\bal
u_2^\rho&({\bf p},\lambda)={\cal R}(\theta)
u_1^\rho({\bf p},\lambda)\nonumber\\
&\equiv 
\,e^{-i\pi/2}{\cal R}_y(\theta)
{\cal R}_z(\pi){\cal R}_y(\pi){\cal R}^{-1}_y(\theta) 
u_1^\rho({\bf p},\lambda) \label{E5}
\end{align}
where ${\cal R}_y=e^{-i\theta{\cal J}_y}$ is the rotation through angle $\theta$ about the $\hat y$ axis.  Using the decomposition (\ref{A4}) the rotation operates only on the spinors $\chi$, and 
\bea
{\cal R}(\theta)\chi_\lambda(\theta)&\equiv& e^{-i\pi/2}
R_y(\theta) R_z(\pi)R_y(\pi)R^{-1}_y(\theta) \chi_\lambda(\theta)
\nonumber\\
&=&R_y(\theta)\chi_{-\lambda}(0)=\chi_{-\lambda}(\theta)
\eea
where the $\chi$ are as defined in (\ref{A3}).  Hence the spinors for particle 2 are 
\bea
u^\rho_2({\bf p},\lambda)&=&N_\rho(p\lambda)\otimes \chi_{_{-\lambda}}(\theta)=
(-1)^{1/2-\lambda}{\cal R}_y(\pi)
u^\rho_1({\bf p},\lambda)
\nonumber\\
&\equiv& u^\rho_2([p,\theta],\lambda) 
={\cal R}_y(\theta) u^\rho_2([p,0],\lambda)  . \label{A6}
\eea
These relations agree with Eq.~(A9) of Ref.~\cite{GVOH} and Eq.~(A6) of Ref.~\cite{Adam:2002cn}, and the definition given in Ref.~\cite{Stadler:1997iu}.  It also follows that
\bea
u^\rho_1({\bf p},\lambda)=
(-1)^{1/2+\lambda}{\cal R}_y(\pi)
u^\rho_2({\bf p},\lambda).\label{A8}
\eea

Although the scattering will be restricted to the $\hat x\hat z$ plane, the definition of angular momentum states requires treatment of rotations about the $\hat z$ axis.  Here we depart from the conventions of Jacob and Wick, followed in Ref.~I, and define the states for momentum in a arbitrary direction by 
\bal
&\left|(p,\theta,\phi)\lambda_1\lambda_2;\rho_1\rho_2\right>\equiv e^{-i\phi{\cal J}_z} u_{1\,\alpha}^{\rho_1}({\bf p},\lambda_1)\,u_{2\,\beta}^{\rho_2}({\bf p},\lambda_2) \nonumber\\
&\qquad= {\cal R}(\phi,\theta,0)\left|(p,0,0)\lambda_1\lambda_2;\rho_1\rho_2\right>, \label{A9a}
\end{align}
where ${\cal R}(\phi,\theta,\gamma)=e^{-i\phi{\cal J}_z}e^{-i\theta{\cal J}_y}e^{-i\gamma{\cal J}_z}$ is the general rotation through Euler angles $\phi$, $\theta$, and $\gamma$, and $\alpha$ and $\beta$ are the Dirac indices  on subspace 1 and 2, respectively (usually suppressed).  In Ref.~I and Jacob and Wick the states were defined using the rotation ${\cal R}(\phi,\theta,-\phi)$; the convention (\ref{A9a}) is favored for extensions of this formalism to three-body states, as discussed in \cite{Stadler:1997iu}.  One of the objectives of this appendix is to show that all of the results of Ref.~I also follow from the definition (\ref{A9a}).

\subsection{Separation of the off-shell particle into $\rho$-spin states} \label{sec:offG}

In the cm system, where $k_2=\{W-E_k, -{\bf k}\}$, with ${\bf k}$ the spatial components of the relative four-momentum $k=\sfrac12 (k_1-k_2)$, the propagator for the off-shell particle 2 can be written
\bal
&\frac{1}{m-\not\!k_2}\nonumber\\
&\quad=\frac{m}{E_k}\sum_\lambda\left\{\frac{u_2^+({\bf k},\lambda)\bar u_2^+({\bf k}, \lambda)}{2E_k-W}  - \frac{u_2^-({\bf k},\lambda)\bar u_2^-({\bf k}, \lambda)}{W}  \right\}\, ,\label{E9}
\end{align}
where $u_2^\rho$ are the spinors defined in Eq.~(\ref{E5}).  Substituting this into  Eq.~(\ref{eq:spec}), allowing for relative momenta in all directions as defined in Eq.~(\ref{A9a}),  and keeping all the indices, gives the following equations
\bal
&M^{\rho_1\rho_2,\rho'_1\rho'_2}_{\lambda_1\lambda_2,\lambda'_1\lambda'_2}(p,p';P)=\overline{V}^{\rho_1\rho_2,\rho'_1\rho'_2}_{\lambda_1\lambda_2,\lambda'_1\lambda'_2}(p,p';P)
\nonumber\\
& - \sum_{\mu_1\mu_2\rho\,k} \overline{V}^{\rho_1\rho_2,(+)\rho}_{\lambda_1\lambda_2,\mu_1\mu_2}(p,k;P) \;  G^\rho(k)\;  M^{(+)\rho,\rho'_1\rho'_2}_{\mu_1\mu_2,\lambda'_1\lambda'_2}(k,p';P) \label{E10}
\end{align}
where 
\bea
\sum_{\mu_1\mu_2\rho\,k} =\int \frac{d^3k}{(2\pi)^3}\sum_{\mu_1=-1/2}^{1/2}\sum_{\mu_2=-1/2}^{1/2}\sum_{\rho=-}^+
\eea
and, using the notation of Eq.~(\ref{A9a}), the matrix elements of the kernel (and scattering amplitude) are
\bal
&\overline{V}^{\rho_1\rho_2,\rho'_1\rho'_2}_{\lambda_1\lambda_2,\lambda'_1\lambda'_2}(p,p';P)=\frac{m^2}{E_pE_{p'}}
\nonumber\\
& \times\left<(p,\theta,\phi)\lambda_1\lambda_2;\rho_1\rho_2\right|\overline{{\cal V}}(p, p'; P)\left|(p',\theta',\phi')\lambda'_1\lambda'_2;\rho'_1\rho'_2\right> 
\label{E12}
\end{align}
and the propagators for $\rho=+$ and $-$ states are
\bea
G^+(k)=\frac{1}{2E_k-W}\, , \qquad\quad G^-(k) =-\frac{1}{W}\, . \label{Gprop}
\eea
Note that the factors of $m/E_k$ in (\ref{E9}) and the volume integral in (\ref{eq:spec}) are absorbed if the kernel and scattering amplitude are normalized as in Eq.~(\ref{E12}).  [The propagators (\ref{Gprop}) differ by a factor of $k^2/(2\pi)^3$ from the $g^\pm(k)$ of Eq.~(2.89) of Ref.~I; in this paper this factor is written explicitly in all equations.]

Equation (\ref{E10}) is simplified further by expanding the kernel and scattering amplitude in states with good angular momentum.  These states are defined in the next section.

\subsection{Angular momentum states}\label{sec:ang}

States of good angular momentum are projected  from the general two-particle states (\ref{A9a}) by integrating over the polar and azimuthal angles, as was done in Ref.~\cite{Stadler:1997iu}.  The result is
\bal
\left|pJM_J,\lambda_1\lambda_2;\rho_1\rho_2\right>\equiv\eta_J & \int  d\Omega_p\,{D}^{J\,*}_{M_J,\lambda}(\phi,\theta,0)\qquad\qquad
\nonumber\\
&\times \left|(p,\theta,\phi)\lambda_1\lambda_2;\rho_1\rho_2\right>, 
\label{A10}
\end{align}
where the following shorthand notation 
\bea
\lambda&=&\lambda_1-\lambda_2\nonumber\\
\eta_J&=&\sqrt{\frac{2J+1}{4\pi}}\nonumber\\
\int d\Omega_p&\equiv&\int_0^{2\pi} d\phi \int_0^\pi
d\theta\,\sin\theta
\eea
will be used repeatedly below.  Equation (\ref{A10})  agrees with Eq.~(4.8) of Ref.~\cite{Stadler:1997iu}.

The coupling coefficients (compare with Eq.~(2.84) of Ref.~I) are
\bal
&\left<{\bf p}|JM_J,\lambda_1\lambda_2;\rho_1\rho_2\right>\nonumber\\
&\qquad\equiv\left<(p,\theta,\phi)\lambda_1\lambda_2;\rho_1\rho_2|pJM_J,\lambda_1\lambda_2;\rho_1\rho_2\right>
\nonumber\\
&\qquad= \eta_JD^{J\,*}_{M_J,\lambda}(\phi,\theta,0)
\end{align}
and the partial wave expansion of the states becomes
\bal
&\left|(p,\theta,\phi)\lambda_1\lambda_2;\rho_1\rho_2\right>\nonumber\\
&\qquad=\sum_{JM_J} \eta_JD^{J}_{M_J,\lambda}(\phi,\theta,0)\left |pJM_J,\lambda_1\lambda_2;\rho_1\rho_2\right>. \label{A12a}
\end{align}
The normalization condition
\bal
\eta_J^2\int d\Omega_p\,&{D}^{J\,*}_{M_J,\lambda}(\phi,\theta,0) D^{J'}_{M_{J'},\lambda}(\phi,\theta,0)
=\delta_{JJ'}\delta_{M_JM_{J'}} \label{norm}
\end{align}
insures that (\ref{A10}) and (\ref{A12a}) are consistent.

\subsection{Symmetries of the angular momentum states}\label{sec:sym}

\subsubsection{Parity}

Under the ${\cal Y}=\exp({-i\pi{\cal J}_y}){\cal P}$ transformation (parity followed by rotation through angle $\pi$ about the $\hat y$ axis), the nucleon helicity spinors transform to
\bea
{\cal Y}\,u_1^\rho({\bf p},\lambda)&=&\gamma^0 N_\rho(p\lambda)(-i\sigma_y)\chi_{_\lambda}(\theta)
\nonumber\\
&=&\rho(-1)^{1/2+\lambda}
u_1^\rho({\bf p},-\!\lambda)
\nonumber\\
{\cal Y}\,u_2^\rho({\bf p},\lambda)&=&\gamma^0 N_\rho(p\lambda)(-i\sigma_y)\chi_{-\lambda}(\theta)
\nonumber\\
&=&\rho(-1)^{1/2-\lambda}
u_2^\rho({\bf p},-\!\lambda) .
\eea
In the notation of Eq.~(\ref{A9a}), the parity relation for $NN$ states can be written
\bal
{\cal Y}&\,\left|(p,\theta,0)\lambda_1\lambda_2;\rho_1\rho_2\right>
\nonumber\\
&\quad=\rho_1\rho_2(-1)^{1+\lambda}\left|(p,\theta,0)-\!\lambda_1\,-\!\lambda_2;\rho_1\rho_2\right>.
\label{A11}
\end{align}
 Note that $(-1)^{2\lambda} =1$, showing that these results are identical to those previously given in Eqs.~(A14) and (A17) of Ref.~I.

The effect of parity on the states of good angular momentum follows from (\ref{A10}) and (\ref{A11})
\bal
{\cal P}&\left|pJM_J,\lambda_1\lambda_2;\rho_1\rho_2\right>
=\eta_J\int d\Omega_p\,{D}^{J\,*}_{M_J,\lambda}(\phi,\theta,0)
\nonumber\\&
\quad\times{\cal R}(\phi,\theta,0) {\cal R}^{-1}(0,\pi,0){\cal Y}\left|(p,0,0)\lambda_1\lambda_2;\rho_1\rho_2\right>
\nonumber\\
&=\eta_J\int d\Omega_{p'}
\sum_{\lambda'}{D}^{J\,*}_{M_J,\lambda'}(\phi',\theta',0)D^{J\,*}_{\lambda',\lambda}(0,\pi,0)
\nonumber\\
&\quad\times{\cal R}(\phi',\theta',0)
\rho_1\rho_2(-1)^{1+\lambda}\left|(p,0,0)\,-\!\lambda_1\,-\!\lambda_2;\rho_1\rho_2\right>
\nonumber\\
&=\rho_1\rho_2(-1)^{J-1}\left|pJM_J,\,-\!\lambda_1\,-\!\lambda_2;\rho_1\rho_2\right>.
\label{A13}
\end{align}
The second line follows from the first by
introducing the rotation ${\cal R}(\phi',\theta',0)$ with $0\le\phi'\le2\pi$ and $0\le\theta'\le\pi$ such that ${\cal R}(\phi',\theta',0)={\cal R}(\phi,\theta,0){\cal R}^{-1}(0,\pi,0)$, and using the group representation properties of the $D$ matrices.  The second follows from the relations $D^J_{\lambda'\lambda}(0,\pi,0)=d^J_{\lambda'\lambda}(\pi)=(-1)^{J-\lambda}\delta_{\lambda',-\lambda}$.  

It is convenient to work with good angular momentum states that also have definite parity.  From (\ref{A13}) these are 
\bal
&\left|pJM_J,\lambda_1\lambda_2;\rho_1\rho_2;\delta_P\right>\equiv\left|pJM_J,\lambda_1\lambda_2;\rho_1\rho_2\right>\nonumber\\
&\quad+\delta_P\,\rho_1\rho_2(-1)^{J-1}\left|pJM_J,\,-\!\lambda_1\,-\!\lambda_2;\rho_1\rho_2\right>\qquad
\end{align}
where
\bal
&{\cal P}\left|pJM_J,\lambda_1\lambda_2;\rho_1\rho_2;\delta_P\right>
=\delta_P\left|pJM_J,\lambda_1\lambda_2;\rho_1\rho_2;\delta_P\right>. \label{A18}
\end{align}

\subsubsection{Particle interchange}

Under particle interchange ${\cal P}_{12}$, {\it either\/} the Dirac indices of the two spinors are exchanged, {\it or\/} the helicity and $\rho$-spin labels are exchanged.  These two forms of the interchange operator are 
\bal
&{\cal P}_{12}\left|(p,\theta,0)\lambda_1\lambda_2;\rho_1\rho_2\right>=
u_{1\,\beta}^{\rho_1}({\bf p},\lambda_1)
u_{2\,\alpha}^{\rho_2}({\bf p},\lambda_2)\nonumber\\
&\qquad=(-1)^{1+\lambda} {\cal R}_y(\pi)\left|(p,\theta,0)\lambda_2\lambda_1;\rho_2\rho_1\right>\, ,
\label{A12}
\end{align}
where the relations (\ref{A6}) and (\ref{A8}) were used in the last step.  Since ${\cal R}_y(2\pi)=1$, this result is identical to (A27) and the equation following (A31) of Ref.~I.

The effect of the interchange operator on the states of good angular momentum is computed in the same way as the effect of the parity operator.  Using (\ref{A12}) with ${\cal R}_y(\pi)$ replaced by 
${\cal R}^{-1}_y(\pi)$ gives
\bal
{\cal P}_{12}&\left|pJM_J,\lambda_1\lambda_2;\rho_1\rho_2\right>
=\eta_J\int d\Omega_p\,{D}^{J\,*}_{M_J,\lambda}(\phi,\theta,0)
\nonumber\\
&\times{\cal R}(\phi,\theta,0) {\cal R}^{-1}(0,\pi,0)(-1)^{1+\lambda}\left|(p,0,0)\lambda_2\lambda_1;\rho_2\rho_1\right>
\nonumber\\
&=\eta_J\int d\Omega_{p'}
\sum_{\lambda'}{D}^{J\,*}_{M_J,\lambda'}(\phi',\theta',0)D^{J\,*}_{\lambda',\lambda}(0,\pi,0)
\nonumber\\
&\times{\cal R}(\phi',\theta',0)
(-1)^{1+\lambda}\left|(p,0,0)\,\lambda_2\lambda_1;\rho_2\rho_1\right>
\nonumber\\
&=(-1)^{J-1}\left|pJM_J,\lambda_2\lambda_1;\rho_2\rho_1\right>.
\label{A13a}
\end{align}
This agrees with Eq.~(A32) of Ref.~I.


\subsection{Partial wave expansion of the kernel}\label{sec:me}

Using the fact that the kernel (and scattering amplitude) conserves angular momentum, the matrix elements of the kernel in angular momentum space are 
\bal
&\left<pJM_J,\lambda_1\lambda_2;\rho_1\rho_2|\overline{V}(p,p';P)|p'J'M_{J'}\lambda'_1\lambda'_2;\rho'_1\rho'_2\right>
\nonumber\\
&\qquad\qquad=\overline{V}^{J\,\rho_1\rho_2,\rho'_1\rho'_2}_{\lambda_1\lambda_2,\lambda'_1\lambda'_2}(p,p';P)\,\delta_{JJ'}\delta_{M_JM_{J'}} \, , \label{E26}
\end{align}
where the angular momentum states were defined in Eq.~(\ref{A10}).
Setting $J=J'$ and $M_J=M_{J'}$ gives the following partial wave projections
\bal
&\overline{V}^{J\,\rho_1\rho_2,\rho'_1\rho'_2}_{\lambda_1\lambda_2,\lambda'_1\lambda'_2}(p,p';P)
\nonumber\\
&=\eta_J^2
\int\int d\Omega_p\, d\Omega_{p'}\,D^{J}_{M_J\lambda}(\phi_{p},\theta_{p},0) D^{J\,*}_{M_J\lambda'}(\phi_{p'},\theta_{p'},0)
\nonumber\\
&\qquad\times
\overline{V}^{\rho_1\rho_2,\rho'_1\rho'_2}_{\lambda_1\lambda_2,\lambda'_1\lambda'_2}(p,p';P)\, . \label{E28}
\end{align}
The most general partial wave expansion  wave expansion for $\overline{V}$ is therefore
\bal
&\overline{V}^{\rho_1\rho_2,\rho'_1\rho'_2}_{\lambda_1\lambda_2,\lambda'_1\lambda'_2}(p,p';P)
= \sum_{J M_J}  
\overline{V}^{J\,\rho_1\rho_2,\rho'_1\rho'_2}_{\lambda_1\lambda_2,\lambda'_1\lambda'_2}(p,p';P)\nonumber\\
&\qquad\times {\eta_J^2} \; D^{J\,*}_{M_J\lambda}(\phi_{p},\theta_{p},0) D^{J}_{M_J\lambda'}(\phi_{p'},\theta_{p'},0)\, . \label{E29}
\end{align}
The consistency of (\ref{E28}) and (\ref{E29}) is assured by the orthonormality relation (\ref{norm}).
The expansion (\ref{E29}) is very convenient for the general derivation of the partial wave equations.

However, in order to carry out practical calculations, it is more convenient to express the partial wave relations in their simplest form.  Equation (\ref{E29}) can be simplified by carrying out the sum over the angular momentum projections $M_J$, using the addition theorem for the $d$ functions,
\bea
d^J_{\lambda'\lambda}(\theta)=\sum_{M_J} D^{J}_{M_J\lambda}(\phi_{p},\theta_p,0) D^{J\,^*}_{M_J\lambda'}(\phi_{p'},\theta_{p'},0),\qquad \label{E30}
\eea
where $\cos\theta={\bf \hat p}\cdot {\bf \hat p}'$.  This gives an alternative form for the partial wave expansion
\bal
&\overline{V}^{\rho_1\rho_2,\rho'_1\rho'_2}_{\lambda_1\lambda_2,\lambda'_1\lambda'_2}(p\,\theta,p';P)
= \sum_J \eta_J^2 d^J_{\lambda'\lambda}(\theta) V^{J\,\rho_1\rho_2,\rho'_1\rho'_2}_{\lambda_1\lambda_2,\lambda'_1\lambda'_2}(p,p';P), \qquad\label{A28}
\end{align}
which agrees with Eq.~(2.85) of Ref.~I.   Using the orthogonality of the $d$ matrices
\bea
2\pi\,\eta_J^2\int_0^\infty \sin\theta d \theta \; d^J_{\lambda'\lambda}(\theta) \; d^{J'}_{\lambda'\lambda}(\theta)=\delta_{JJ'}.
\eea 
gives the simple result for the partial wave projection of the kernel
\bal
\overline{V}^{J\,\rho_1\rho_2,\rho'_1\rho'_2}_{\lambda_1\lambda_2,\lambda'_1\lambda'_2}(p,p';P)=& 
2\pi \int_0^\pi \sin\theta\,d\theta\,d^J_{\lambda'\lambda}(\theta)\nonumber\\
&\times
\overline{V}^{\rho_1\rho_2,\rho'_1\rho'_2}_{\lambda_1\lambda_2,\lambda'_1\lambda'_2}(p,p';P).
 \label{A26}
\end{align}
This result can also be obtained directly from (\ref{E28}) by averaging over the spin projections $M_J$, using the addition theorem (\ref{E30}), and integrating over the redundant angles.  If we choose the scattering to be in the $\hat x\hat z$ plane, with ${\bf p}'=\{0,0,p'\}$ and ${\bf p}=\{p\sin\theta,0,p\cos\theta\}$, the matrix elements and the integrals (\ref{A26}) are easily evaluated.


\begin{widetext}

\subsection{Partial wave expansion of the equations} \label{sec:wave}

Equation~(\ref{E10})  can now be expanded in partial waves using the general expansion (\ref{E29}).   The integral with particle 1 propagating on-shell in the intermediate state,

\bal
&\left<\overline{V}GM\right>
=
\sum_{\mu_1\mu_2\rho\,k} \overline{V}^{\rho_1\rho_2,(+)\rho}_{\lambda_1\lambda_2,\mu_1\mu_2}(p,k;P) \;  G^\rho(k)\;  M^{(+)\rho,\rho'_1\rho'_2}_{\mu_1\mu_2,\lambda'_1\lambda'_2}(k,p';P),
\end{align}
can be easily reduced if both $\overline{V}$ and $M$ are expanded in partial waves using (\ref{E29}), and the integral over $\Omega_k$ carried out using the orthogonality relations (\ref{norm}).  Recalling that $\lambda=\lambda_1-\lambda_2$ (and similarly for $\lambda'$ and $\mu$), the result becomes

\bal
\left<\overline{V}GM\right>&=\int_0^\infty \frac{k^2dk}{(2\pi)^3} \int d\Omega_k \sum_{{\mu_1\mu_2\rho}\atop{JJ'M_JM_{J'}} }  \overline{V}^{J\,\rho_1\rho_2,(+)\rho}_{\lambda_1\lambda_2,\mu_1\mu_2}(p,k;P)\;G^\rho(k)\;{M}^{J'\,(+)\rho,\rho'_1\rho'_2}_{\mu_1\mu_2,\lambda'_1\lambda'_2}(k,p';P)
\nonumber\\
&\qquad\qquad\times  \eta^2_J\eta^2_{J'}  D^{J\,*}_{M_J,\lambda}(\phi,\theta,0) D^{J}_{M_{J},\mu}(\phi_k,\theta_k,0)  D^{J'\,*}_{M_{J'},\mu}(\phi_k,\theta_k,0) D^{J'}_{M_{J'},\lambda'}(\phi',\theta',0)
\nonumber\\
&=\int_0^\infty \frac{k^2dk}{(2\pi)^3}  \sum_{{\mu_1\mu_2\rho}\atop{JM_J} }  \overline{V}^{J\,\rho_1\rho_2,(+)\rho}_{\lambda_1\lambda_2,\mu_1\mu_2}(p,k;P)\;G^\rho(k)\;{M}^{J\,(+)\rho,\rho'_1\rho'_2}_{\mu_1\mu_2,\lambda'_1\lambda'_2}(k,p';P)
\nonumber\\
&\qquad\qquad\times  \eta^2_J   D^{J\,*}_{M_J,\lambda}(\phi,\theta,0) D^{J}_{M_{J},\lambda'}(\phi',\theta',0).
 \label{A21}
\end{align}
Comparing this with the partial wave expansions for $V(p, p'; P)$ and $M(p, p'; P)$ leads directly to the partial wave equations 
\bea
M^{J\,\rho_1\rho_2,\rho'_1\rho'_2}_{\lambda_1\lambda_2,\lambda'_1\lambda'_2}(p, p';P)&=&
\overline{V}^{J\,\rho_1\rho_2,\rho'_1\rho'_2}_{\lambda_1\lambda_2,\lambda'_1\lambda'_2}(p, p';P)
\nonumber\\
&&-\int_0^\infty \frac{k^2dk}{(2\pi)^3} \sum_{\mu_1\mu_2\rho} \overline{V}^{J\,\rho_1\rho_2,(+)\rho}_{\lambda_1\lambda_2,\mu_1\mu_2}(p,k;P)G^\rho(k;P)
 M^{J\,(+)\rho,\rho'_1\rho'_2}_{\mu_1\mu_2,\lambda'_1\lambda'_2}(k, p';P)\, . \label{E25}
\eea
\end{widetext}

These equations allow us to find $M^{J\,\rho_1\rho_2,\rho'_1\rho'_2}$ once we know $M^{J\,(+)\rho_2,\rho'_1\rho'_2}$.   To find $M^{J\,(+)\rho_2,\rho'_1\rho'_2}$ we need only set $\rho_1=+$ in Eq.~(\ref{E25}).  But these are not the equations we will solve; they must first be simplified using parity and particle interchange symmetry.

\begin{table*}
\begin{minipage}{4.5in}
\caption{The 16 independent helicity amplitudes ${\bf  V}_{\lambda_2\lambda'_2}(\delta_{_{p_0}},\delta_S)\equiv{\bf V}^{J\,++,++}_{++,\lambda_2\lambda'_2}(\delta_{_{p_0}},\delta_S)$.  Only the sign of the helicity is specified: $\lambda_2=+$ means $\lambda_2=+1/2$.  The parity of each potential is $\eta\delta_S$, where $\eta\equiv(-1)^{J-1}$.  The  symmetry ${\cal P}_{_{12}}$ under particle interchange is shown. }
\label{tab:I}
\begin{ruledtabular}
\begin{tabular}{ccc|ccc}
name & amplitude & ${\cal P}_{12}$  & name & amplitude & ${\cal P}_{12}$    \\ [0.05in]
\tableline
$v_1$ & ${\bf V}_{++}(+,-)$  & $\eta$ & 
                    $v_2$ &${\bf V}_{--}(-,-)$ & $\eta$ \cr
$v_3$ &${\bf V}_{+-}(+,-)$ & $\eta$& 
                    $v_{4}$ &$ {\bf V}_{-+}(-,-)$&$\eta$\cr
             \tableline
$v_5$ &${\bf V}_{++}(-,-)$ &$-\eta$ & 
                    $v_6$ &$ {\bf V}_{--}(+,-)$&$-\eta$ \cr
$v_7$ &${\bf V}_{+-}(-,-)$ & $-\eta$& 
                    $v_{8}$ &$ {\bf V}_{-+}(+,-)$&$-\eta$ \cr
             \tableline
$v_9$ &${\bf V}_{++}(+,+)$ & $\eta$ & 
                 $v_{10}$ &$ {\bf V}_{--}(+,+)$&$\eta$ \cr
$v_{11}$ &${\bf V}_{+-}(+,+)$& $\eta$& 
                 $v_{12}$ &$ {\bf V}_{-+}(+,+)$&$\eta$ \cr
            \tableline
$v_{13}$ &${\bf V}_{++}(-,+)$& $-\eta$& 
                 $v_{14}$ &$ {\bf V}_{--}(-,+)$ &$-\eta$ \cr
$v_{15}$ &${\bf V}_{+-}(-,+)$ & $-\eta$ & 
               $v_{16}$ &$ {\bf V}_{-+}(-,+)$& $-\eta$ \cr
\end{tabular}
\end{ruledtabular}
\end{minipage}
\end{table*}



\subsection{Separation into channels with good parity and interchange symmetry}\label{sec:sep}

\begin{table*}
\begin{minipage}{4.5in}
\caption{Symmetries of the 16 independent kernels with $\rho_1=\rho'_1=+$ and four different combinations of $\{\rho_2, \rho'_2\}$.  For convenience, the values of $\delta_{_{p_0}}$ and $\delta_S$ taken from Table \ref{tab:I} are given.  In each case the parity is $\rho'_2\eta\delta_S$ and the the exchange symmetry is $\eta\delta_{_{p_0}}(\rho_2\rho'_2\delta_S)^{(\lambda_1-\lambda_2)}$.   All phases except $\delta_{_{p_0}}$ and $\delta_S$  are in terms of $\eta$; for $\pm$ read $\pm\eta$.}
\label{tab:III}
\begin{ruledtabular}
\begin{tabular}{lcc|cc|cc|cc|cc} &\multicolumn{2}{c}{
$\{\rho_2\rho'_2\}=$}&\multicolumn{2}{c}{$++$} &\multicolumn{2}{c}{$+-$}&\multicolumn{2}{c}{$-+$}&\multicolumn{2}{c}{$--$}\cr
\tableline
 $\;$name &$\delta_S$&$\delta_{_{p_0}}$&${\cal P}$ & ${\cal P}_{_{12}}$ &${\cal P}$ & ${\cal P}_{_{12}}$ &${\cal P}$ & ${\cal P}_{_{12}}$ &${\cal P}$ & ${\cal P}_{_{12}}$ \\
\tableline
$\quad v_1$ &$-$& +&$-$&+&+&+&$-$&+&+&+\cr
$\quad v_2$  &$-$&$-$ &$-$&+&+&$-$&$-$&$-$&$+$&+\cr
$\quad v_3$  &$-$&+ &$-$&+&+&+&$-$&+&+&+\cr
$\quad {v_4}$\footnote{The phase of these kernels under ${\cal P}_{12}$ in the $+-$ and $-+$ sectors differs from Ref.~I.}  &$-$& $-$
 &$-$&+&+&$-$&$-$&$-$&$+$&+\cr
\tableline
$\quad v_5$  &$-$& $-$&$-$&$-$&+&$-$&$-$&$-$&+&$-$\cr
$\quad v_6$  &$-$&+ &$-$&$-$&+&+&$-$&+&+&$-$\cr
$\quad v_7$  &$-$&$-$ &$-$&$-$&+&$-$&$-$&$-$&+&$-$\cr
$\quad {v_8}$\footnotemark[1]  &$-$& +&$-$&$-$&+&+&$-$&+&+&$-$\cr
\tableline
$\quad v_9$       &+&+ &+&+&$-$&+&+&+&$-$&+\cr
$\quad v_{10}$  &+& +&+&+&$-$&$-$&+&$-$&$-$&+\cr
$\quad v_{11}$  &+&+ &+&+&$-$&+&+&+&$-$&+\cr
$\quad {v_{12}}$\footnotemark[1]  &+&+ &+&+&$-$&$-$&+&$-$&$-$&+\cr
\tableline
$\quad v_{13}$  &+&$-$ &+&$-$&$-$&$-$&+&$-$&$-$&$-$\cr
$\quad v_{14}$  &+& $-$&+&$-$&$-$&+&+&+&$-$&$-$\cr
$\quad v_{15}$  &+&$-$ &+&$-$&$-$&$-$&+&$-$&$-$&$-$\cr
$\quad {v_{16}}$\footnotemark[1]  &+& $-$&+&$-$&$-$&+&+&+&$-$&$-$\cr
\end{tabular}
\end{ruledtabular}
\end{minipage}
\end{table*}


\begin{table*}
\begin{minipage}{4.5in}
\caption{Symmetries of the 16 independent off-shell kernels with $\rho_1=-$ and $\rho'_1=+$ and four different combinations of $\{\rho_2, \rho'_2\}$.  In each case the parity is $\rho'_2\eta\delta_S$ and the the exchange symmetry is $\eta\delta_{_{p_0}}(-\rho_2\rho'_2\delta_S)^{(\lambda_1-\lambda_2)}$.   All phases are in terms of $\eta$; for $\pm$ read $\pm\eta$.}
\label{tab:IIIa}
\begin{ruledtabular}
\begin{tabular}{ccc|cc|cc|cc}
$\{\rho_2\rho'_2\}=$&\multicolumn{2}{c}{$++$} &\multicolumn{2}{c}{$+-$}&\multicolumn{2}{c}{$-+$}&\multicolumn{2}{c}{$--$}\cr
 name$\quad$ &${\cal P}$ & ${\cal P}_{_{12}}$ &${\cal P}$ & ${\cal P}_{_{12}}$ &${\cal P}$ & ${\cal P}_{_{12}}$ &${\cal P}$ & ${\cal P}_{_{12}}$ \\
\tableline
$v_1$ &$-$&+&+&+&$-$&+&+&+\cr
$v_2$ &$-$&$-$&+&$+$&$-$&$+$&$+$&$-$\cr
$v_3$ &$-$&+&+&+&$-$&+&+&+\cr
$v_4$ &$-$&$-$&+&$+$&$-$&$+$&$+$&$-$\cr
\tableline
$v_5$ &$-$&$-$&+&$-$&$-$&$-$&+&$-$\cr
$v_6$ &$-$&$+$&+&$-$&$-$&$-$&+&$+$\cr
$v_7$ &$-$&$-$&+&$-$&$-$&$-$&+&$-$\cr
$v_8$ &$-$&$+$&+&$-$&$-$&$-$&+&$+$\cr
\tableline
$v_9$      &+&+&$-$&+&+&+&$-$&+\cr
$v_{10}$ &+&$-$&$-$&$+$&+&$+$&$-$&$-$\cr
$v_{11}$ &+&+&$-$&+&+&+&$-$&+\cr
$v_{12}$ &+&$-$&$-$&$+$&+&$+$&$-$&$-$\cr
\tableline
$v_{13}$ &+&$-$&$-$&$-$&+&$-$&$-$&$-$\cr
$v_{14}$ &+&$+$&$-$&$-$&+&$-$&$-$&$+$\cr
$v_{15}$ &+&$-$&$-$&$-$&+&$-$&$-$&$-$\cr
$v_{16}$ &+&$+$&$-$&$-$&+&$-$&$-$&$+$\cr
\end{tabular}
\end{ruledtabular}
\end{minipage}
\end{table*}


Parity is conserved and, as discussed in Appendix~\ref{App:A}, interchange symmetry is imposed by explicitly antisymmetrizing the kernel.  Therefore, the equations can be decoupled into channels with good parity and interchange symmetry. 

Suppressing the momentum  dependence (except for the relative energy, $p_0$, of the final state pair), the partial wave expression of the antisymmetrized kernel is
\bal
&\overline{V}^{J\,\rho_1\rho_2,\rho'_1\rho'_2}_{\lambda_1\lambda_2,\lambda'_1\lambda'_2}
\nonumber\\
&\quad= \frac12\Big\{V^{J\,\rho_1\rho_2,\rho'_1\rho'_2}_{{\rm dir}\;\lambda_1\lambda_2,\lambda'_1\lambda'_2}(p_0) +\delta_I V^{J\,\rho_1\rho_2,\rho'_1\rho'_2}_{{\rm ex}\;\lambda_1\lambda_2,\lambda'_1\lambda'_2}(p_0) \Big\}
\nonumber\\
&\quad= \frac12\Big\{V^{J\,\rho_1\rho_2,\rho'_1\rho'_2}_{{\rm dir}\;\lambda_1\lambda_2,\lambda'_1\lambda'_2}(p_0) 
+\eta\,\delta_IV^{J\,\rho_2\rho_1,\rho'_1\rho'_2}_{{\rm dir}\;\lambda_2\lambda_1,\lambda'_1\lambda'_2}(-p_0)  \Big\}\, ,
\label{E37}
\end{align}
where $V_{\rm dir}$ are the Dirac matrix elements of the direct kernel defined in Eq.~(\ref{direct}) above, $\delta_I=(-1)^I$,  and $\eta=(-1)^{J-1}$.  In the last line Eq.~(\ref{A13a}) was used to express the exchange diagram in terms of the direct diagram, as illustrated in Fig.~\ref{fig:ex}.   Equation (\ref{E37}) displays the fact that the exchange of particles in the final state (equivalent to $\lambda_1\leftrightarrow \lambda_2$,  $\rho_1\leftrightarrow \rho_2$, and $p_0\to-p_0$) multiplies the kernel by the phase $\delta_I$, as expected.  

\begin{figure}
\centerline{
\mbox{
\includegraphics[width=3in]{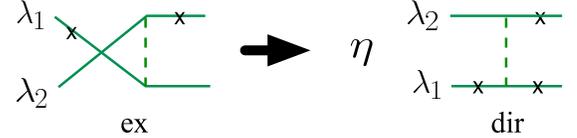}
}
}
\caption{\footnotesize\baselineskip=10pt (Color on line) Diagramatic representation of the transformation of the exchange term, with relative energy $p_0$ into a direct term with relative energy $-p_0$ (called an ``alternating'' term in Ref.~I).  The phase $\eta$ arises from the exchange operation derived in Eq.~(\ref{A13a}). }
\label{fig:ex}
\end{figure} 

An antisymmetrized kernel with good parity can be constructed from (\ref{E37}) by combining terms with different helicities.   We could define (but will not use)
\begin{equation}
\overline{\overline{V}}^{\,J\,\rho_1\rho_2,\rho'_1\rho'_2}_{\lambda_1\lambda_2,\lambda'_1\lambda'_2}
= \Big\{\overline{V}^{J\,\rho_1\rho_2,\rho'_1\rho'_2}_{\lambda_1\lambda_2,\lambda'_1\lambda'_2} +\delta_S \overline{V}^{J\,\rho_1\rho_2,\rho'_1\rho'_2}_{\lambda_1\lambda_2,-\!\lambda'_1\;-\!\lambda'_2} \Big\}. \label{E38}
\end{equation}
where $\delta_S=\delta_P\,\rho'_1\rho'_2\,\eta$.    Under parity, the phase of this kernel and the good parity state (\ref{A18}) are both $\delta_P$.

In this paper, instead of using the kernels (\ref{E38}), we  introduce the related family
\bal
&{\bf  V}^{J\,\rho_1\rho_2,\rho'_1\rho'_2}_{\lambda_1\lambda_2,\lambda'_1\lambda'_2}(\delta_{_{p_0}},\delta_S)
\nonumber\\
&\quad=
\frac{1}{2}\Big\{V^{J\,\rho_1\rho_2,\rho'_1\rho'_2}_{{\rm dir}\;\lambda_1\lambda_2,\lambda'_1\lambda'_2}(p_0) + \delta_S V^{J\,\rho_1\rho_2,\rho'_1\rho'_2}_{{\rm dir}\;\lambda_1\lambda_2,\,-\!\lambda'_1\,-\!\lambda'_2}(p_0) 
\nonumber\\
&+\delta_{_{p_0}}V^{J\,\rho_2\rho_1,\rho'_1\rho'_2}_{{\rm dir}\;\lambda_1\lambda_2,\lambda'_1\lambda'_2}(-p_0)
+\delta_{_{p_0}}\delta_S V^{J\,\rho_2\rho_1,\rho'_1\rho'_2}_{{\rm dir}\;\lambda_1\lambda_2,\,-\!\lambda'_1\,-\!\lambda'_2}(-p_0)\Big\}
, \qquad\label{A31a}
\end{align}
where $\delta_{_{p_0}}$ is the phase of ${\bf V}$ under $p_0\to -p_0$ and $\rho_1\leftrightarrow \rho_2$, and $\delta_S$ is phase introduced in Eq.~(\ref{E38}).  {\it These linear combinations are very similar to those used in Ref.~I, with the few differences identified in Table \ref{tab:III}.  The notational changes introduced in this paper simplify some of the discussion but do not change any of the numerical results found in Ref.~I\/}.   

The phase of the kernels (\ref{A31a}) under the parity transformation is still $\delta_P=\rho'_1\rho'_2\,\eta\,\delta_S$, but their exchange symmetries are more complicated.  Before looking at these symmetries, note that these kernels have the special feature that the helicities of the outgoing particles are identical in every term.  It turns out that (\ref{A31a}) are in one-to-one correspondence with (\ref{E38}) provided the phases are matched correctly.  If $\lambda_1=\lambda_2$, Eqs.~(\ref{E38}) and (\ref{A31a}) are identical (provided $\delta_I= \eta\,\delta_{_{p_0}}$).  If $\lambda_1=-\lambda_2$ (the only other possibility) we can use the parity relation (\ref{A13}) to transform each of the last two terms in (\ref{A31a}) into
\bal
V^{J\,\rho_2\rho_1,\rho'_1\rho'_2}_{{\rm dir}\,\lambda_1\lambda_2,\lambda'_1\lambda'_2}(-p_0)&=\rho_{\rm T}
V^{J\,\rho_2\rho_1,\rho'_1\rho'_2}_{{\rm dir}\,-\!\lambda_1\,-\!\lambda_2,\,-\!\lambda'_1\,-\!\lambda'_2}(-p_0)
\nonumber\\
&=\rho_{\rm T}
V^{J\,\rho_2\rho_1,\rho'_1\rho'_2}_{{\rm dir}\,\lambda_2\,\lambda_1,\,-\!\lambda'_1\,-\!\lambda'_2}(-p_0)
, \label{A30}
\end{align}
where $\rho_{\rm T}\equiv\rho_1\rho_2\rho'_1\rho'_2$, and the last line is true only because $\lambda_1\ne\lambda_2$.  This has the effect of interchanging $\lambda_1$ and $\lambda_2$, but also changes the signs of the initial helicities, and hence has the additional effect of interchanging the  the last two terms of Eq.~(\ref{A31a}).  Hence, the result will be identical to (\ref{E38}) if  $\delta_I=\eta\,\delta_{_{p_0}}\rho_{\rm T}\,\delta_S$.  The results for both equal and unequal $\lambda_i$ can therefor be combined into the compact formula $\delta_I=\eta\,\delta_{_{p_0}}(\rho_{\rm T}\,\delta_S)^{\lambda_1-\lambda_2}$.

 So the only difference between the sets of kernels (\ref{A31a}) and (\ref{E38}) are the definitions of the phases.  We found that the the set (\ref{A31a}) were more convenient to work with in practical calculations. and the preceding argument shows that the phases of these kernels are
\bea
\begin{array}{ll}
\delta_P=\rho'_1\rho'_2\,\eta\,\delta_S &\qquad {\rm parity}\cr
\delta_I=\eta\,\delta_{_{p_0}}(\rho_{\rm T}\,\delta_S)^{\lambda_1-\lambda_2} &\qquad{\rm exchange}\end{array} \label{E41}
\eea 

If particle 1 is on-shell,  so that $\rho_1=\rho'_1=+$, there are 16 independent kernels for each $J, \rho_2$ and $\rho'_2$.   Using parity and exchange symmetry, the amplitudes can always be arranged so that $\lambda_1=\lambda'_1=+1/2$, leaving the helicities of particle 2 and the phases $\delta_S$ and $\delta_{_{p_0}}$ unconstrained.   Following the notation of Ref.~I, the 16 independent amplitudes, denoted $v_i$, are defined in Table~\ref{tab:I}.  The parity and interchange symmetry of these kernels is also given in the table.

Similar symmetries exist for the kernels with one or two negative rho-spin indices, but as given in Eq.~(\ref{E41}), the negative parity of the negative energy spinors alters the relations.  The symmetries of all the kernels with $\rho_1=\rho'_1=+$ are summarized in Table \ref{tab:III}.

In some applications (not discussed in this paper) kernels (and scattering amplitudes) are needed in which {\it both\/} particles in the final state off-shell.  This requires the use of kernels with $\rho_1=-$.  The symmetries of these kernels are given in Table \ref{tab:IIIa}.  Since the parity is defined by the structure of the initial state, the parity of these potentials is identical to those with $\rho_1=+$.  The exchange symmetry, defined by the final state, is unchanged when $\lambda_1=\lambda_2$, but the use of (\ref{E41}) introduces an extra minus sign when $\lambda_1\ne\lambda_2$.

\begin{table}
\begin{minipage}{3in}
\caption{Symmetries of the four independent $NN$ scattering channels.  The isospin assigmnents are given for $J=0,1,2,\cdots$. }
\label{tab:II}
\begin{ruledtabular}
\begin{tabular}{lccc}
channel  &${\cal P}$ & ${\cal P}_{_{12}}$ & isospin  \\
\tableline
singlet (S) & $-\eta$ & $\eta$ &  $0,1,0,\cdots$\cr
triplet (T) & $-\eta$ & $-\eta$ & $1,0,1,\cdots$ \cr
coupled (C) & $\eta$ & $\eta$ & $0,1,0,\cdots$ \cr
virtual (V) & $\eta$ & $-\eta$ & \cr
\end{tabular}
\end{ruledtabular}
\end{minipage}
\end{table}


Scattering is divided into four distinct channels, depending on the symmetry under parity and interchange.  These four groups are defined in Table~\ref{tab:II}.  Singlet and triplet channels have $L=J$, and thus, for the positive $\rho$-spin sector, must have parity equal to $(-1)^J$, leading to $\delta_S=-1$.  Under particle interchange, the singlet channel with total spin equal to zero gives a phase equal to  $\eta$; requiring this to be antisymmetric leads to the isospin  assignment $I=(1-(-1)^J)/2$.  For the triplet channel the symmetry is $-\eta$ and the isospin is $I=(1+(-1)^J)/2$.  The coupled channel has $L=J\pm1$, giving $\delta_S=+1$, symmetry equal to $\eta$, and isospin assignment $I=(1-(-1)^J)/2$.  Finally, it is possible to construct an unphysical virtual coupled channel state with parity $\eta$ and interchange symmetry $-\eta$, but this state has $\delta_{_{p_0}}=-1$ in all four kernels, giving no scattering when $p_0=0$ (the on-shell condition).  While they completely decouple from physical two-body scattering, they do contribute, in principle, to three-body scattering \cite{Stadler:1997iu}.

Organizing the helicity and $\rho$-spin of the initial state in a column vector with labels $\left|\lambda'_2,\rho'_2\right>$ (by assumption, $\lambda'_1=+$ and $\rho'_1=+$)
\bea
\left|\Gamma \right>=\left( \begin{array}{c}
 \left| +,+\right>\cr
 \left| +,-\right>\cr
 \left| -,+\right>\cr
 \left| -,-\right>
\end{array}\right)\, ,
\eea
and picking potentials from Table \ref{tab:III} with the correct symmetry properties for each channel, 
the singlet  matrix is
\bea
{\bf V}_S=\left(\begin{array}{cc|cc} 
v_1^{++} & v_9^{+-}  &v^{++}_3 &v_{11}^{+-}    \cr
 v_{1}^{-+} &  v_{9}^{--} & v_{3}^{-+}  &v_{11}^{--} \cr
\tableline
 v^{++}_4 & v_{16}^{+-}  & v_2^{++} & v_{14}^{+-} \cr
 v_{8}^{-+} & v_{12}^{--}&v_{6}^{-+}& v_{10}^{--} 
\end{array}\right)\label{singlet}
\eea
the triplet matrix is 
\bea
{\bf V}_T=\left(\begin{array}{cc|cc} 
v_5^{++} & v_{13}^{+-}  &
v^{++}_7 &v_{15}^{+-}    \cr
v_{5}^{-+} &  v_{13}^{--} &
v_{7}^{-+}  &v_{15}^{--} \cr
\tableline
 v^{++}_8 & v_{12}^{+-}  & 
 v_6^{++} & v_{10}^{+-} \cr
 v_{4}^{-+} & v_{16}^{--}&
 v_{2}^{-+}& v_{14}^{--} 
\end{array}\right) \label{triplet}
\eea
and the coupled matrix is
\bea
{\bf V}_C=\left(\begin{array}{cc|cc} 
v_9^{++} & v_1^{+-}  &v^{++}_{11} &v_{3}^{+-}    \cr
 v_{9}^{-+} &  v_{1}^{--} & v_{11}^{-+}  &v_{3}^{--} \cr
\tableline
 v^{++}_{12} & v_{8}^{+-}  & v_{10}^{++} & v_{6}^{+-} \cr
 v_{16}^{-+} & v_{4}^{--}&v_{14}^{-+}& v_{2}^{--} 
\end{array}\right)\, . \label{coupled}
\eea
In this paper, the kernels $v_4, v_8, v_{12},$ and $v_{16}$ in the $\rho$-spin sectors $\{+,-\}$ and $\{-,+\}$ are defined with a different phase from Ref.~I (see the footnote to Table \ref{tab:III}).  This phase change leads to the following correspondence between this paper and Ref.~I:  in the $+-$ and $-+$ sectors {\it only\/}, substitute $v_4\leftrightarrow v_{16}$ and $v_{8}\leftrightarrow v_{12}$.  These substitutions bring equations (\ref{singlet}), (\ref{triplet}), and (\ref{coupled}) into agreement with equations (2.106), (2.107), and (2.108) of Ref.~I.  In conclusion: {\it the matrices given in this paper are identical to the matrices given in Ref.~1; only the names of some of the elements have changed.\/}

The matrices ${\bf V}^{\rm off}$ that couple an on-shell initial state (with $\rho'_1=+$) to a final state when $\rho_1=-$ (present only if {\it both\/} particles in the final state are off-shell) can be constructed from the symmetries in Table \ref{tab:IIIa}.  The singlet kernel is
\bea
{\bf V}^{\rm off}_S=\left(\begin{array}{cc|cc} 
v_1^{++} & v_9^{+-}  &v^{++}_3 &v_{11}^{+-}    \cr
 v_{1}^{-+} &  v_{9}^{--} & v_{3}^{-+}  &v_{11}^{--} \cr
\tableline
 v^{++}_8 & v_{12}^{+-}  & v_6^{++} & v_{10}^{+-} \cr
 v_{4}^{-+} & v_{16}^{--}&v_{2}^{-+}& v_{14}^{--} 
\end{array}\right)\label{singletoff}\, ,
\eea
the triplet kernel is
\bea
{\bf V}^{\rm off}_T=\left(\begin{array}{cc|cc} 
v_5^{++} & v_{13}^{+-}  &
v^{++}_7 &v_{15}^{+-}    \cr
v_{5}^{-+} &  v_{13}^{--} &
v_{7}^{-+}  &v_{15}^{--} \cr
\tableline
 v^{++}_4 & v_{16}^{+-}  & 
 v_2^{++} & v_{14}^{+-} \cr
 v_{8}^{-+} & v_{12}^{--}&
 v_{6}^{-+}& v_{10}^{--} 
\end{array}\right) \label{tripletoff}
\eea
and the coupled matrix is
\bea
{\bf V}^{\rm off}_C=\left(\begin{array}{cc|cc} 
v_9^{++} & v_1^{+-}  &v^{++}_{11} &v_{3}^{+-}    \cr
 v_{9}^{-+} &  v_{1}^{--} & v_{11}^{-+}  &v_{3}^{--} \cr
\tableline
 v^{++}_{16} & v_{4}^{+-}  & v_{14}^{++} & v_{2}^{+-} \cr
 v_{12}^{-+} & v_{8}^{--}&v_{10}^{-+}& v_{6}^{--} 
\end{array}\right)\, . \label{coupledoff}
\eea

With this notation the partial wave equations (\ref{E25}) with parity and particle interchange symmetry as defined in Table \ref{tab:II} can be written
\bea
{\bf M}_I={\bf V}_I  -\int_0^\infty \frac{k^2dk}{(2\pi)^3} {\bf V}_I \; {\bf G}\;  {\bf M}_I \label{E49}
\eea
where $I=\{S, T, C\}$ and the propagator matrix is
\bea
{\bf G} =\left(\begin{array}{cc|cc} 
G^{+} & 0  & 0& 0   \cr
0 &  G^{-} & 0 &0 \cr
\tableline
 0 & 0 & G^+& 0\cr
 0&0 &0 &G^-
\end{array}\right)
\eea
with $G^\pm$ defined in Eq.~(\ref{Gprop}).  These are the equations that are solved numerically.


\end{document}